\documentclass[pre,notitlepage, twocolumn]{revtex4-1}
\usepackage{amsmath}
\usepackage{amssymb}
\usepackage{graphicx}
\usepackage{enumerate}
\usepackage{color}
\usepackage[colorlinks = true,
            linkcolor = blue,
            urlcolor  = blue,
            citecolor = blue,
            anchorcolor = blue]{hyperref}
\usepackage{subcaption}
\usepackage{physics}
\usepackage{placeins}
\usepackage{tikz}
\usepackage{booktabs, tabularx}

\def\doi#1#2{\href{https://doi.org/#2}{#1}}

\newcommand{\pd}[2]{\frac{\partial #1}{\partial #2}}
\newcommand{\pds}[2]{\frac{\partial^2 #1}{\partial #2^2}}
\newcommand{\eq}[1]{\begin{equation} #1 \end{equation}}

\newcommand{\nn}{\nonumber}
\newcommand{\psis}{\psi_\text{s}}
\newcommand{\psii}{\psi_\text{i}}
\newcommand{\us}{u_\text{s}}

\newcolumntype{Y}{>{\centering\arraybackslash}X}

\AtBeginDocument{
\heavyrulewidth=.08em
\lightrulewidth=.05em
\cmidrulewidth=.03em
\belowrulesep=.65ex
\belowbottomsep=0pt
\aboverulesep=.4ex
\abovetopsep=0pt
\cmidrulesep=\doublerulesep
\cmidrulekern=.5em
\defaultaddspace=.5em
}

\begin{document}

\title{Multisector parabolic-equation approach to compute acoustic scattering by noncanonically shaped impenetrable objects}

\author{Adith Ramamurti}
\email[]{adith.ramamurti@nrl.navy.mil}
\author{David C. Calvo}
\email[]{david.calvo@nrl.navy.mil}

\affiliation{Acoustics Division, Code 7165, U.S. Naval Research Laboratory, Washington, DC 20375, USA}

\date{December 18, 2019}

\begin{abstract}
Parabolic equation (PE) methods have long been used to efficiently and accurately model wave phenomena described by hyperbolic partial differential equations. 
A lesser-known but powerful application of parabolic equation methods is to the target scattering problem. 
In this paper, we use noncanonically shaped objects to establish the limits of applicability of the traditional approach, and introduce wide-angle and multiple-scattering approaches to allow accurate treatment of concave scatterers. 
The PE calculations are benchmarked against finite-element results, with good agreement obtained for convex scatterers in the traditional approach, and for concave scatterers with our modified approach.
We demonstrate that the PE-based method is significantly more computationally efficient than the finite-element method at higher frequencies where objects are several or more wavelengths long. 

\end{abstract}
\maketitle

\section{Introduction}

Parabolic equation (PE) methods are a powerful technique to model long-range acoustic propagation in complex environments \cite{jensen2000computational,collins200097}.  
While, historically, wave propagation has been the primary application of parabolic equation methods in acoustics, a PE technique was demonstrated by Levy and Zaporozhets for mid- to high-frequency target scattering calculations \cite{zaporozhets1996modelling,levy1998target,zaporozhets1999application,zaporozhets1999bistatic,levy2000parabolic}. 
The primary advantages of this approach relative to finite-element methods are computational efficiency  --- particularly for higher frequencies and limited angular sectors in the far field --- and ease of implementation \cite{levy2000parabolicbook, jensen2000computational}. 

Parabolic equations have been applied to acoustic target scattering in two ways: through direct computation, where the scattered field is marched across the object in different directions, with the incident field acting as a source on the boundary of the scatterer \cite{levy1998target,zaporozhets1999application,zaporozhets1999bistatic,levy2000parabolic}; and the so-called on-surface radiation condition, which computes the scattered pressure field or its normal derivatives on the surface of the object to solve for the far-field directivity \cite{kriegsmann1987new, calvo2003higher, calvo2004wide}.
Acoustic target scattering calculations using the former approach were only benchmarked against objects with easily obtainable analytic solutions, and issues relating to wide-angle and multiple-scattering phenomena limited the maximum concavity of objects to which either method could be applied \cite{levy1998target,calvo2004wide}. 

The primary goal of this work is to further benchmark the direct-computation (which we will call the multisector PE) algorithm against now-available finite-element method (FEM) calculations to discern its accuracy and limits of efficacy, as well as implement improvements that make it applicable to a larger variety of objects, including highly concave scatterers. 
To make the latter improvement, we take inspiration from both wide-angle \cite{saad1986new,collins1991higher} as well as iterative and multiple-scattering \cite{collins1992two,mills2000two,lingevitch2002two,lingevitch2010parabolic} approaches to propagation using parabolic equations.

In Sec. \ref{sec:PEmethod}, we overview how the parabolic equation can be used to compute the target strength of a scatterer. 
In Sec. \ref{sec:verapp}, we benchmark the narrow-angle PE method against FEM calculations for a variety of convex objects and boundary conditions. 
Finally, in Sec. \ref{sec:wams}, we probe where the narrow-angle PE formulation breaks down and propose and demonstrate wide-angle and multiple-scattering approaches that make possible target scattering calculations for concave scatterers. 

\section{Parabolic equations and scattering}\label{sec:PEmethod}

The one-way two-dimensional parabolic equation describing acoustic waves propagating in the paraxial direction $x$ is
\eq{
\pd{u}{x} = -i k (1 - Q) u
\,, \label{eq:pe}
} 
where $u= \psi e^{-ikx}$; $\psi$ is the pressure field;
\eq{
\begin{split}
Q &= \sqrt{\frac{1}{k^2} \pds{}{z}+n^2} \equiv \sqrt{1+q}\,; \\
q &= \frac{1}{k^2} \pds{}{z}+n^2 - 1\,;
\end{split}
\nn
}
$k$ the reference wave number; and $n$ the index of refraction \cite{levy2000parabolic}. 
We assume the pressure field has standard $\exp{-i\omega t}$ time dependence. 
For simplicity and clarity, the index of refraction will be taken to be unity in this work, although, in practice, the ability to propagate the scattered field into a medium with a slowly varying index of refraction profile is a powerful advantage of the method.

\begin{figure}
    \begin{subfigure}[b]{.5\linewidth}
        \centering
\begin{tikzpicture}
\draw[thick, ->] (0,1) -- (1,1) ;
\draw (.3,1.2) -- (.3,0.8) ;
\draw (.5,1.2) -- (.5,0.8) ;
\draw (.7,1.2) -- (.7,0.8) ;
\node at (.5,1.5) {\fontsize{7}{7}\selectfont$\psii(x',z')$};
\draw[->] (2,1) -- (2,2) ;
\draw[->] (2,1) -- (3,1) ;
\draw[->] (.25,-.25) -- (1,-.25);
\draw[->] (.25,-.25) -- (.25,.5);
\node at (1.1,-.45) {$x'$};
\node at (.05,.6) {$z'$};
\node at (3.1,0.9) {$x$};
\node at (1.8,2.1) {$z$};
\draw (2,1) ellipse (0.6cm and 0.3cm);
\draw (2,1) -- (4,1.4) ;
\draw (2,1) -- (4,0.6) ;
\draw[<-] (3.5,1) arc (0:-25:2);
\node at (3.2,0) {\fontsize{6}{6}\selectfont Paraxial cone};
\end{tikzpicture}
    \caption{}
        \label{subfig:left}
    \end{subfigure}%
    \begin{subfigure}[b]{.5\linewidth}
        \centering
\begin{tikzpicture}
\draw[thick, ->, rotate around = {-30:(2,1)}] (0,1) -- (1,1) ;
\draw[rotate around = {-30:(2,1)}] (.3,1.2) -- (.3,0.8) ;
\draw[rotate around = {-30:(2,1)}] (.5,1.2) -- (.5,0.8) ;
\draw[rotate around = {-30:(2,1)}] (.7,1.2) -- (.7,0.8) ;
\node[rotate around = {-30:(2,1)}]  at (1.2,1) {\fontsize{7}{7}\selectfont$\psii(x',z')$};
\draw[->] (2,1) -- (2,2);
\draw[->] (2,1) -- (3,1);
\node at (3.1,0.9) {$x$};
\node at (1.8,2.1) {$z$};
\node[->,rotate around = {-30:(2,1)}] at (.8,-1) {$x'$};
\node[->,rotate around = {-30:(2,1)}] at (.3,.5) {$z'$};
\draw[->,rotate around = {-30:(2,1)}] (.25,-.25) -- (1,-.25) ;
\draw[->,rotate around = {-30:(2,1)}] (.25,-.25) -- (.25,.5);
\draw[rotate around = {-30:(2,1)}] (2,1) ellipse (0.6cm and 0.3cm);
\draw (2,1) -- (4,1.4) ;
\draw (2,1) -- (4,0.6) ;
\draw[<-] (3.5,1) arc (0:-25:2);
\node at (3.2,0) {\fontsize{6}{6}\selectfont Paraxial cone};
\end{tikzpicture}
        \caption{}
        \label{subfig:right}
    \end{subfigure}
    \caption{Schematic of coordinate systems for the multisector parabolic equation method. Subfigures show the cases where the scattered paraxial direction $x$ is at an angle of (a) $\phi=0$ and (b) $\phi=\pi/6$ with respect to the incident direction $x'$. The marching occurs in the $xz$-plane (defined by the paraxial direction $x$), with the object and incident wave defined in the $x'z'$-plane. The paraxial cone designates the angular range around the paraxial direction for which the PE is valid.}
    \label{fig:coords}
\end{figure}
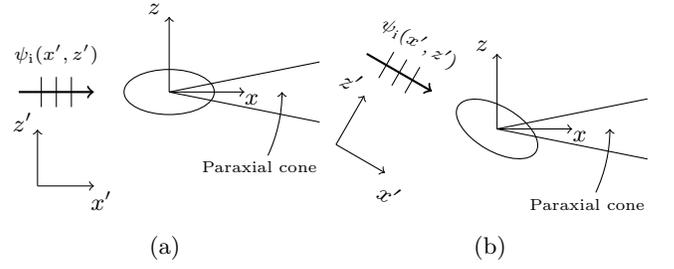

The total field $\psi$ can be decomposed into its incident $\psii$ and scattered $\psis$ components. 
The PE-scattering method solves for the scattered field, using the incident field as a source on the boundary of the object. 
We will primarily be working with a reduced scattered field, which varies slowly with space, $\us = \psis e^{-ikx}$, where the paraxial direction $x$ is {\em independent of the direction of propagation of the incident wave}. 
A schematic detailing the relevant coordinate systems is shown in Fig.~\ref{fig:coords}.
The parabolic equation for the forward-scattered field is identical to that of the total field,
\eq{
\pd{u_\text{s}}{x} = -i k (1 - Q) u_\text{s}\,. \label{eq:pe}
} 

A general form of the boundary condition on the object is given by
\eq{
\alpha \pd{\psi}{\vec{n}} + \beta \psi = 0\,,
}
where $\alpha$ and $\beta$ are free parameters and $\vec{n}$ is the vector normal to the boundary of the object; $(\alpha=0,\,\beta=1)$ and $(\alpha=1,\,\beta=0)$ correspond to objects with soft (pressure release) and hard (rigid) boundaries, respectively. 
In terms of the incident and scattered fields, we have
\eq{
\alpha \pd{\psis}{\vec{n}} + \beta \psis = - \alpha \pd{\psii}{\vec{n}} - \beta \psii \,. \nn
}
Rewriting the boundary condition for the $\us$ field,
\eq{
\begin{split}
\alpha n_x \left(\pd{\us}{x} + ik \us \right) + \alpha n_z \pd{\us}{z}  + \beta \us  \\ = - \alpha e^{-ikx}\pd{\psii}{\vec{n}}- \beta e^{-ikx}\psii \,,
\end{split} \label{eq:bc}
}
where $n_x,n_z$ are the components of the normal vector to the object.

To implement this boundary condition in the parabolic equation formulation, we first must make an approximation for the operator $Q$ in Eq. (\ref{eq:pe}). 
The simplest approximation is to simply Taylor expand the square root in $q$ to first order: $Q \approx 1+ q/2$. 
This yields the well-known narrow-angle formulation of the parabolic equation (with index of refraction $n=1$), 
\eq{
\pd{\us}{x} =  \frac{i}{2k}\pds{\us}{z} \,. \label{eq:narrowpe}
}
The narrow-angle parabolic equation is valid in a cone of opening angle $\sim$$\pi/12$ around the paraxial direction \cite{ jensen2000computational}, shown schematically in Fig.~\ref{fig:coords}.
Substituting the right-hand side of Eq. (\ref{eq:narrowpe}) for the $x$ derivative of Eq. (\ref{eq:bc}) yields the boundary condition
\eq{
\begin{split}
\frac{i\alpha n_x}{2k}\pds{\us}{z} + \alpha n_z \pd{\us}{z}  + (\alpha n_x ik + \beta) \us  \\= - \alpha e^{-ikx}\pd{\psii}{\vec{n}}- \beta e^{-ikx}\psii \,,
\end{split} \label{eq:pebc}
}
which has no range derivative dependence.

Numerical solutions using the PE-scattering method are implemented via a finite-difference algorithm on a Cartesian grid. 
The scatterer is discretized in a stair-step manner, and the field is marched in different paraxial directions (multiple sectors) relative to the scatterer using the parabolic equation, with the scattered field sourced by the appropriate boundary conditions as per Eq. (\ref{eq:pebc}). 
On the boundary of the scatterer, we use one-sided first-order finite-difference approximations; second-order approximations, especially in three dimensions, induce instabilities.

The formal solution for the parabolic equation above is
\eq{
u(x+\delta x,z) = \exp(-ik\delta x)\exp(ik\delta x \sqrt{Q})u(x,z)\,.
}
In general, the operator $Q$ or the solution itself can be better approximated using Pad\'{e} approximants, rather than a first-order Taylor expansion; the parabolic equation is then applicable in a wider angular range (dependent on the degree of the approximation used) around the paraxial direction. 

The discretized form of the solution up to a second-order Pad\'{e} approximant of the exponential (detailed in Ref.~\cite{saad1986new}) can be written in the form
\eq{
\begin{split}
u^m &+ \frac{b_0}{k^2} Z u^{m}+\frac{b_1}{k^4} Z^2 u^{m} \\ &= u^{m-1} + \frac{a_0}{k^2} Z u^{m-1} +\frac{a_1}{k^4} Z^2 u^{m-1} \,, 
\end{split}
}
where $Z$ is the matrix operator corresponding to the discretized second derivative $\partial^2/\partial z^2$, and $m$ designates the marching step or index in the $x$ direction with step size $\Delta x$.  
Values of coefficients are in Table \ref{tab:padecoeffs}.  
In this work, the second derivative is discretized as
\eq{
\pds{u^m_j}{z}  = \frac{u^m_{j-1} - 2 u^m_j + u^m_{j+1}}{\Delta z^2} \,,  \nn
}
where $j$ is the index in the $z$ coordinate.

\begin{table}[h!]
\caption{Coefficients for various Pad\'{e} approximations of order $(m,n)$ of the square root of the PE. $\Delta \equiv ik \Delta x$.}
\begin{tabularx}{\columnwidth}{l Y Y Y}
\toprule
	Coeff. & (2,2) \cite{saad1986new} &  (2,1) \cite{zaporozhets1996modelling} & (1,0)\\ \colrule
	        $a_0$ & $\frac{3+\Delta}{4}$ & $\frac{\Delta^2+3\Delta+3}{6(\Delta+1)}$ & 0\\  
                $a_1$ & $\frac{\Delta^2 + 6\Delta +3}{48}$ & 0 & 0 \\ 
                $b_0$  & $\frac{3 - \Delta}{4}$ & $\frac{3 - 2\Delta^2}{6(\Delta+1)}$ & $-\frac{\Delta}{2}$ \\  
                $b_1$ & $\frac{\Delta^2 - 6\Delta+3}{48}$ & $\frac{\Delta(\Delta^2-3)}{24(\Delta+1)}$ & 0 \\ \botrule 
\end{tabularx}

\label{tab:padecoeffs}
\end{table}

In Section \ref{sec:verapp}, we only use the narrow-angle [i.e., Pad\'{e}-(1,0)] formulation, as using wide-angle formulations on the boundary of the scatterer result in spurious oscillations; we will return to an implementation of wide-angle approximations later in the work. 

In two dimensions, the discretized Pad\'{e}-(1,0) approximation with the above discretization of the second derivative --- traditionally known as the backward-time centered-space method when used in finite-difference time-domain simulations --- gives a system of equations represented by a tridiagonal banded matrix at each range step, while in three dimensions, one has a sparse matrix with five nonzero diagonals. 
These systems can typically be solved very quickly with modern sparse matrix direct solvers. 
When using higher-degree Pad\'{e} approximations, the sparsity structure of the matrix becomes more complex, and an iterative solver is most efficient.

The target strength of an object in the far-field can be calculated from the near-field pressure just beyond the scatterer. 
For an incident plane wave of unit amplitude and reference length 1 m, where the ratio of reflected to incident intensities is given by $I_r/I_i = \sigma/4\pi$, with $\sigma$ the differential scattering cross section \cite{urick1967principles, levy1998target},
\eq{
\text{TS}(\phi) = 10 \log \left[ \frac{k \cos^2\phi}{2 \pi} \left\lvert  \int_{-\infty}^\infty  dz' \psi_s(x',z') e^{-ik\sin\phi z'} \right\rvert^2 \right]\,.
} 
We note that due to the reference length, target strength is valid for $ka \gg 1$, i.e., when the object size $a$ (in meters) is much larger than the wavelength of the incident plane wave. 
We also note that this two-dimensional target strength expression assumes global cylindrical spreading, and as such is used as a computational test for benchmarking the PE algorithm.
The angular range of validity of the target strength calculation is $\pm \pi/12$ for the narrow-angle formulation of the PE, and thus, in two-dimensions, 12 runs in different paraxial directions are necessary to characterize the full angular spectrum of an asymmetric object.

All the discussion above is identical in three dimensions, with
\eq{
Q = \sqrt{\frac{1}{k^2} \pds{}{y} + \frac{1}{k^2} \pds{}{z} +n^2} \,. \nn
}
 The narrow-angle parabolic equation is then (with index of refraction $n=1$)
\eq{
\pd{\us}{x} = \frac{i}{2k}\left(\pds{\us}{y} + \pds{\us}{z}\right)\,,
}
with boundary conditions given by
\eq{
\begin{split}
\frac{i\alpha n_x}{2k}\bigg(\pds{\us}{z}  +& \pds{\us}{y}\bigg) \\+ \alpha \bigg(n_z \pd{\us}{z} &+  n_y \pd{\us}{y}\bigg)  + (\alpha n_x ik + \beta) \us \\&= - \alpha e^{-ikx}\pd{\psii}{\vec{n}}- \beta e^{-ikx}\psii \,.
\end{split}
}
The target strength of an object in three dimensions (3D) for an incident plane wave of unit amplitude with reference length 1 m is
\eq{
\begin{split}
\text{TS}(\theta,\phi) = 10 \log \Bigg[ \frac{k^2 \cos^2\theta}{4 \pi^2} \Bigg\lvert  \int_{-\infty}^\infty \int_{-\infty}^\infty  dy' dz'  \psi_s(x',y',z') \\ \times e^{-ik\sin\theta (y'\cos\phi +z'\sin\phi)} \Bigg\rvert^2 \Bigg]\,.
\end{split}
}
Once again, as with the 2D case, the target strength is a valid function of $ka$ when $ka\gg1$.

\section{Verification for convex scatterers}\label{sec:verapp}

To verify the method, we will examine a variety of convex scatterers in two and three dimensions and compare the PE target strength calculations to the finite-element method results computed using COMSOL Multiphysics\textsuperscript{\tiny\textregistered}
 \cite{comsol}.  
For all of these simulations, we consider an incident plane wave of unit amplitude, with  --- unless noted otherwise --- sound speed $c_0 = 1500 \text{ m/s}$ and frequency $f = 1500 \text{~Hz}$, corresponding to wavelength $\lambda = 1$ m and wave number $k = 2 \pi$ m$^{-1}$. 
The density of the medium is taken to be $\rho = 1000 $ kg/m$^3$, which plays a role when scattering from objects with impedance boundary conditions.
As stated above, we take the index of refraction to be unity, $n=1$. 
In all PE simulations in this work, the grid spacing is $\lambda/20$ in the paraxial (marching) direction and $\lambda/10$ in the transverse direction(s), while FEM simulations have maximum element size of $\lambda/6$.

\begin{figure}[h!]
\begin{subfigure}[b]{.38\textwidth}
\includegraphics[width=\textwidth]{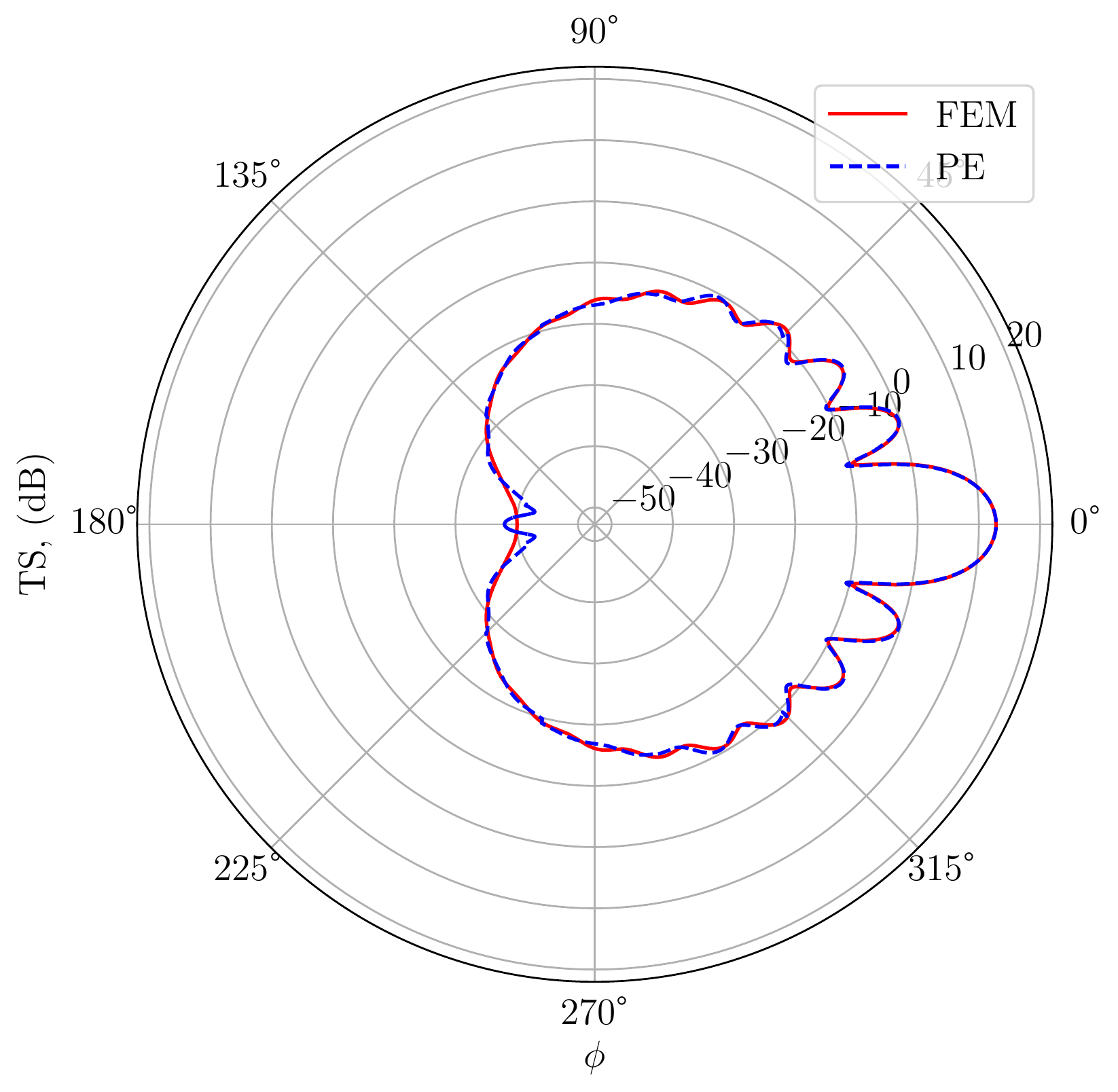}
\caption{}
\end{subfigure}\\
\begin{subfigure}[b]{.38\textwidth}
\includegraphics[width=\textwidth]{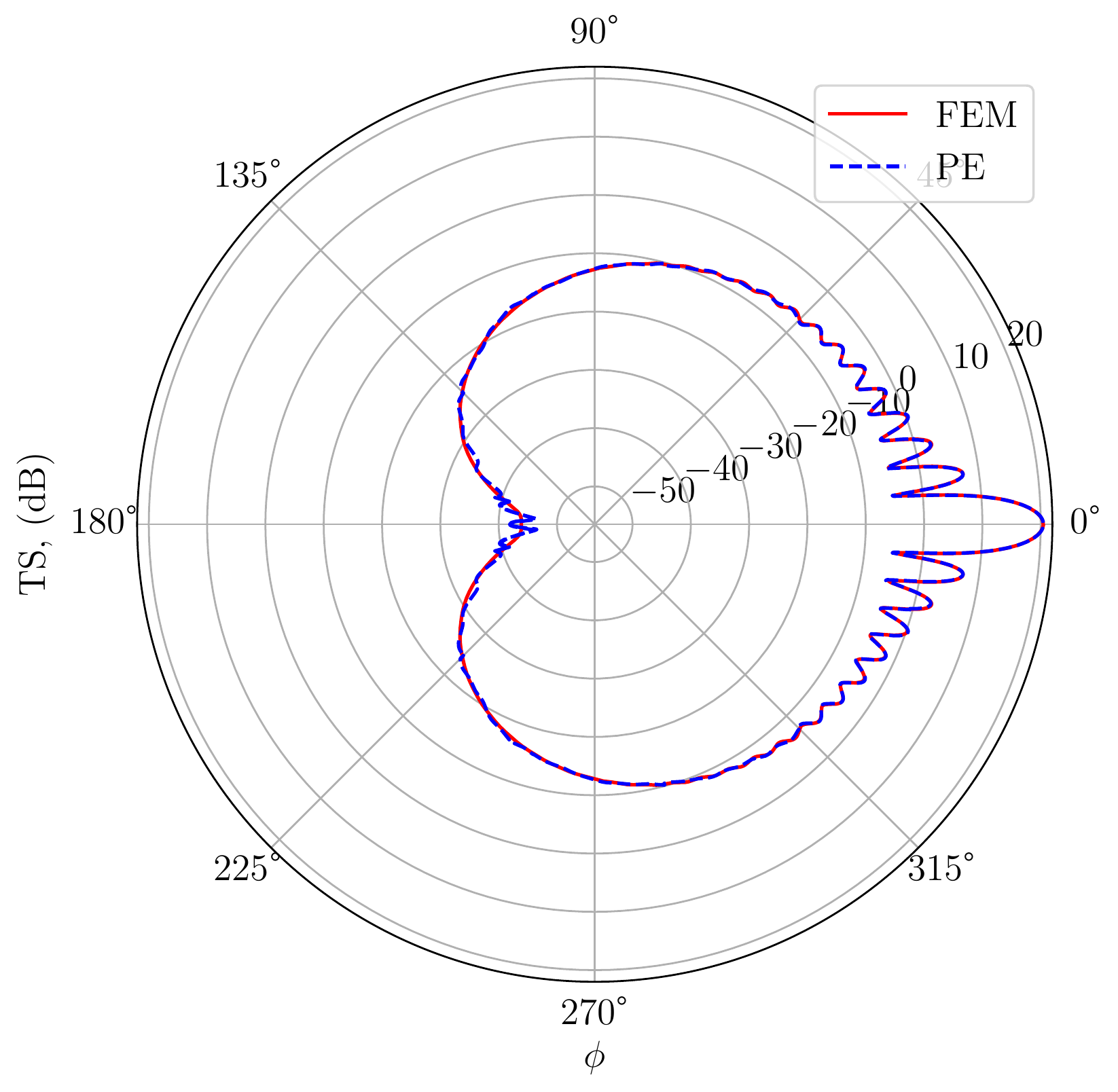}
\caption{}
\end{subfigure} 
\caption{Two-dimensional target strength a circle with impedance [as defined in Eq. (\ref{eq:pebc}) with $\alpha = 1$ and $\beta/ik = 1$] boundary conditions for an incident plane wave for (a) $ka = 4 \pi$ and (b) $ka = 10 \pi$. Dashed blue lines are from the multisector PE method, and solid red are finite-element results.}\label{fig:icircle1}
\end{figure}

\begin{figure}
\begin{subfigure}[b]{.38\textwidth}
\includegraphics[width=\textwidth]{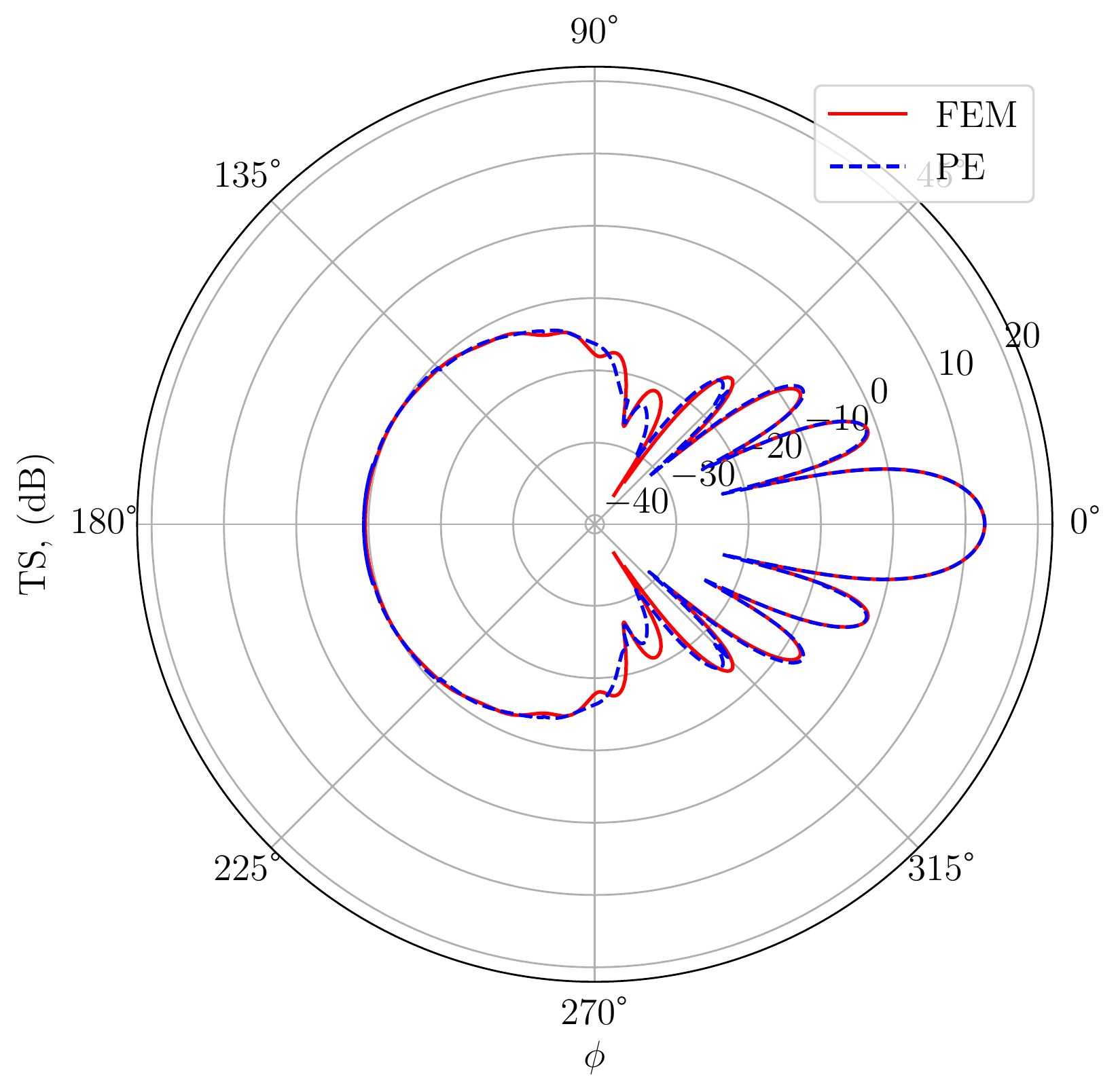}
\caption{}
\end{subfigure}\\
\begin{subfigure}[b]{.38\textwidth}
\includegraphics[width=\textwidth]{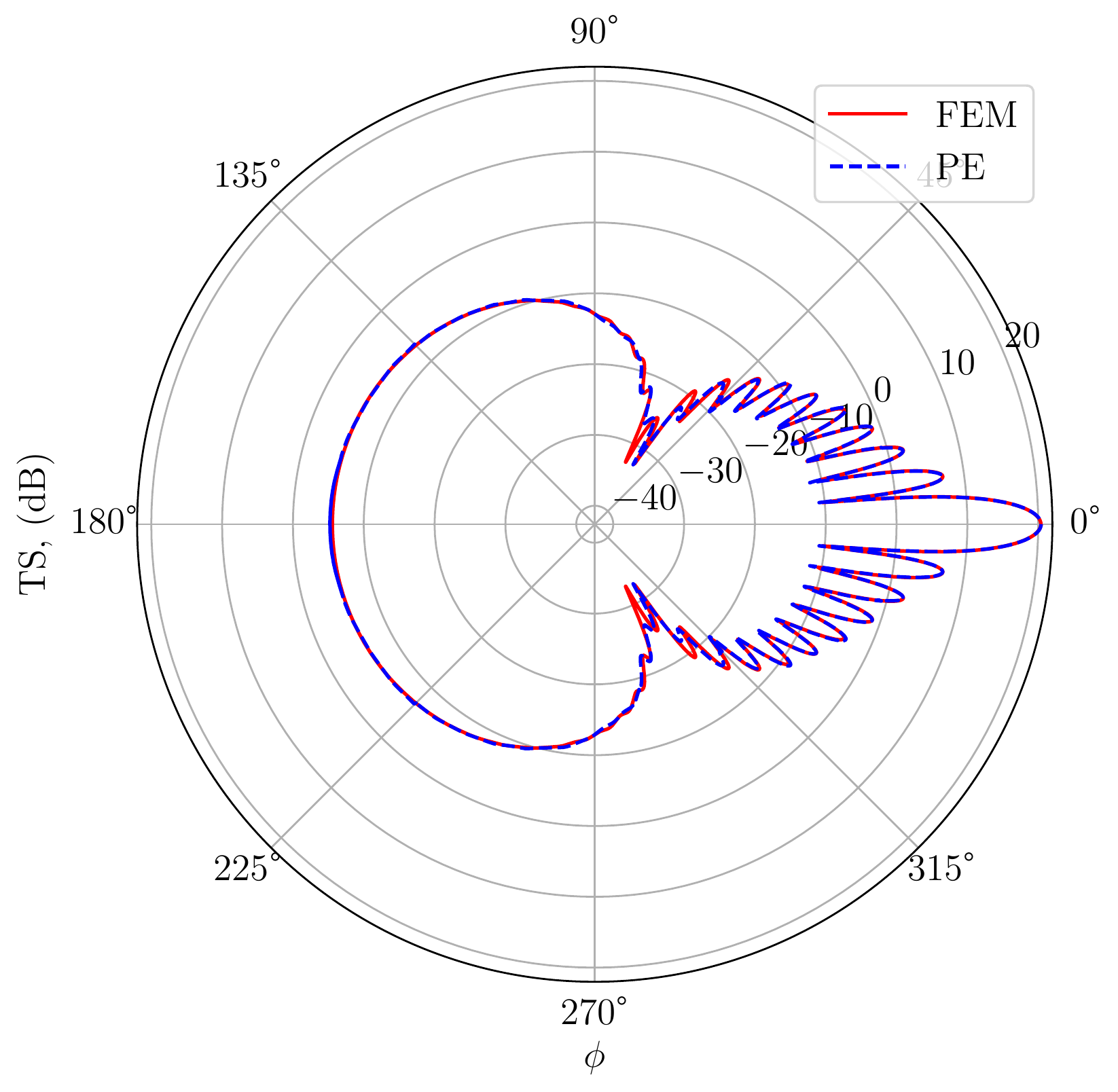}
\caption{}
\end{subfigure}
\caption{Two-dimensional target strength a circle with impedance [as defined in Eq. (\ref{eq:pebc}) with $\alpha = 1$ and $\beta/ik = 0.5$] boundary conditions  for an incident plane wave for (a) $ka = 4 \pi$ and (b) $ka = 10 \pi$. Dashed blue lines are from the multisector PE method, and solid red are finite-element results.}\label{fig:icircle05}
\end{figure}

We begin by expanding on the results presented in Ref.~\cite{levy1998target}. 
In that work, the results from which we have replicated in Appendix \ref{app:repl}, verification was only presented for soft and hard boundary conditions.
These boundary conditions can be considered as the extreme cases; most realistic objects will have boundary conditions with nonzero values for both the wave-field and its normal derivative, which correspond to impedances smaller than 1. 
The results for impedance boundary conditions for a circle are shown in Figs.~\ref{fig:icircle1}~and~\ref{fig:icircle05}. 
The two cases studied are for $(\alpha, \beta) = (1,ik), (1,ik/2)$. The $\beta/ik = 1$ case corresponds to the impedance of an object of density $\rho=1000$ kg/m$^3$ and sound speed $c_0 = 1500$ m/s ($\rho c = 1.5\times 10^6$ kg/m$^2$s), which {\em mimics} an ideally penetrable object, with the backscattered field close to zero. 
The second case, $\beta/ik = 0.5$, corresponds to the impedance of a material with $\rho=1000$ kg/m$^3$ and $c_0 = 3000$ m/s ($\rho c = 3 \times 10^6$ kg/m$^2$s). 
Once again, there is excellent agreement between the PE and FEM solutions.

\FloatBarrier

\begin{figure}[h!]
\begin{subfigure}[b]{.32\textwidth}
\includegraphics[width=\textwidth]{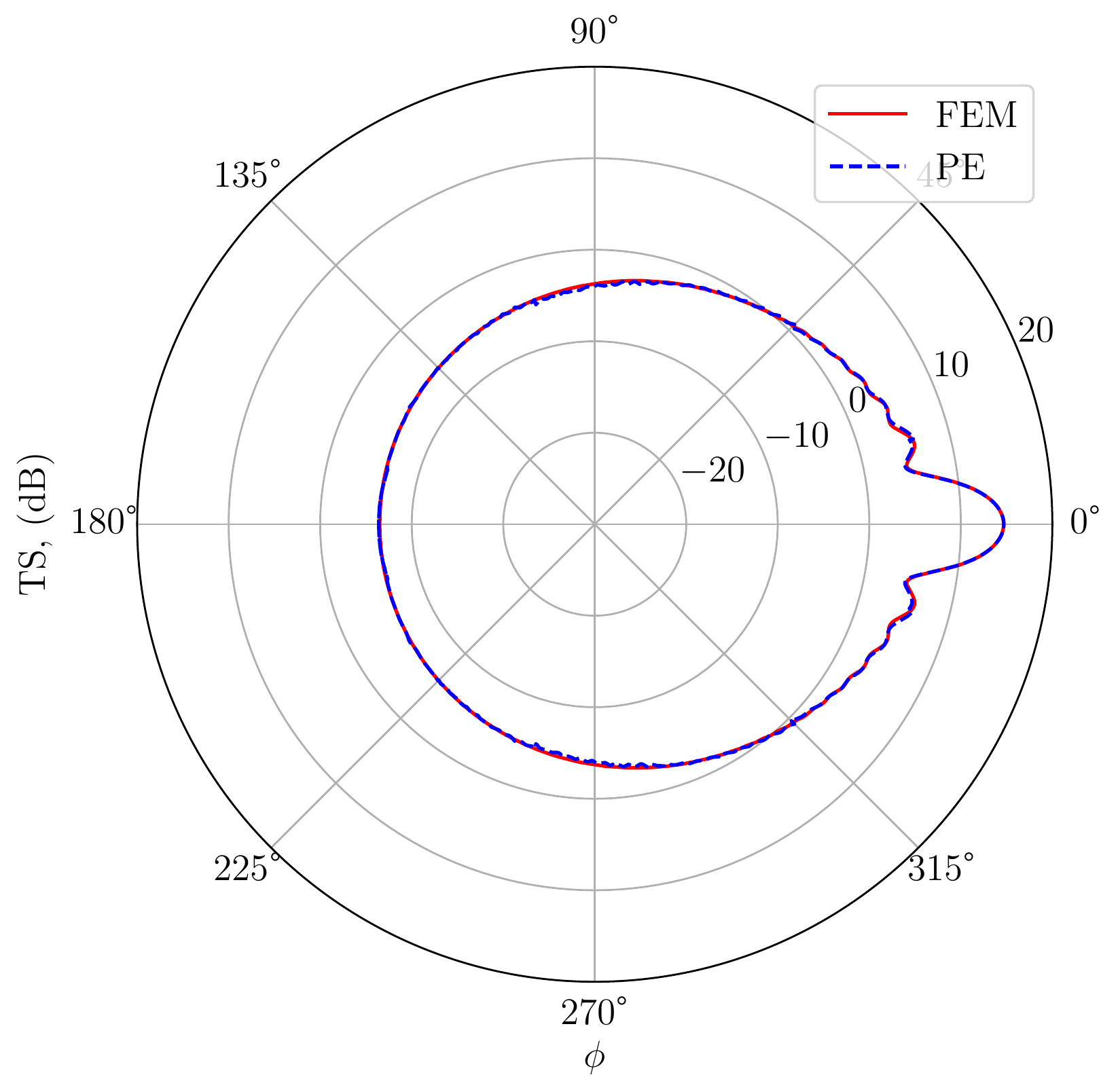}
\caption{}
\end{subfigure}
\begin{subfigure}[b]{.32\textwidth}
\includegraphics[width=\textwidth]{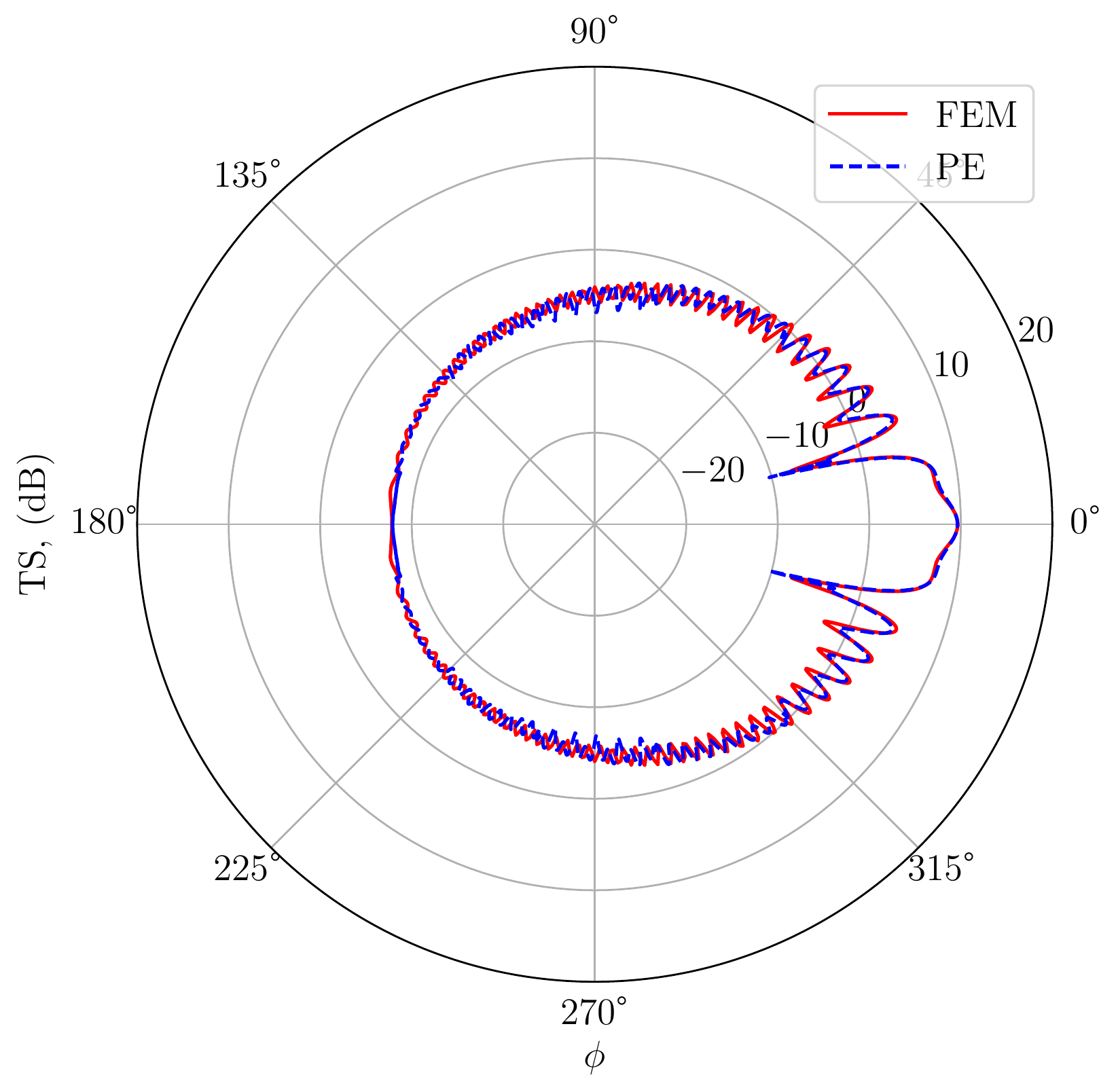}
\caption{}
\end{subfigure}
\begin{subfigure}[b]{.32\textwidth}
\includegraphics[width=\textwidth]{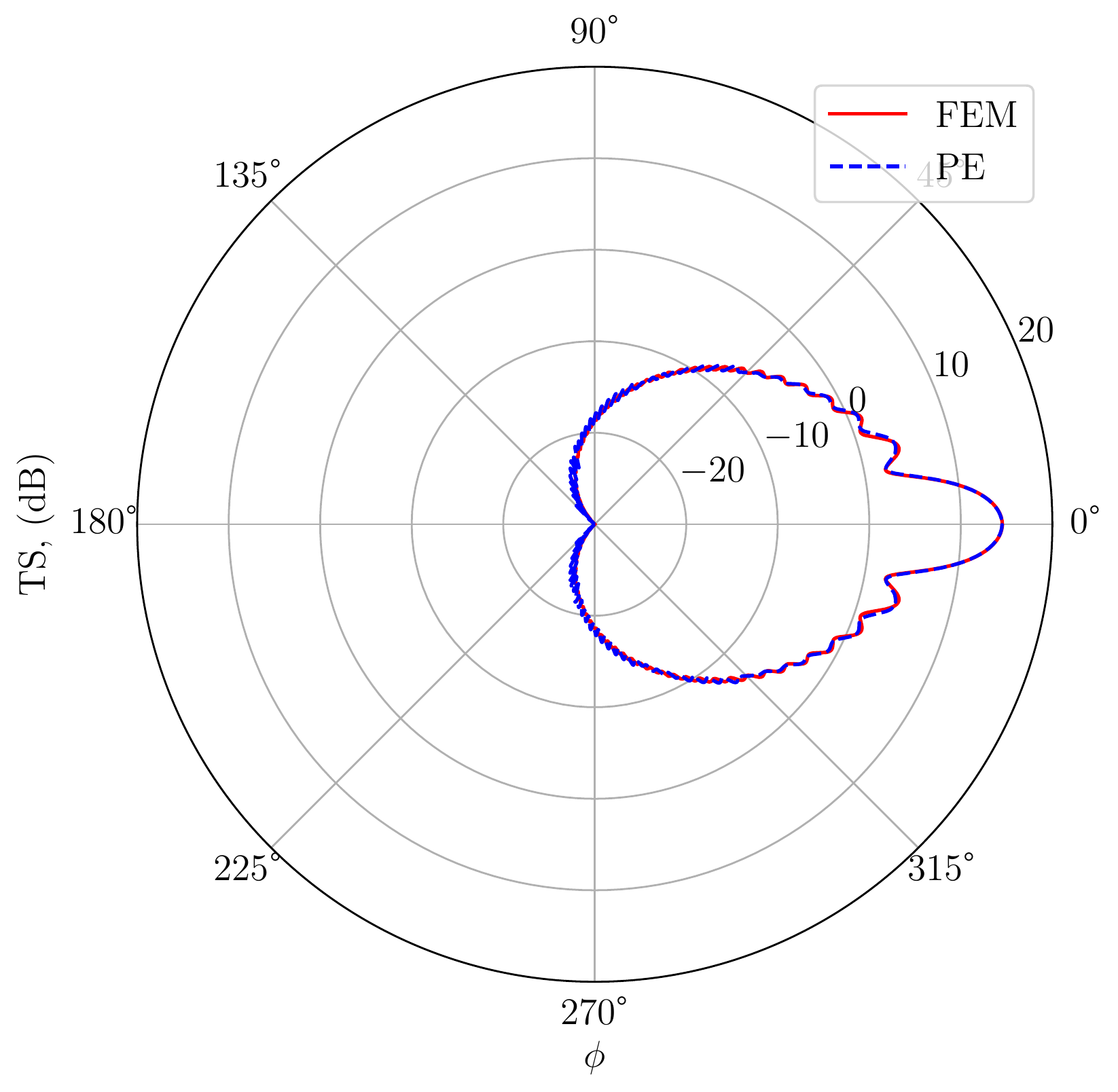}
\caption{}
\end{subfigure}
\caption{Two-dimensional target strength of an ellipse with $a_x=10$ m, $a_z=2$ m ($ka_x = 20\pi$)  with (a) soft, (b) hard, and (c) impedance [as defined in Eq. (\ref{eq:pebc}) with $\alpha = 1$ and $\beta = ik$]  boundary conditions for an incident plane wave. Dashed blue lines are from the multisector PE method, and solid red are finite-element results.}\label{fig:ellipse102TS}
\end{figure}

Following the promising results above for simple 2D objects, we consider slightly more irregular objects. 
Figure~\ref{fig:ellipse102TS} shows the target strength of an ellipse with $a_x = 10$ ($ka_x = 20\pi$) with $a_z = 2$ for end-on incidence of plane waves of unit amplitude. 
Subfigures are for soft, hard and impedance ($\alpha = 1,\beta = ik$) boundary conditions. 
These results are comparable to those presented in Refs.~\cite{calvo2003higher,calvo2004wide} for the wide-angle on-surface radiation condition; both methods provide similar accuracy. 

\begin{figure}[h!]
\begin{subfigure}[b]{.32\textwidth}
\includegraphics[width=\textwidth]{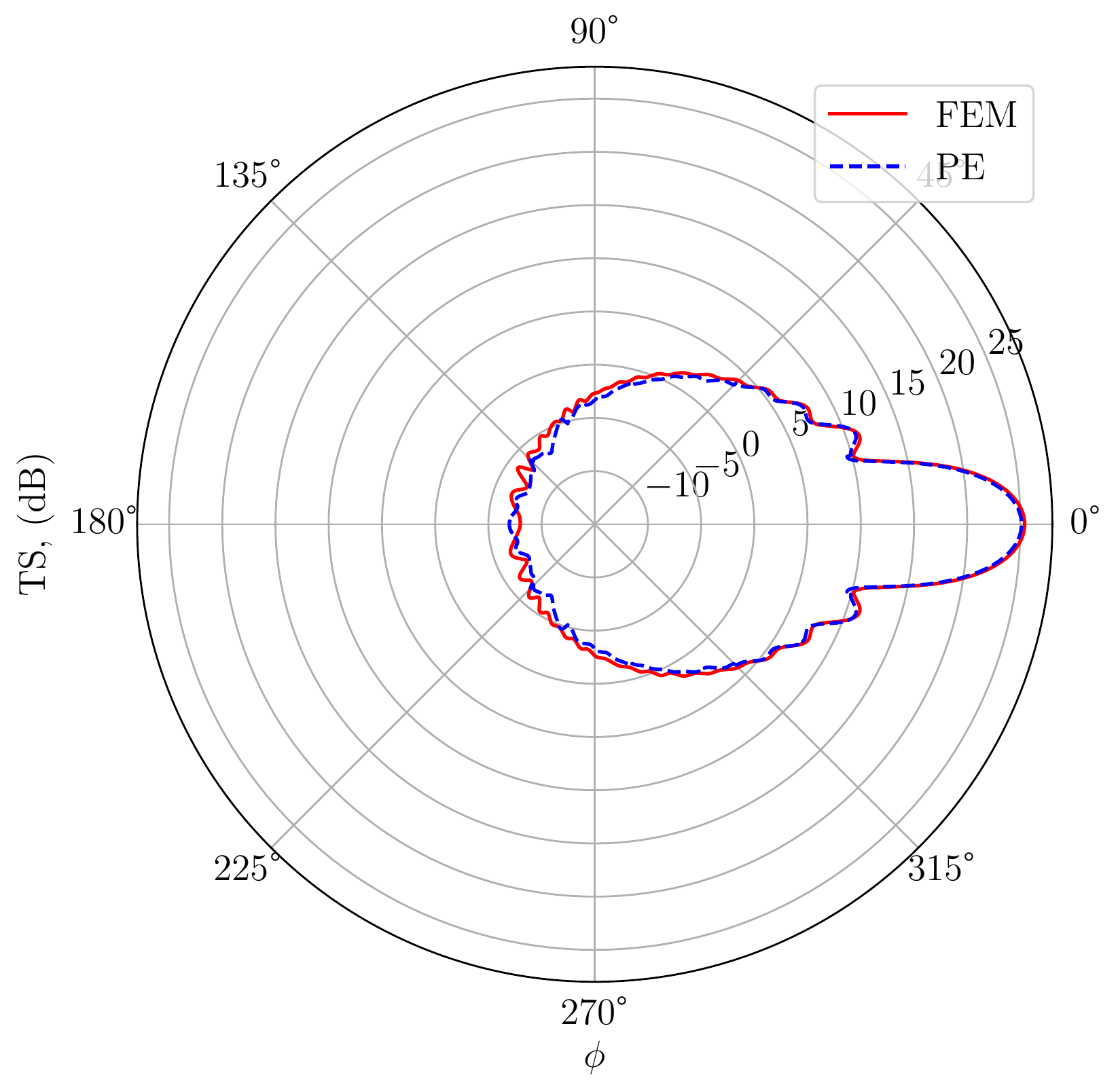}
\caption{}
\end{subfigure}
\begin{subfigure}[b]{.32\textwidth}
\includegraphics[width=\textwidth]{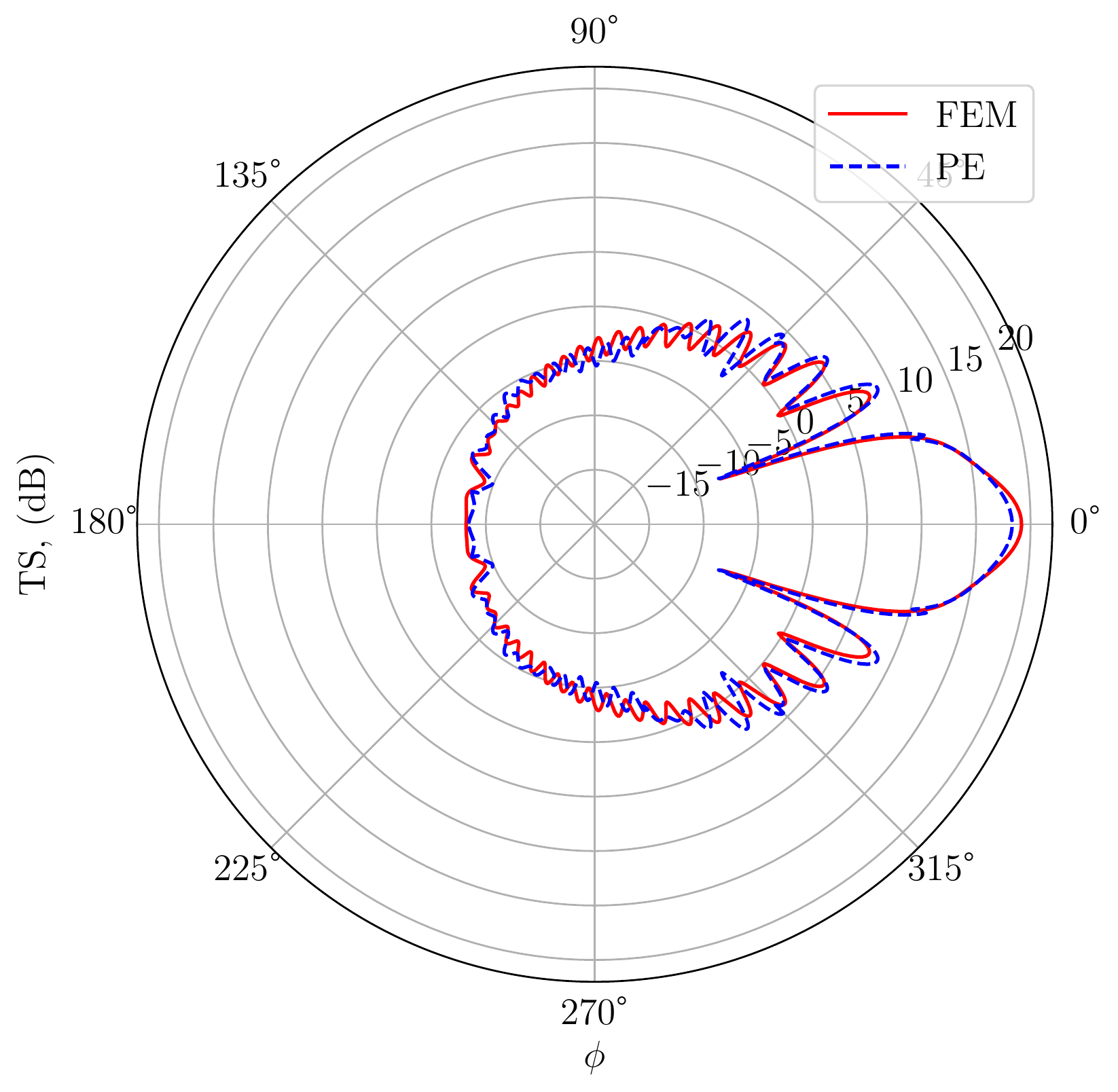}
\caption{}
\end{subfigure}
\begin{subfigure}[b]{.32\textwidth}
\includegraphics[width=\textwidth]{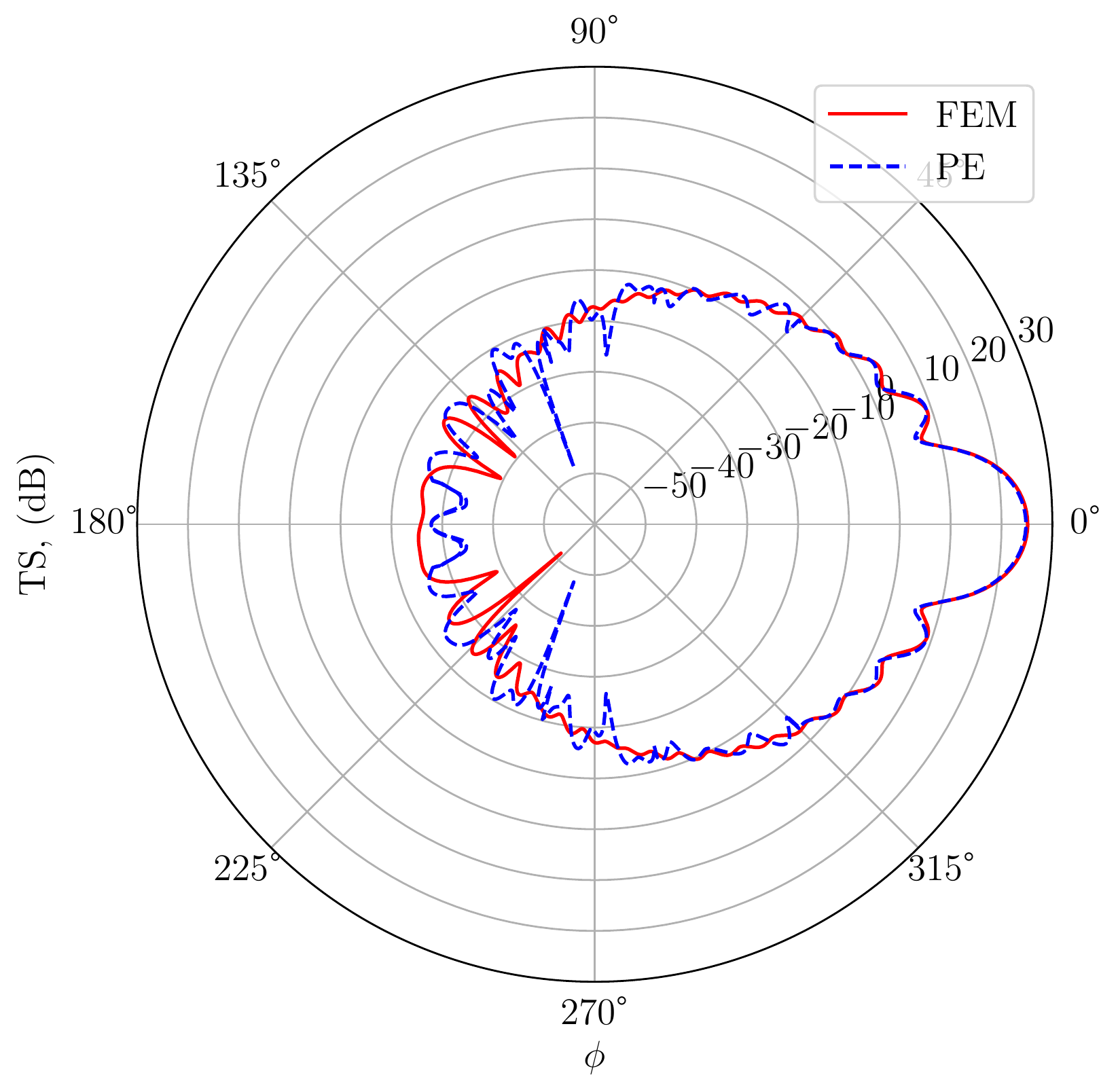}
\caption{}
\end{subfigure}
\caption{Three-dimensional target strength of an ellipsoid with $a_x = 5$ m and  $a_y=a_z = 2$ m  ($ka_x = 10\pi$) with (a) soft, (b) hard, and (c) impedance [as defined in Eq. (\ref{eq:pebc}) with $\alpha = 1$ and $\beta = ik$]  boundary conditions for end-on plane-wave incidence. Dashed blue lines are from the multisector PE method, and solid red are finite-element results.}\label{fig:elpsTS}
\end{figure}

Similarly, Figure \ref{fig:elpsTS} shows the target strength for an ellipsoid with $a_x = 5$ ($ka_x = 10\pi$) and $a_y=a_z = 2$ m, for the same cases as above. 
Even for elongated objects, the PE method matches well with the FEM for all boundary condition cases, although there are small discrepancies in the backscattering in the 3D impedance case. 
The overall signal is around $-40$~dB, however, so the deviation could be due to numerical error in both the PE and FEM calculations.

\begin{figure}[h!]
\begin{subfigure}[b]{.32\textwidth}
\includegraphics[width=\textwidth]{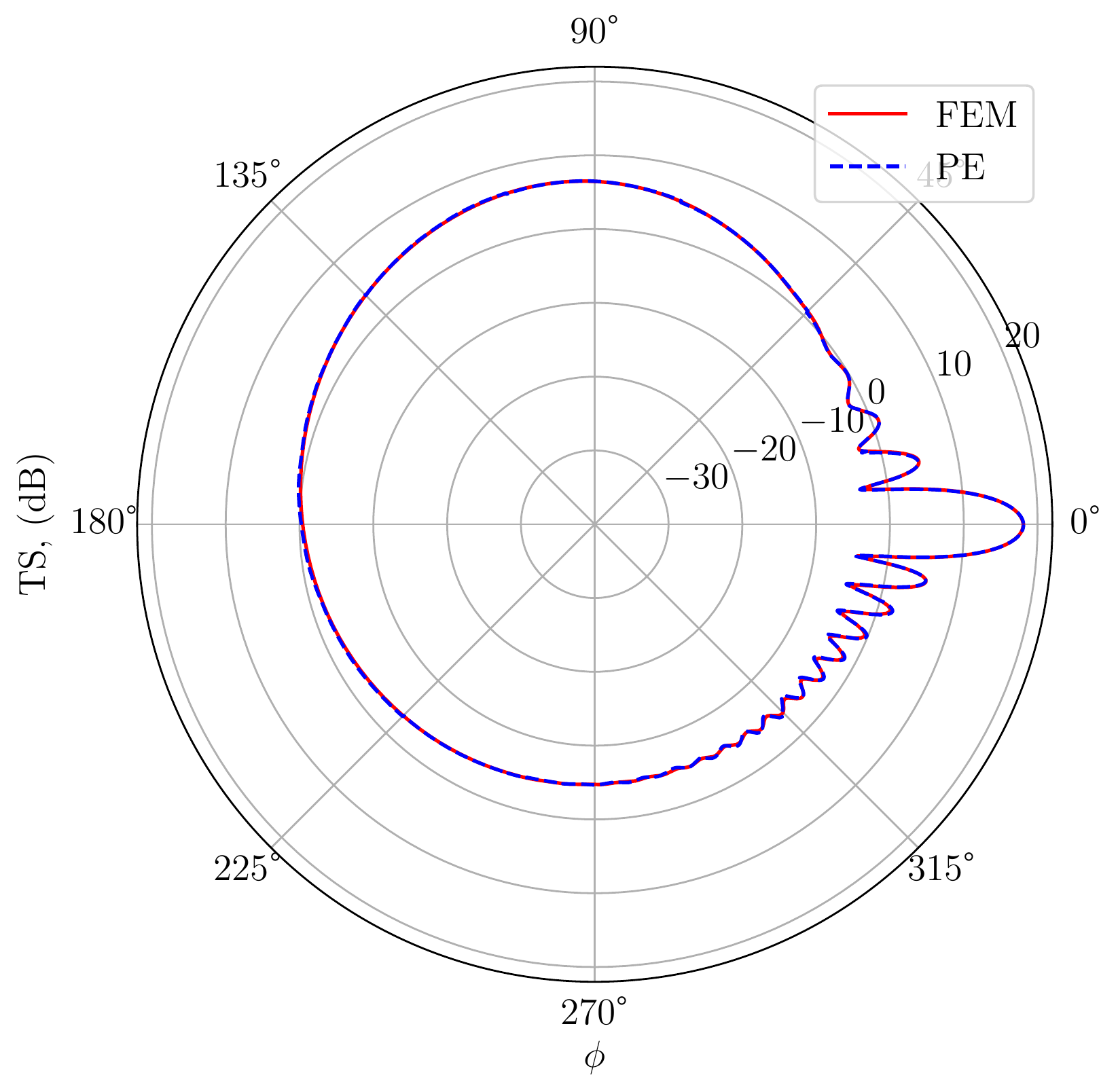}
\caption{}
\end{subfigure}
\begin{subfigure}[b]{.32\textwidth}
\includegraphics[width=\textwidth]{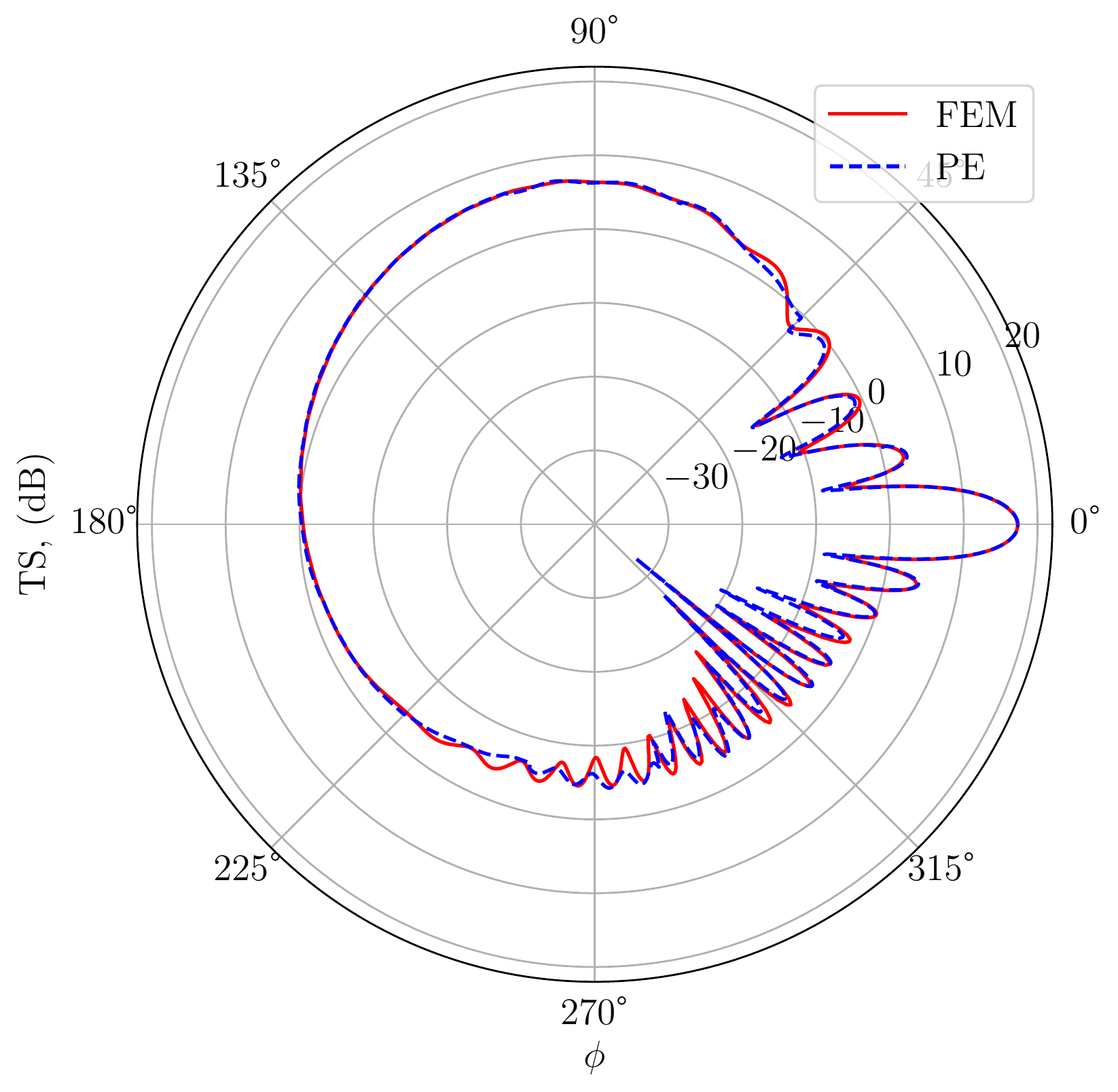}
\caption{}
\end{subfigure}
\begin{subfigure}[b]{.32\textwidth}
\includegraphics[width=\textwidth]{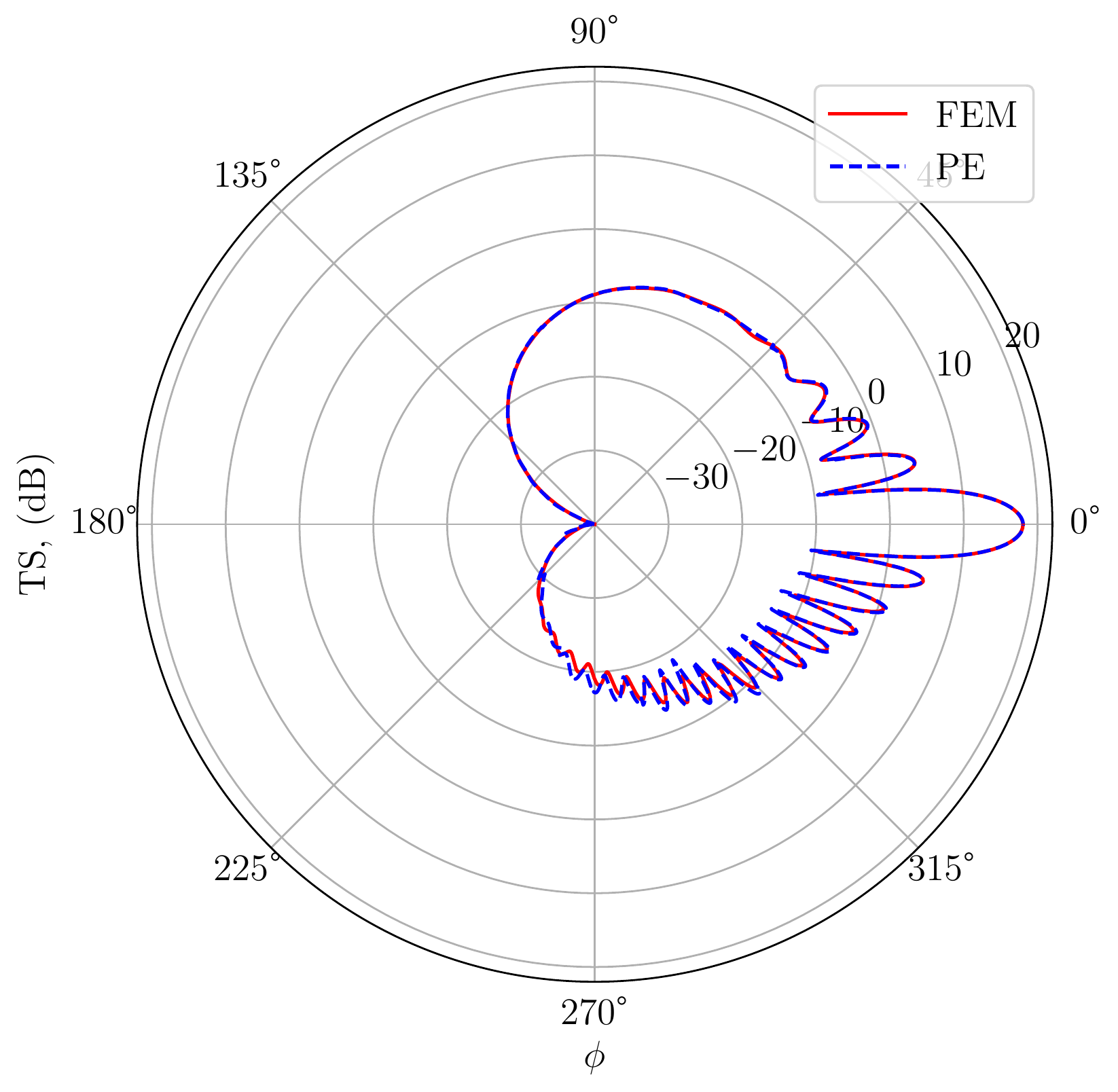}
\caption{}
\end{subfigure}
\caption{Two-dimensional target strength of an ellipse angled at 45$^\circ$ with $a_x=5,a_z=2$ with (a) soft, (b) hard, and (c) impedance [as defined in Eq. (\ref{eq:pebc}) with $\alpha = 1$ and $\beta = ik$] boundary conditions for an incident plane wave. Dashed blue lines are from the multisector PE method, and solid red are finite-element results.}\label{fig:ellipse52r45TS}
\end{figure}

\begin{figure}[h!]
\begin{subfigure}[b]{.33\textwidth}
\includegraphics[width=\textwidth]{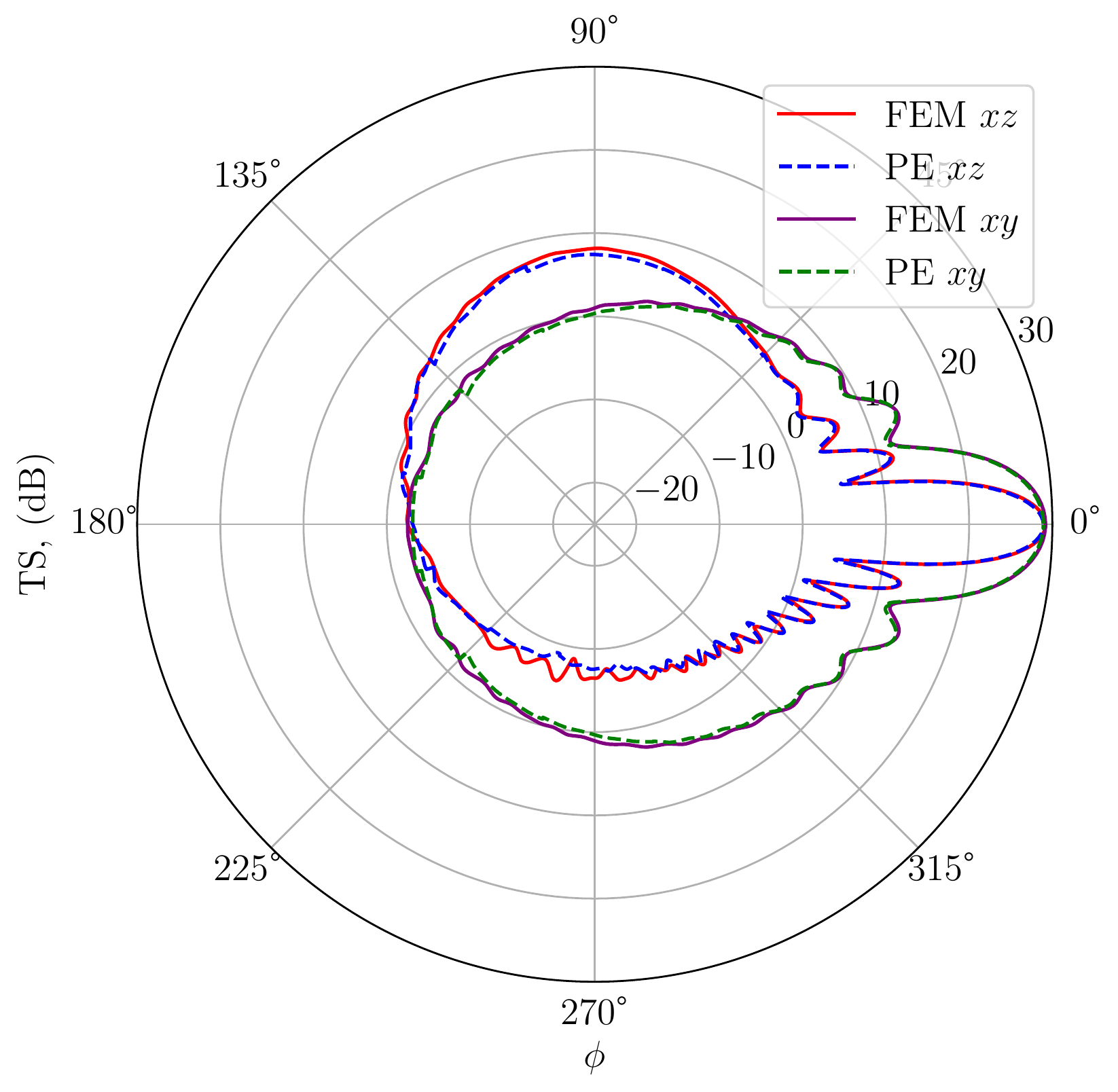}
\caption{}
\end{subfigure}
\begin{subfigure}[b]{.33\textwidth}
\includegraphics[width=\textwidth]{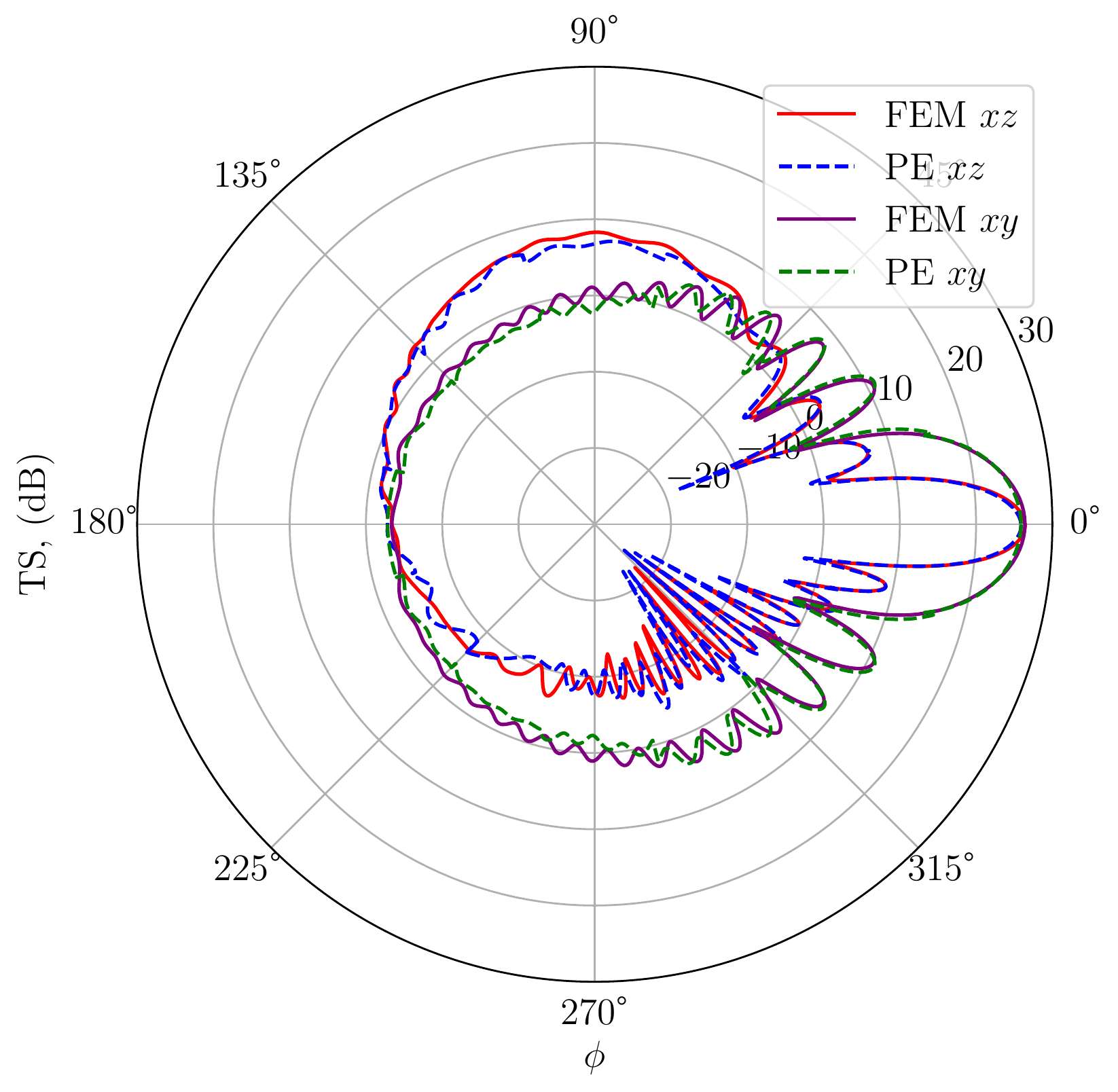}
\caption{}
\end{subfigure}
\begin{subfigure}[b]{.33\textwidth}
\includegraphics[width=\textwidth]{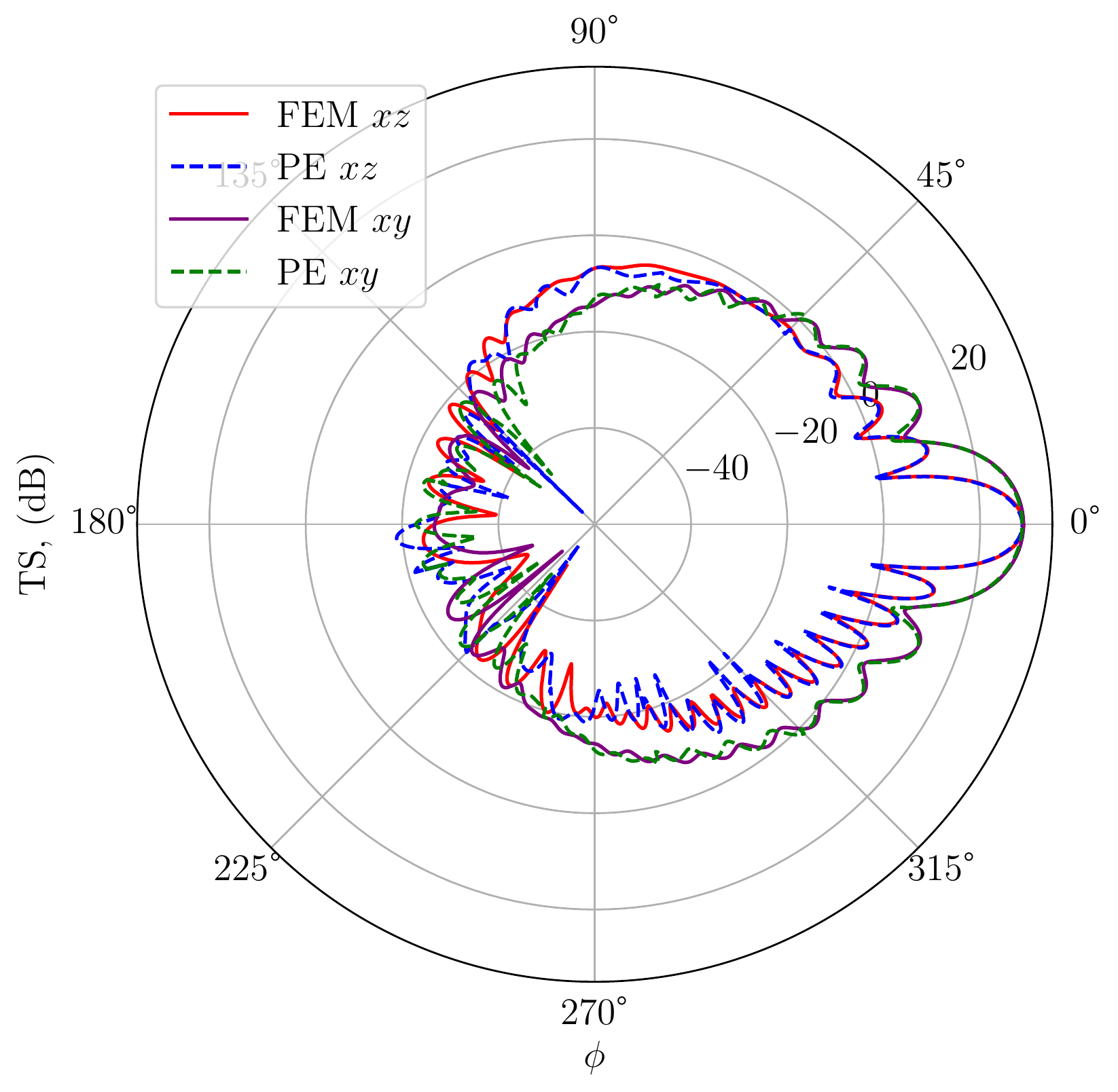}
\caption{}
\end{subfigure}
\caption{Three-dimensional target strength of an ellipsoid with $a_x = 5$ m and  $a_y=a_z = 2$ m with (a) soft, (b) hard, and (c) impedance [as defined in Eq. (\ref{eq:pebc}) with $\alpha = 1$ and $\beta = ik$]  boundary conditions for 45$^\circ$ plane-wave incidence. Dashed blue ($xz$-plane) and green ($xy$-plane) lines are from the multisector PE method, and solid red ($xz$-plane) and purple ($xy$-plane) are finite-element results.}\label{fig:elpsTSr45}
\end{figure}

Next, we consider the cases of ellipses and ellipsoids rotated an angle of 45$^\circ$ with respect to the incident plane wave. 
Figures~\ref{fig:ellipse52r45TS}~and~\ref{fig:elpsTSr45} show the far-field pressure for a plane wave in the $x$-direction scattered from an ellipse and ellipsoid, respectively, with $a_x = 5, a_z (= a_y) =2$ for oblique incidence. 
Subfigures show the results for soft, hard, and impedance ($\alpha=1,\beta=ik$) boundary conditions.  
Similarly to the previous case, there is good agreement between the PE and FEM in 2D, and also in the soft and hard cases in 3D; there is more variance between the FEM and PE in the 3D impedance case, although at very low magnitudes of the target strength. 
These results indicate that the multisector PE method is applicable when studying scattering with asymmetric insonification.

\begin{figure}[h!]
\begin{subfigure}[b]{.4\textwidth}
\includegraphics[width=\textwidth]{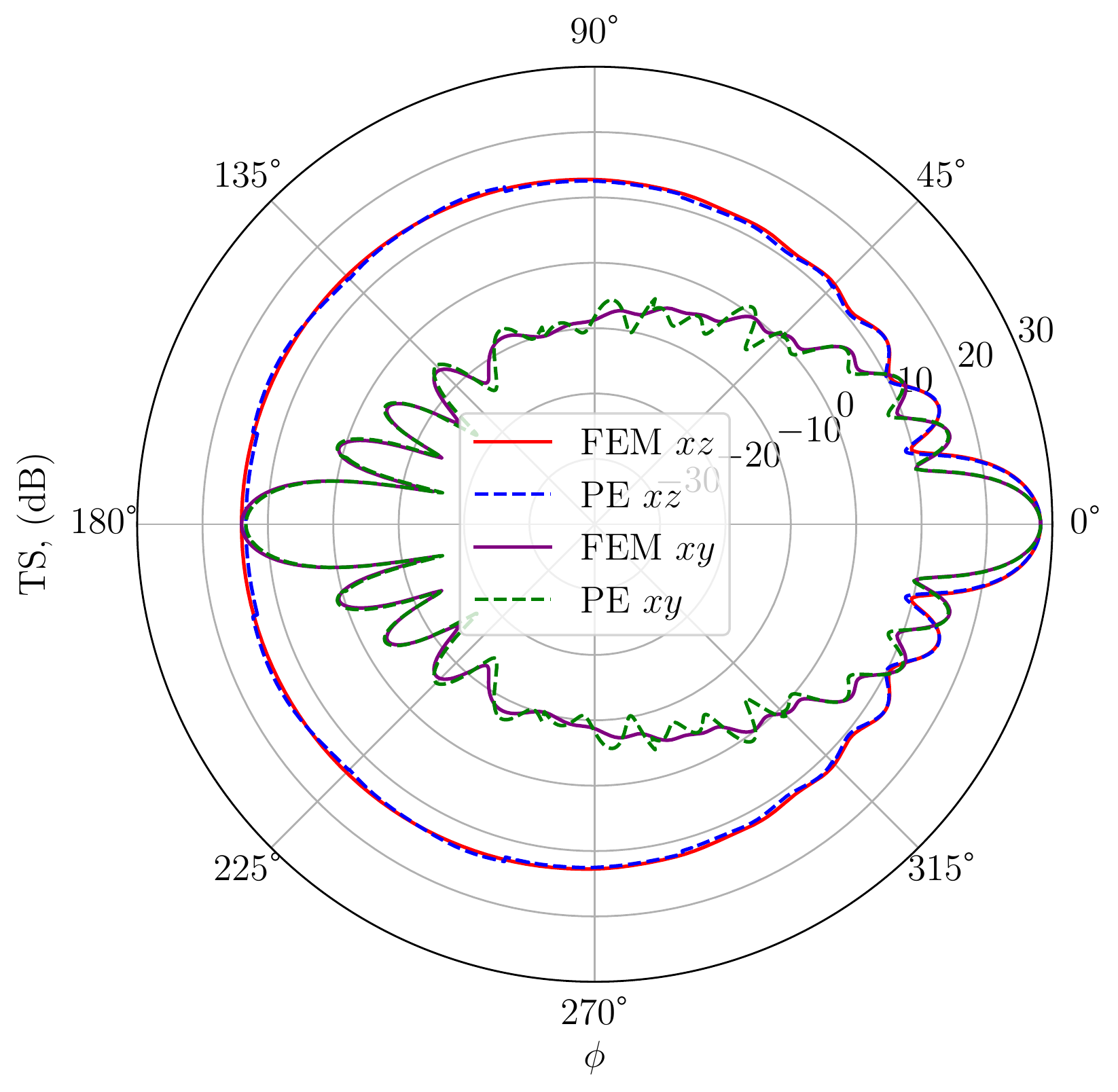}
\caption{}
\end{subfigure} \\
\begin{subfigure}[b]{.4\textwidth}
\includegraphics[width=\textwidth]{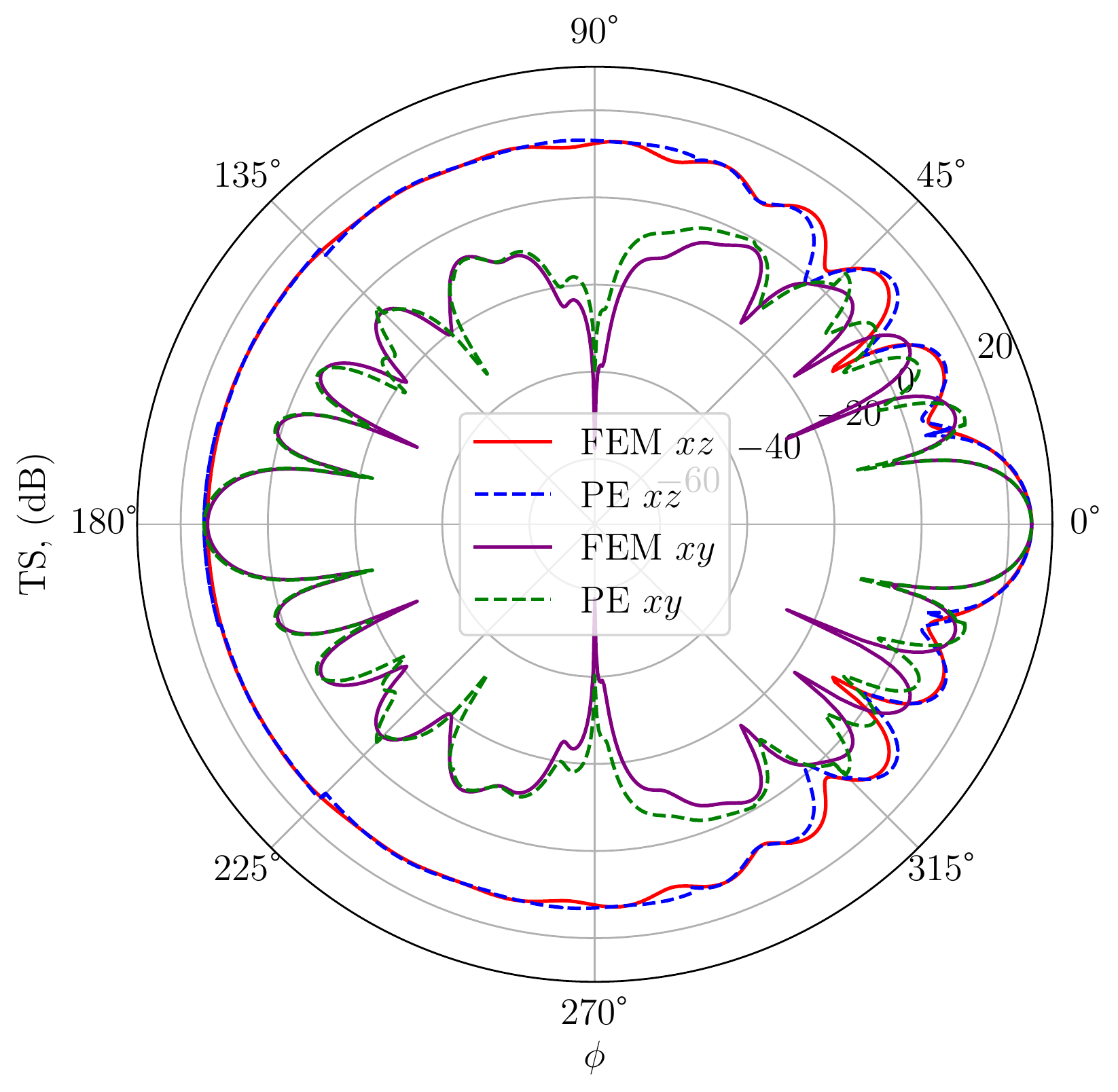}
\caption{}
\end{subfigure}
\caption{Three-dimensional target strength of a finite circular cylinder with $a = 2$ m and $h = 5$ m for broadside plane-wave incidence, with (a) soft and (b) hard boundary conditions. Dashed blue ($xz$ plane) and green ($xy$ plane) lines are from the multisector PE method, and solid red ($xz$ plane) and purple ($xy$ plane) are finite-element results.}\label{fig:cylTS}
\end{figure}

Finally, as an example for an object with sharp edges, we consider the case of a finite cylinder in three dimensions. 
The target strength calculations for a circular cylinder of radius $a_x =a_z =2$ m and height $h=5$ m and broadside plane-wave incidence are shown in Fig.~\ref{fig:cylTS}. 
Once again, we see good agreement between the FEM and PE methods for both the soft and hard objects; the sharp edges of the scatterer do not induce any spurious oscillations or otherwise incorrect behavior in the scattered field.

One important advantage of the parabolic equation method is its speed relative to a finite-element code, especially at higher $ka$. 
To illustrate this, we compare the time it takes to do a full-sweep of the PE (i.e. 72 wedges for the narrow-angle formulation) to the time for the full FEM solution. 
If one is looking at only a few angular sectors for scattering, then this reduces the number of necessary PE wedges. 
In addition, one can utilize symmetry in both the PE and FEM simulations to reduce the computational domain and thus the computation time. 
However, to keep the comparison as direct as possible, we will compare the time taken by both methods to compute the full angular spectrum, utilizing no symmetry, for the $a = 5$ m sphere with hard boundary conditions for a variety of frequencies. 

\begin{figure}[t!]
\includegraphics[width=.5\textwidth]{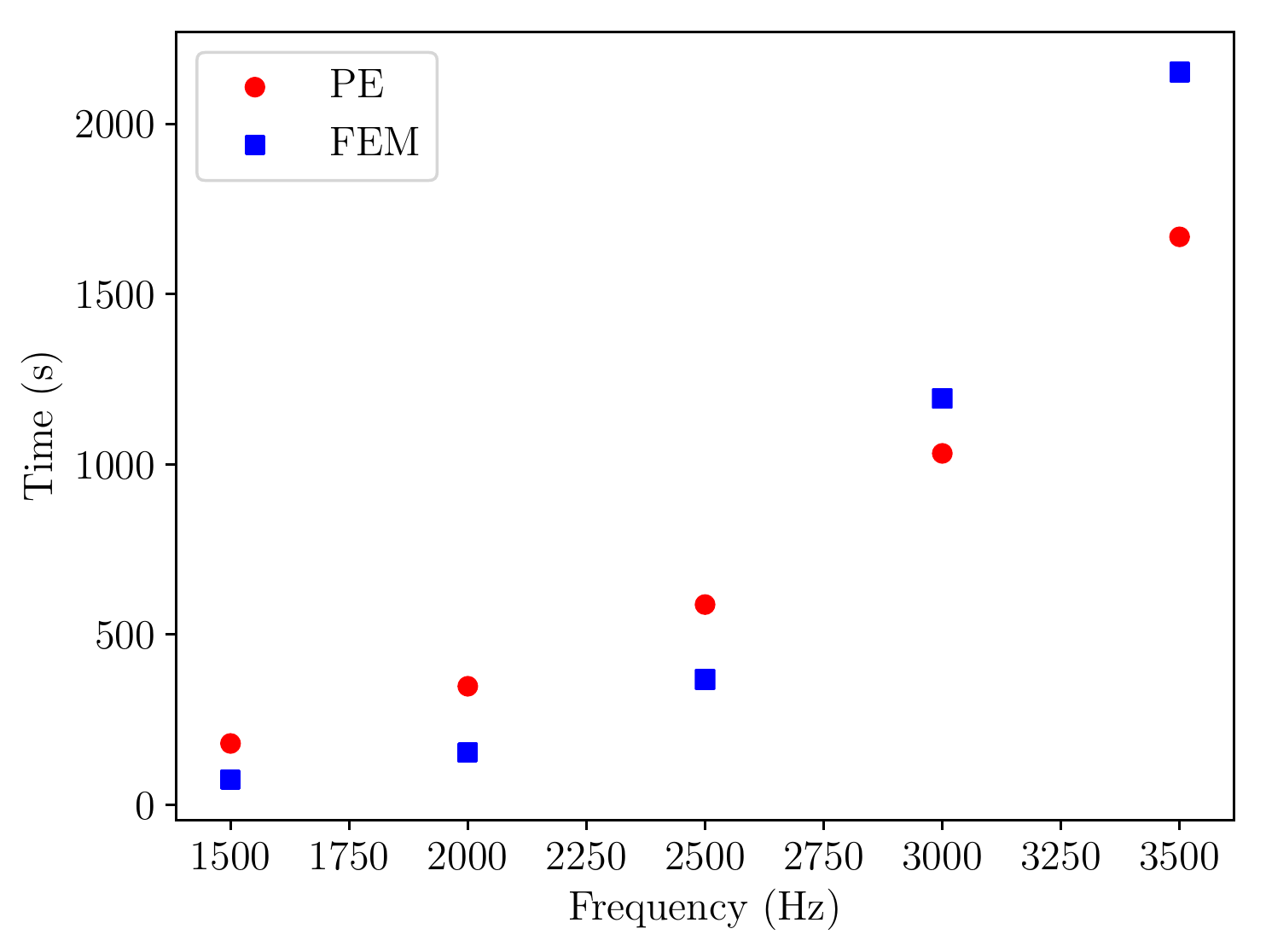}
\caption{Time comparison between finite-element (COMSOL) and PE methods for a full angular spectrum solution of scattering from a hard sphere of radius 5 m. The FEM solution has maximum element size $\lambda/6$, while the PE solution has grid spacing $\lambda/20$ in $x$ and $\lambda/10$ in the transverse directions.}\label{fig:time_sph5}
\end{figure}

The results are detailed in Table \ref{tab:time_sph5} and visualized in Fig.~\ref{fig:time_sph5}. All simulations were run on the same laptop computer with six CPU cores; COMSOL utilized all cores during the computation, and the PE code, written in C++, was parallelized such that each angular sector was run on a single processor. The FEM domain is a sphere of diameter 11 m with a perfectly matched layer (PML) \cite{levy2001perfectly} of thickness 0.5 m, while the PE domain has a size of 11 m $\times$ 15 m $\times$ 15 m with a 1 m thick PML  in the $y$ and $z$ directions. The FEM has maximum element size $\lambda/6$, while the PE method uses grid spacing $\lambda/20$ in the marching direction and $\lambda/10$ in the transverse directions. Note that the times given for the FEM code do not include the time needed for mesh generation, while the PE time includes the (nearly negligible) time needed to calculate boundary condition information. The FEM clearly scales at a rate much greater than the PE, with the PE being more efficient at $ka > 60$ in this example. Note that the PE is always faster when looking at a single angular sector.

\FloatBarrier

\begin{figure}[h!]
\includegraphics[width=.5\textwidth]{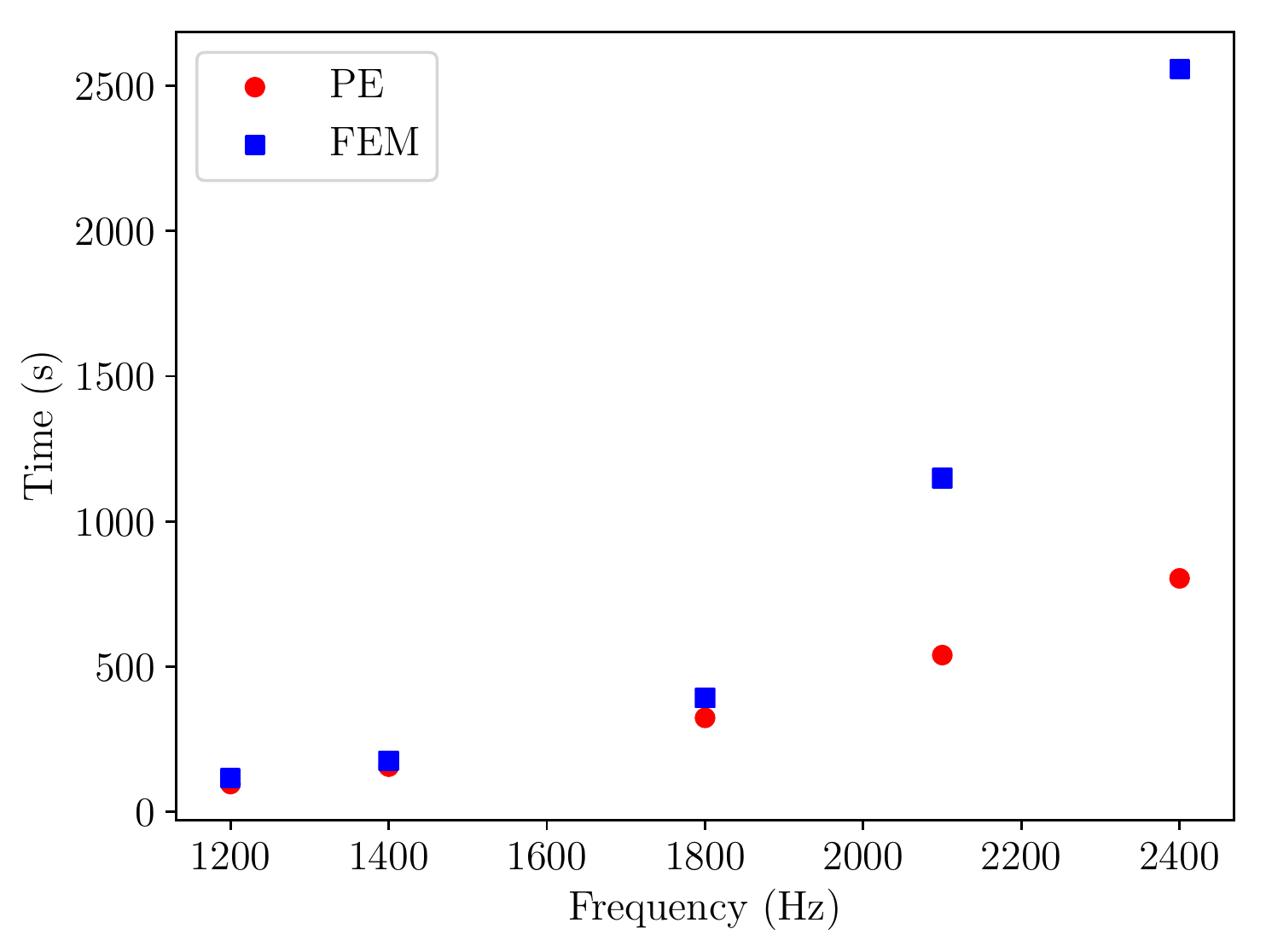}
\caption{Time comparison between finite-element (COMSOL) and PE methods for a full angular spectrum solution of scattering from an ellipsoid of transverse radius 2 m and total length 14 m. The FEM solution times are for the full angular spectrum with maximum element size $\lambda/6$, while the PE calculation times are for the full angular spectrum with grid spacing $\lambda/20$ in $x$ and $\lambda/10$ in the transverse directions.}\label{fig:time_rcyl}
\end{figure}

As we did above with the sphere, we can compare the time it takes for the PE and FEM calculations for a full angular spectrum of an ellipsoid of $a_x = 7$ m, $a_y=a_z=2$ m. 
The domain for the FEM is a box of size 15 m $\times$ 5 m $\times$ 5 m with a PML of 0.5 m on all sides, while the PE method had a domain size of 15 m $\times$ 15 m $\times$ 15 m with a 1 m PML in the transverse directions (kept uniform for all orientations of the object relative to the marching direction).  
The results are detailed in Table \ref{tab:time_rcyl} and visualized in Fig.~\ref{fig:time_rcyl}. 
Note, as before, that the times given for the FEM code do not include the time needed for mesh generation, while the PE time includes the time needed to calculate boundary condition information.

The PE is significantly more efficient than the FEM at frequencies above 1800~Hz, and is comparable at lower frequencies. 
This is unlike the example of the $a = 5$ m sphere, where the FEM was faster until frequencies of approximately 2800~Hz. 
Put into dimensionless units, however, the results are consistent: The PE method is significantly faster than the FEM when $ka  \gtrsim$50.  
This is primarily because when we have elongated objects, the PE marching time goes {\em linearly} with the length of the object when keeping the transverse domain a constant size. 
Of course, the transverse domain can be shrunk when possible, giving a further advantage to the PE method. 
For example, there is no need to have a 15 m $\times$ 15 m transverse domain, as we did above, when marching along the ellipsoid with transverse radius 2 m; this is only necessary when the paraxial direction is perpendicular to the ellipsoid, and even then it is only necessary in one of the transverse directions.

\FloatBarrier

\begin{table}[t!]
\caption{Time comparison between finite-element (COMSOL) and PE methods for scattering from a hard sphere of radius 5 m at selected frequencies. The FEM solution (parallel-processed on six cores) is for the full angular spectrum with maximum element size $\lambda/6$, while the two columns for the PE indicate times for a single angular wedge and the full angular spectrum (parallel-processed on six cores), respectively, with grid spacing $\lambda/20$ in $x$ and $\lambda/10$ in the transverse directions. All times in seconds.}
\begin{tabularx}{\columnwidth}{l Y Y Y Y}
\toprule
Freq. & $ka$ & COMSOL & PE wedge & Full PE \\ \colrule
1500 & 31.4 &74 & 15 & 180 \\ 
2000 &  41.9 &153 & 34 &  408 \\ 
2500 & 52.4& 348 &  49 & 588 \\ 
3000 & 62.8 &1193 &  86 &  1032 \\ 
3500 & 73.3 & 2152 & 139 & 1668 \\ \botrule
\end{tabularx}
\label{tab:time_sph5}
\end{table}

\begin{table}[h!]
\caption{Time comparison between finite-element (COMSOL) and PE methods for scattering from an ellipsoid of transverse radius 2 m and total length 14 m at selected frequencies. The FEM solution (parallel-processed on six cores) is for the full angular spectrum with maximum element size $\lambda/6$, while the two columns for the PE indicate times for a single angular wedge and the full angular spectrum (parallel-processed on six cores), respectively, with grid spacing $\lambda/20$ in $x$ and $\lambda/10$ in the transverse directions. All times in seconds.}
\begin{tabularx}{\columnwidth}{l Y Y Y Y}
\toprule
Freq. & $ka$ &COMSOL & PE wedge & Full PE \\ \colrule
1200 & 35.2 & 117 & 8 & 96 \\ 
1400 & 41.1 & 176 & 13 &  156 \\
1800 & 52.8 & 392 &  27 & 324 \\ 
2100 &  61.6 & 1149 &  45 &  540 \\ 
2400 & 70.4 & 2556 & 67 & 804 \\ \botrule
\end{tabularx}
\label{tab:time_rcyl}
\end{table}

\section{Scattering from concave objects}\label{sec:wams}

Thus far, all results have been using the narrow-angle formulation of the parabolic equation. 
As stated above, we cannot apply the wide-angle PE on the boundary of the object, as the boundary conditions then induce spurious oscillations. 
It can, however, be applied slightly outside the boundary. 
We implement this by using the Pad\'{e}-(2,1) approximation three points outside the boundary of the scatterer, and the Pad\'{e}-(2,2) approximation beyond the scatterer, though the latter is not necessary for target strength calculations. 

To see the effect and extent of improvement from utilizing the wide-angle formulation, we consider a concave object, which we will call the ``bean.''  The object shape is described by \cite{bruno200180}
\eq{
 \begin{split}
\frac{\left(\alpha _1 R \cos \left(\frac{\pi x}{R}\right)+z\right)^2}{b^2 \left(1-\alpha _2 \cos
   \left(\frac{\pi x }{R}\right)\right)} +\frac{x ^2}{c^2}-R^2= 0 \,.
   \end{split}
} 
\begin{figure}[h!]
\includegraphics[width=.45\textwidth]{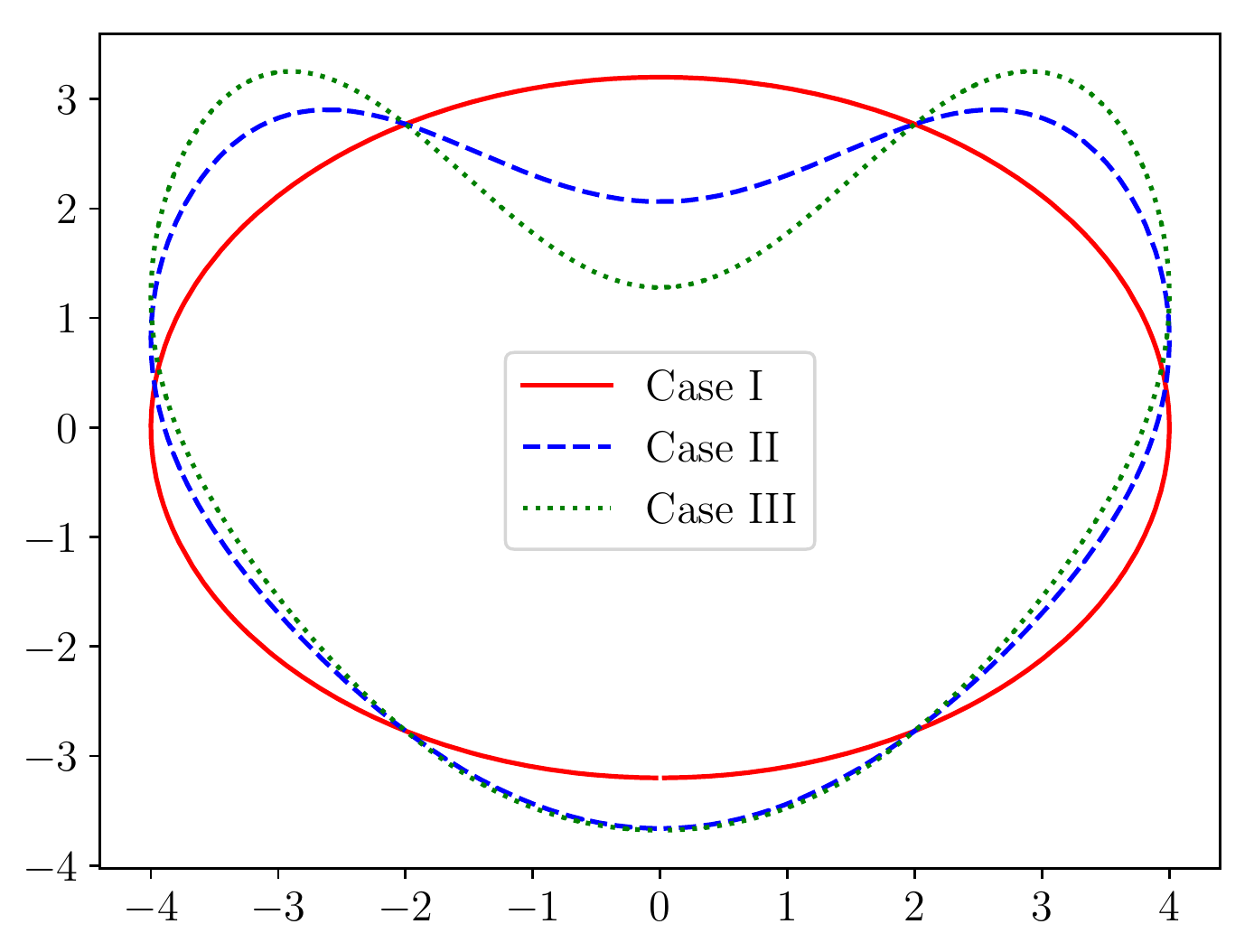}
\caption{Bean geometry for three sets of parameters (see text).}
\label{fig:bean_geom}
\end{figure}
The 2D geometries, with $b = 0.8, c = 1, R = 4$ (all in meters), for three different concavity cases (I: $\alpha_1 = \alpha_2 = 0$, II: $\alpha_1 = \alpha_2 = 0.2$, III: $\alpha_1 = 0.3, \alpha_2 = 0.4$) are shown in Fig.~\ref{fig:bean_geom}.

\FloatBarrier

 \begin{figure}[h!]
\begin{subfigure}[b]{.32\textwidth}
\includegraphics[width=\textwidth]{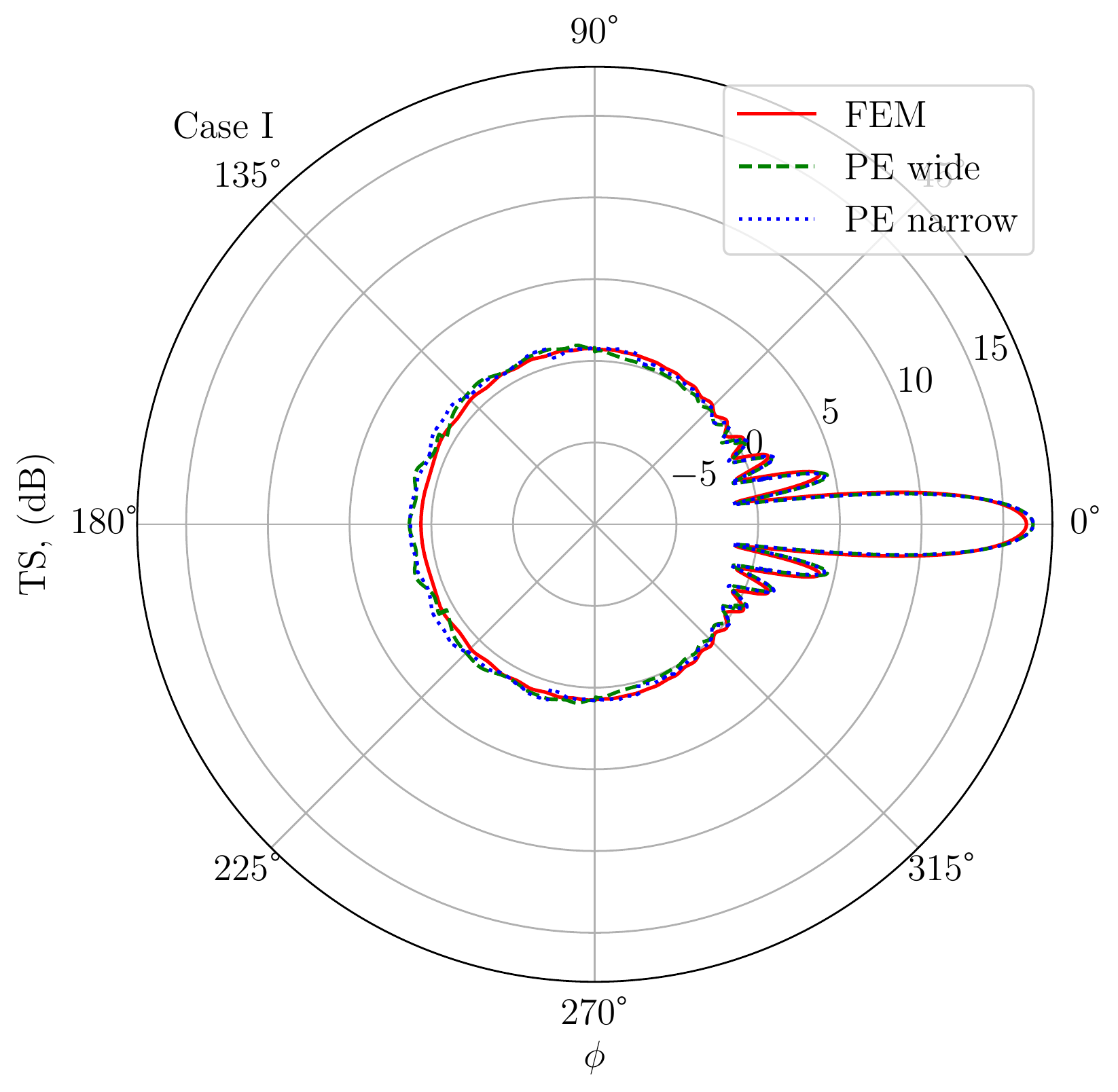}
\end{subfigure}
\begin{subfigure}[b]{.32\textwidth}
\includegraphics[width=\textwidth]{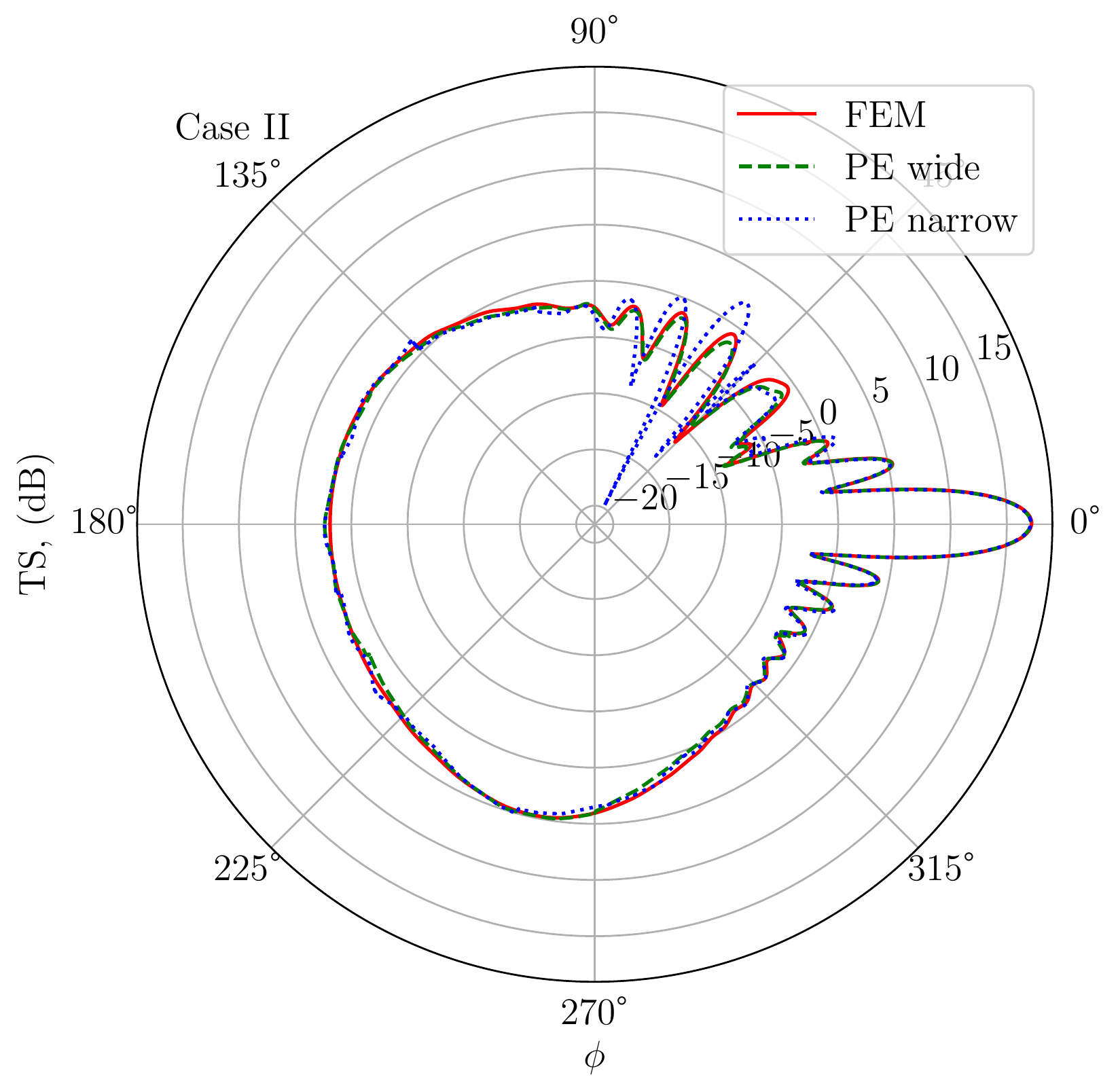}
\end{subfigure}
\begin{subfigure}[b]{.32\textwidth}
\includegraphics[width=\textwidth]{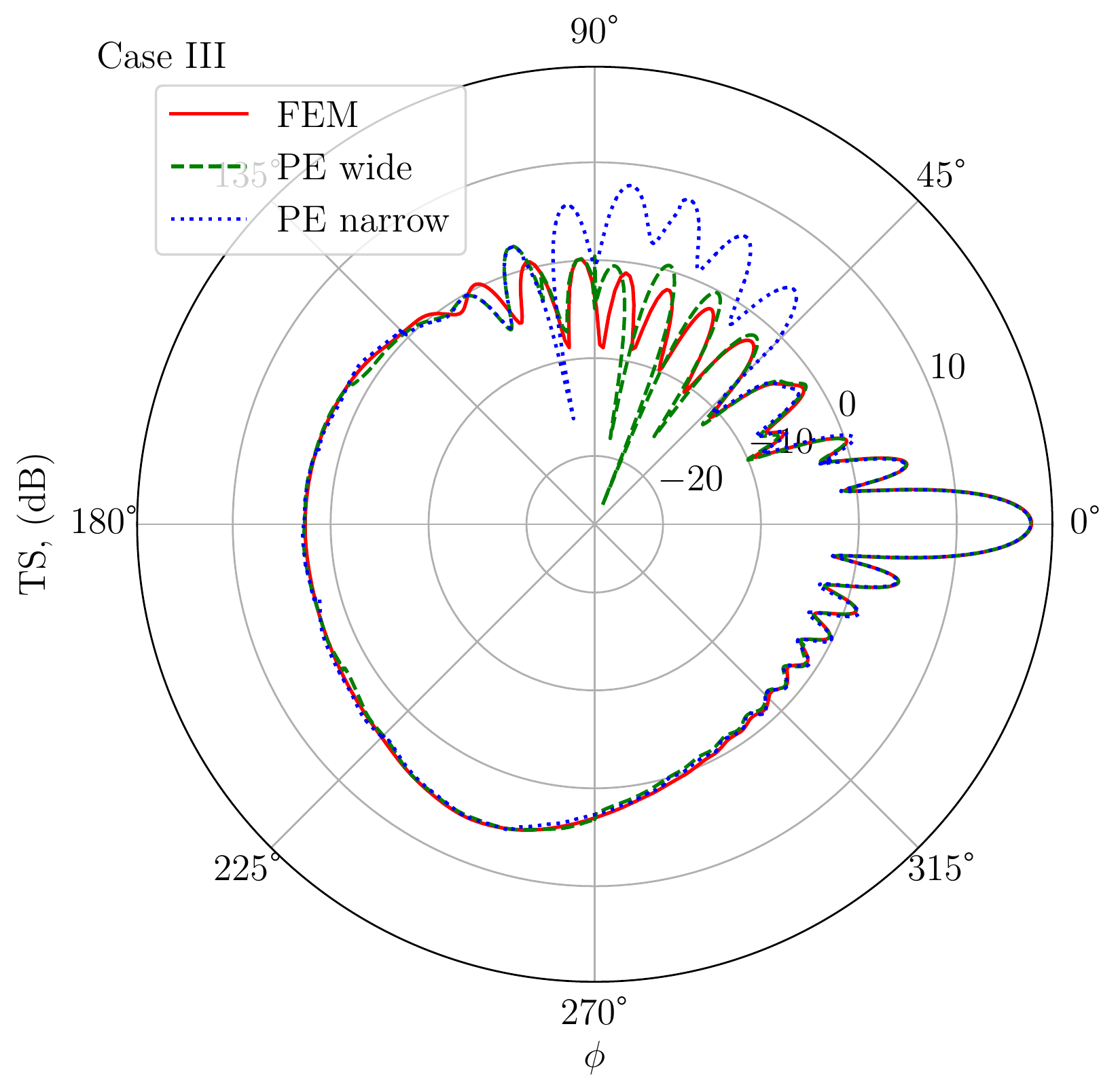}
\end{subfigure} 
\caption{Target strength of a soft bean of $ka=8\pi$ for three different sets of parameters (see text) for an incident plane wave. Red solid lines are FEM calculations, blue dotted are narrow-angle PE, and green dashed are wide-angle PE.}\label{fig:bean_res}
\end{figure}

 \begin{figure}[h!]
\begin{subfigure}[b]{.32\textwidth}
\includegraphics[width=\textwidth]{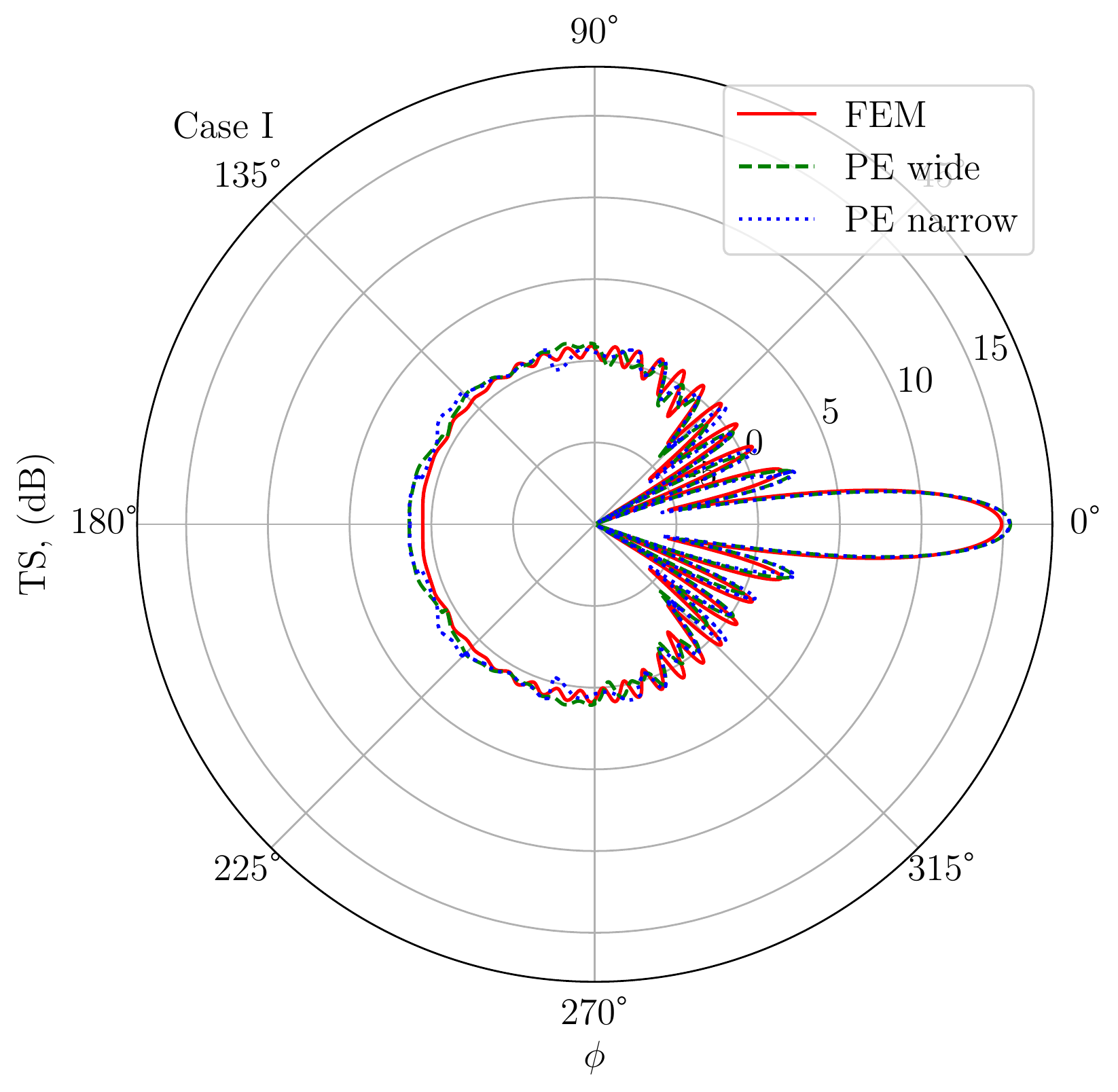}
\end{subfigure}
\begin{subfigure}[b]{.32\textwidth}
\includegraphics[width=\textwidth]{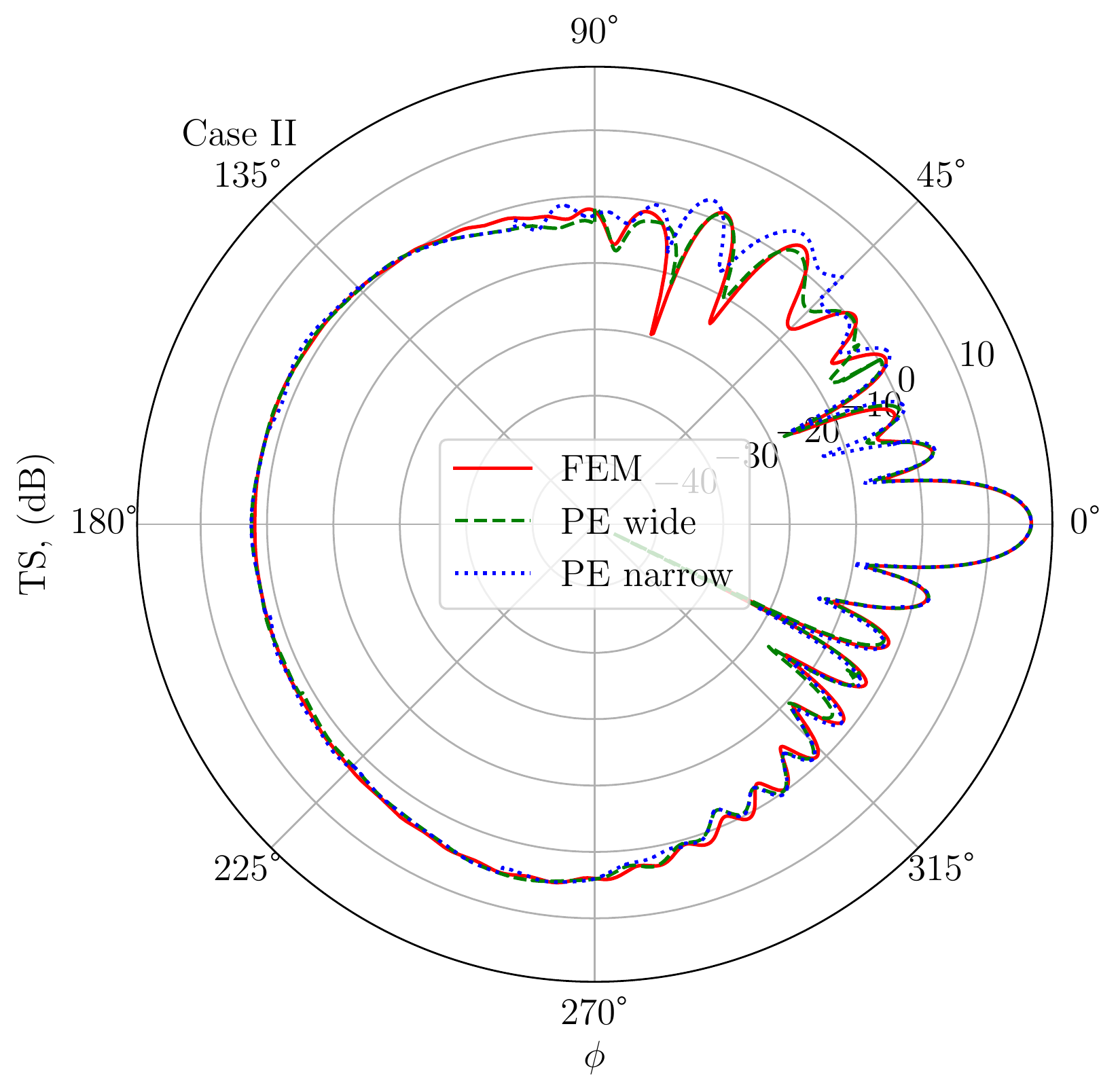}
\end{subfigure}
\begin{subfigure}[b]{.32\textwidth}
\includegraphics[width=\textwidth]{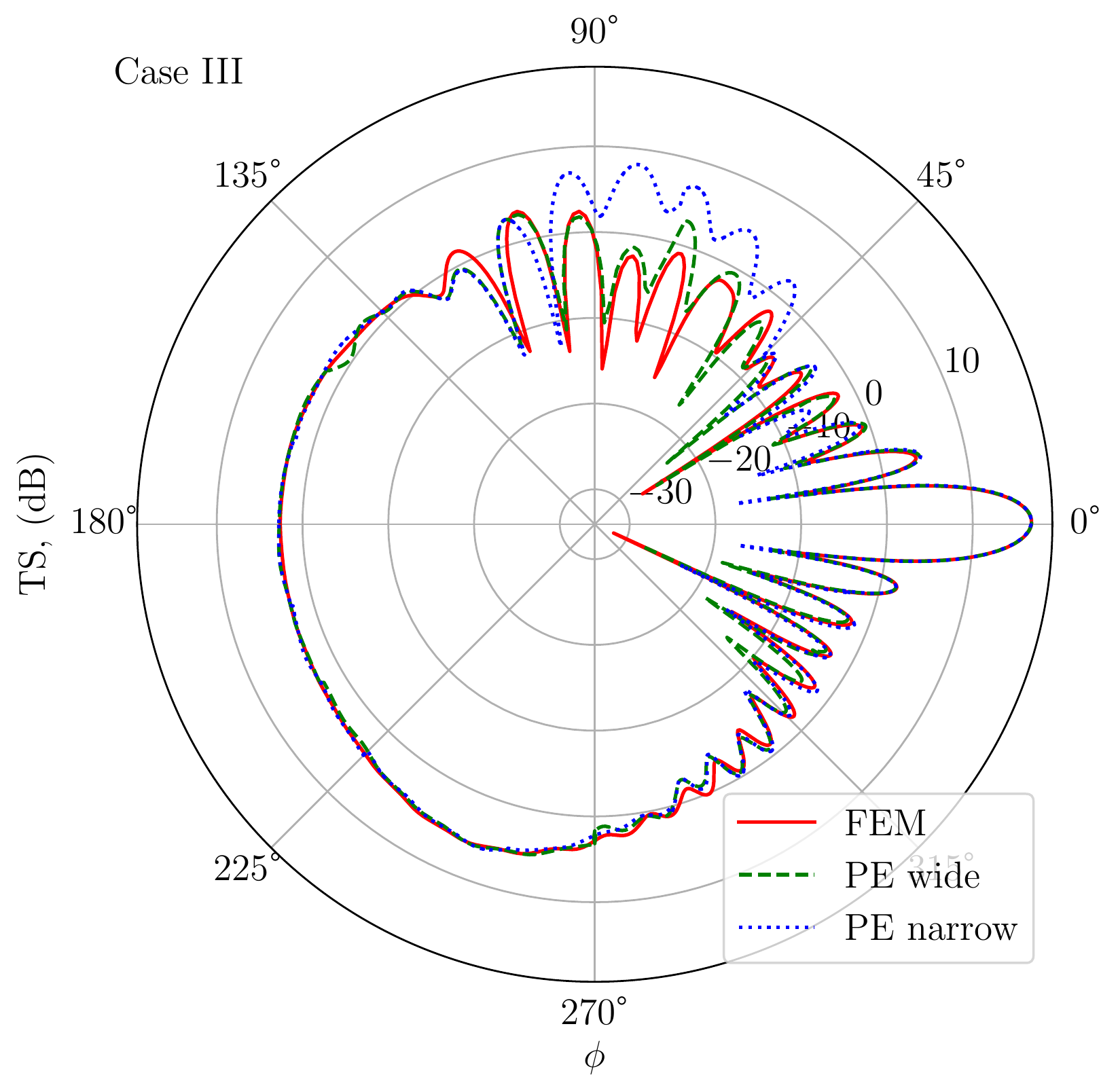}
\end{subfigure} 
\caption{Target strength of a hard bean of $ka = 8\pi$ for three different sets of parameters (see text) for an incident plane wave. Red solid lines are FEM calculations, blue dotted are narrow-angle PE, and green dashed are wide-angle PE.}\label{fig:bean_res_rig}
\end{figure}

The target strength results for an incident plane wave traveling in the positive $x$ direction of frequency 1500~Hz onto the three above objects with soft boundary conditions are shown in Fig.~\ref{fig:bean_res}. 
For Case I, the FEM, narrow-angle PE, and wide-angle PE all give results that are in agreement with each other. 
As the concavity is increased, however, the narrow-angle PE starts to fail. 
Already with Case II, the narrow-angle PE disagrees with the FEM and the wide-angle PE --- the latter two agree --- in the direction where scattered rays are ``coming out'' of the indentation of the object. 
Case III is the most extreme, and we find relatively good agreement between the FEM and the wide-angle PE, although this starts to break down. The same holds true for the bean shape with hard boundary conditions, shown in Fig.~\ref{fig:bean_res_rig}.  
It is possible that using an even wider angle approximation of the square root --- beyond Pad\'{e}-(2,1) --- will allow better agreement in the most extreme cases.

This wide-angle implementation resolves the inaccuracy for the L-shaped geometry studied in Ref.~\cite{levy1998target}. 
If, however, we flip the L-shape horizontally, we run into problems, as the incident wave undergoes {\em multiple scattering} into the perpendicular and backward directions. 
This effect cannot be compensated for by simply implementing the wide-angle equation as above.
\begin{figure}[h!]
\includegraphics[width=.45\textwidth]{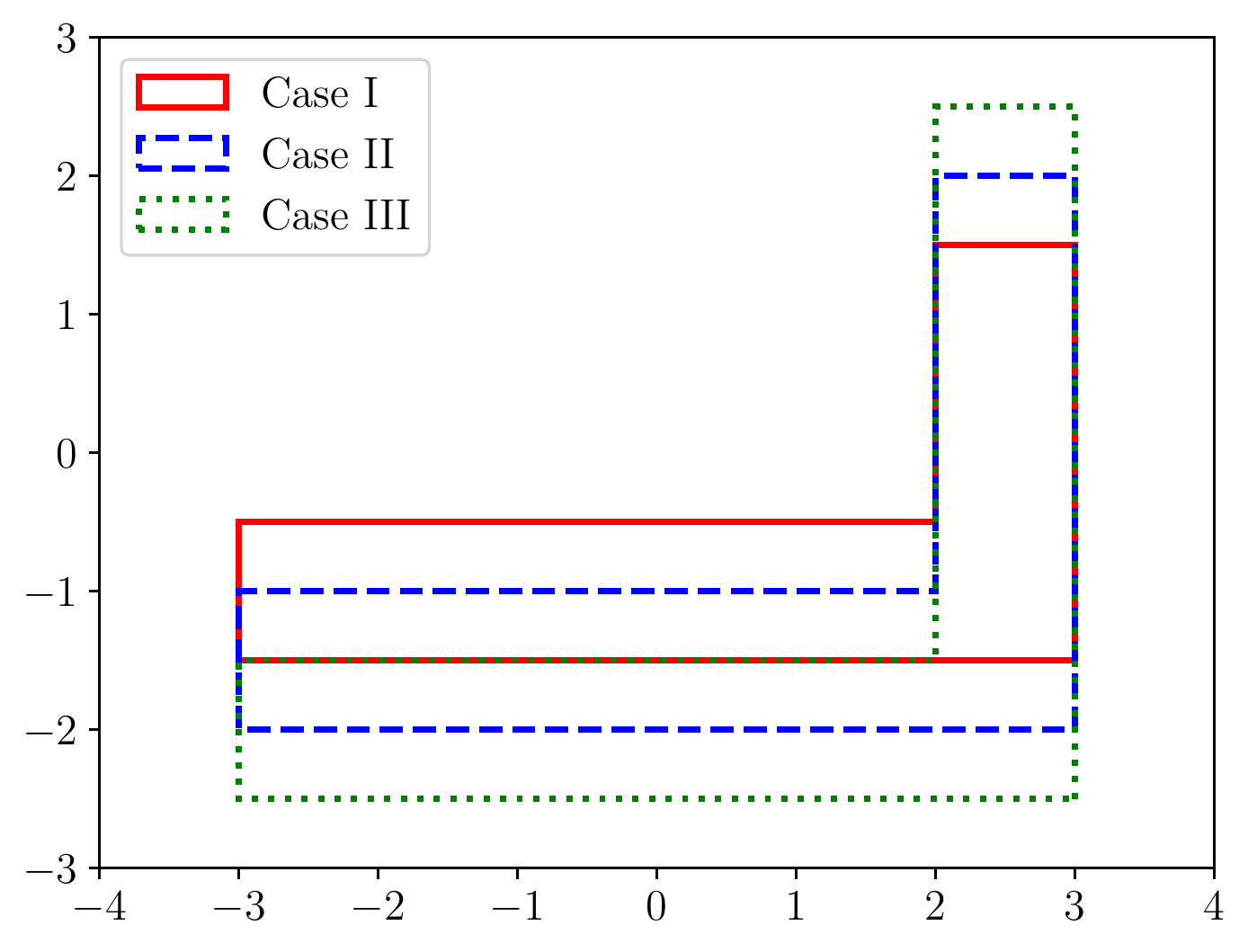}
\caption{L-shaped geometry for three sets of parameters (see text).}
\label{fig:L_geom}
\end{figure}
The geometries under consideration are shown in Fig.~\ref{fig:L_geom}, with the vertical rectangle having width 1 m and heights 3, 4, and 5 m for Cases I, II, and III, respectively, and the lower rectangle having width 5 m and height 1 m. 

\begin{figure}[h!]
\begin{subfigure}[b]{.32\textwidth}
\includegraphics[width=\textwidth]{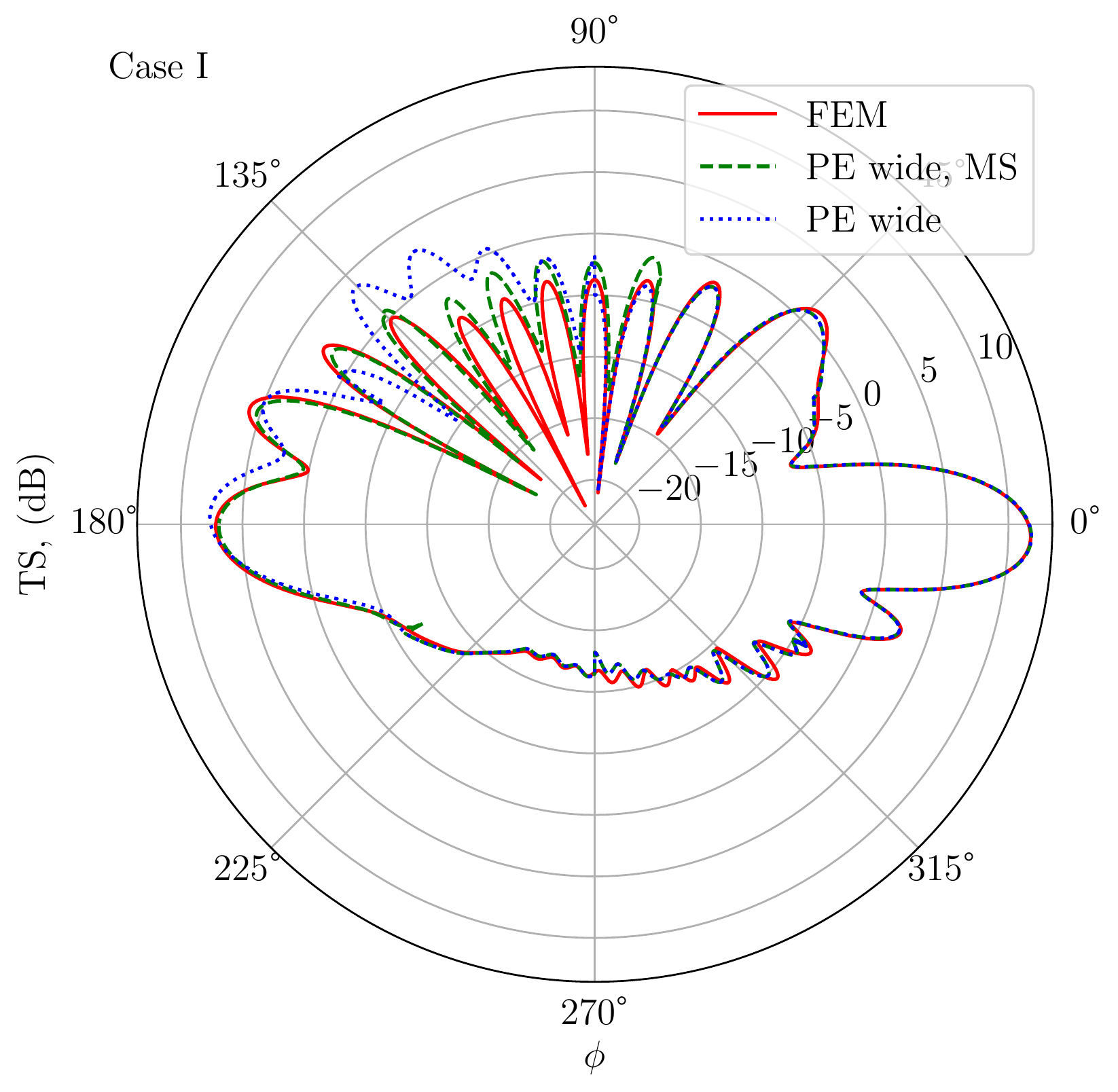}
\end{subfigure}
\begin{subfigure}[b]{.32\textwidth}
\includegraphics[width=\textwidth]{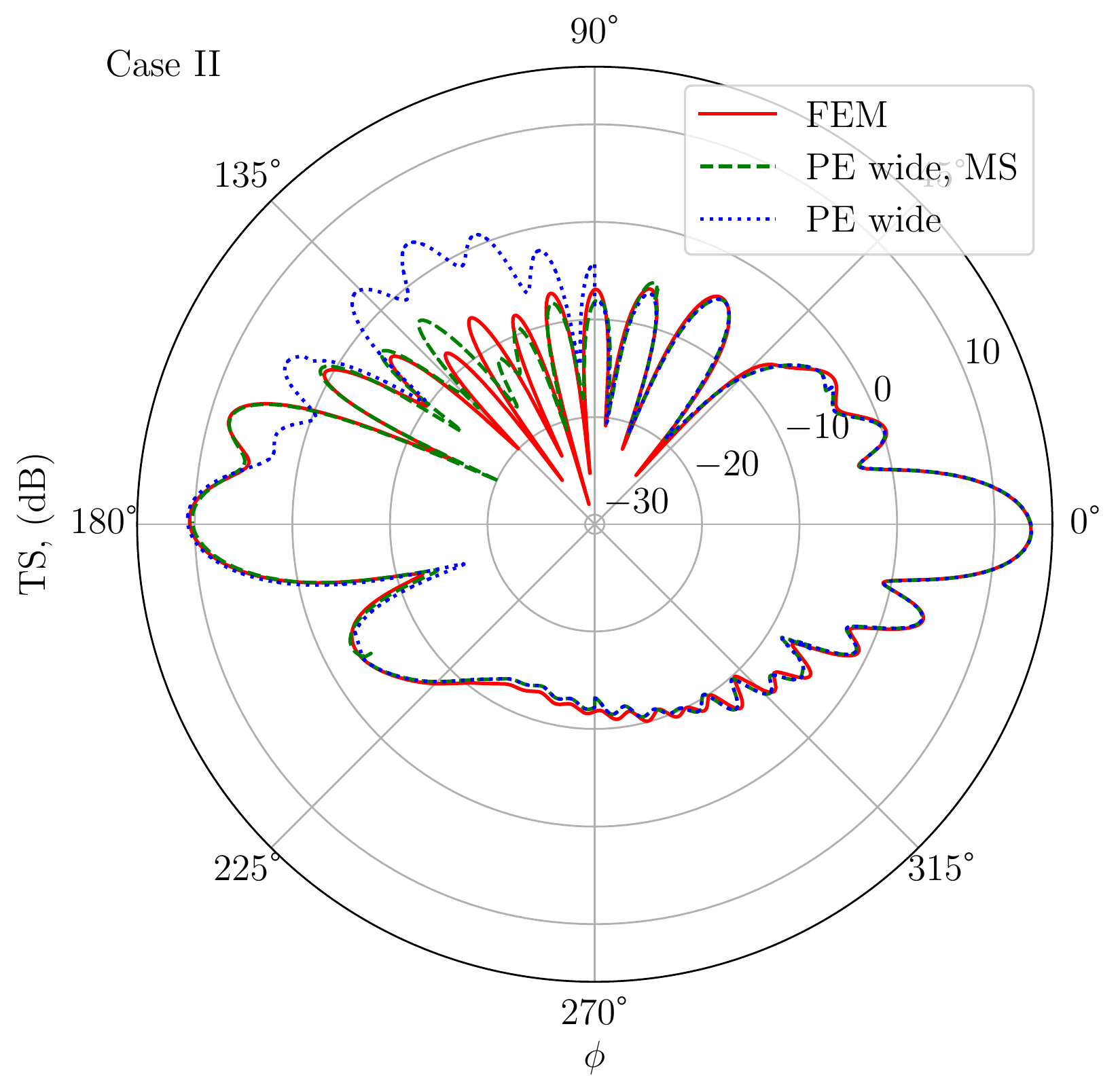}
\end{subfigure}
\begin{subfigure}[b]{.32\textwidth}
\includegraphics[width=\textwidth]{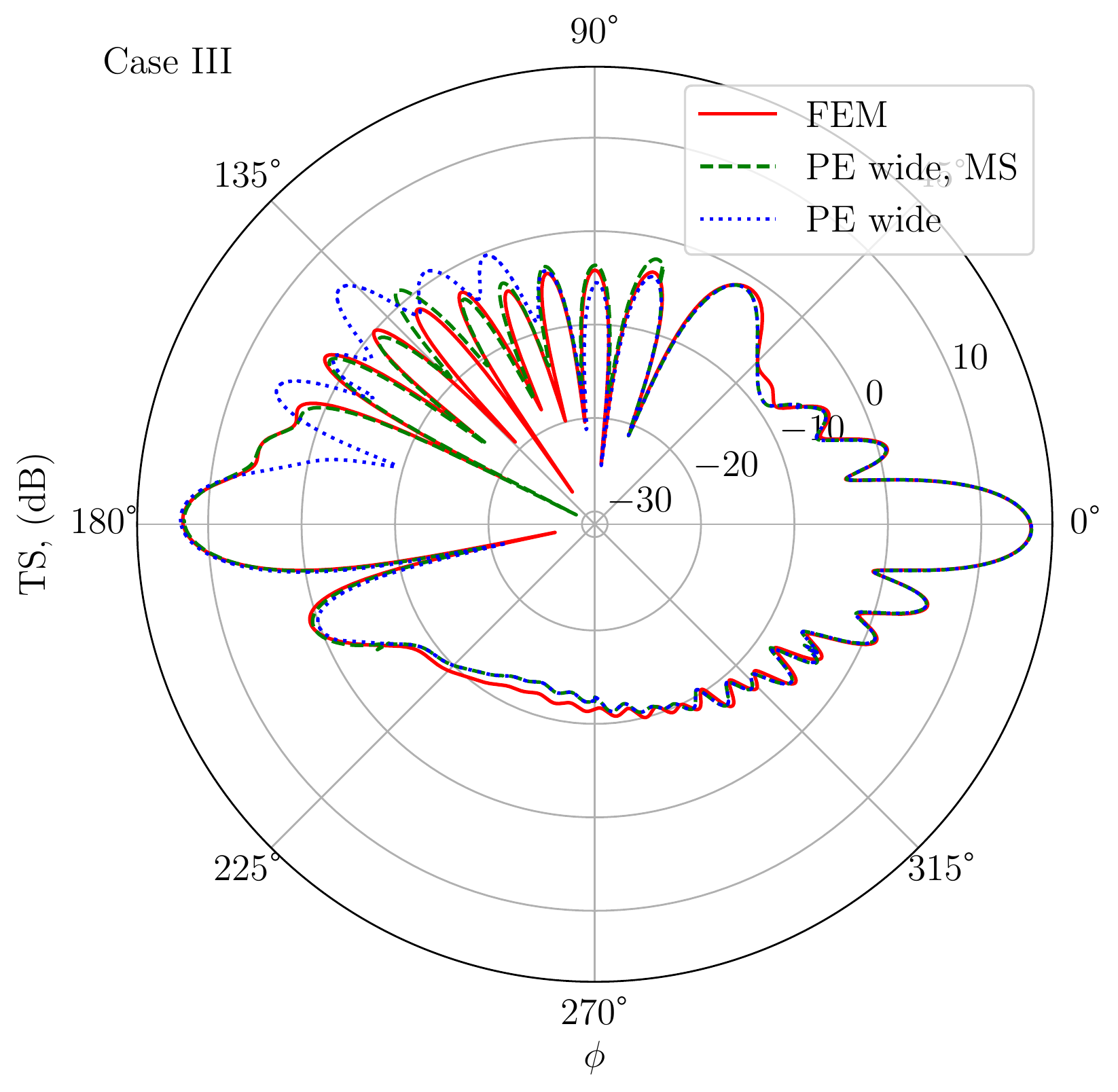}
\end{subfigure} 
\caption{Target strength of a soft L-shape for a plane wave of frequency 1500~Hz ($ka \approx 19$) for three different sets of parameters (see text).  Red solid lines are FEM calculations, blue dotted are wide-angle PE, and green dashed are wide-angle PE with multiple-scattering contributions.}\label{fig:LB_res}
\end{figure}

\begin{figure}[h!]
\begin{subfigure}[b]{.32\textwidth}
\includegraphics[width=\textwidth]{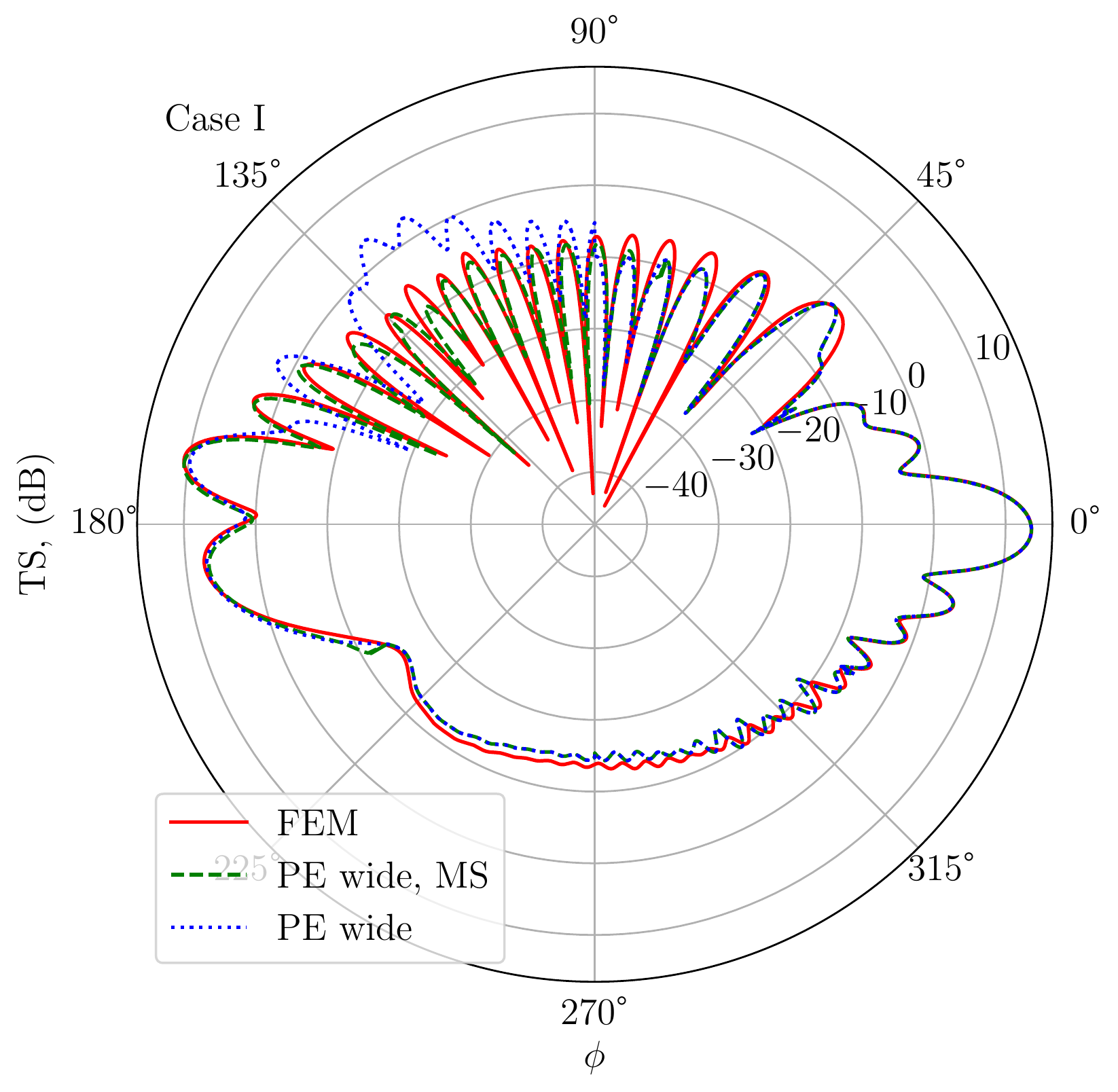}
\end{subfigure}
\begin{subfigure}[b]{.32\textwidth}
\includegraphics[width=\textwidth]{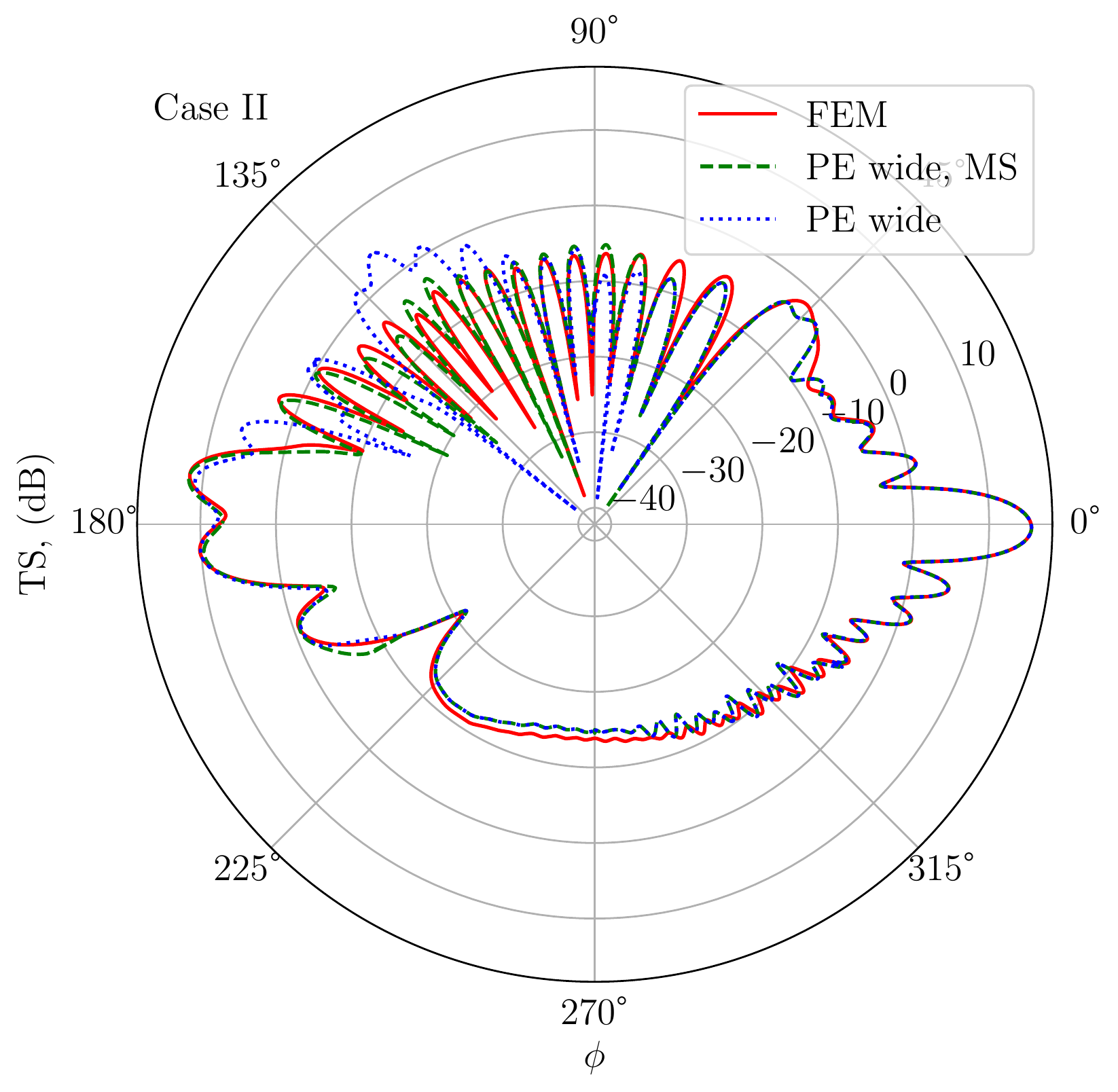}
\end{subfigure}
\begin{subfigure}[b]{.32\textwidth}
\includegraphics[width=\textwidth]{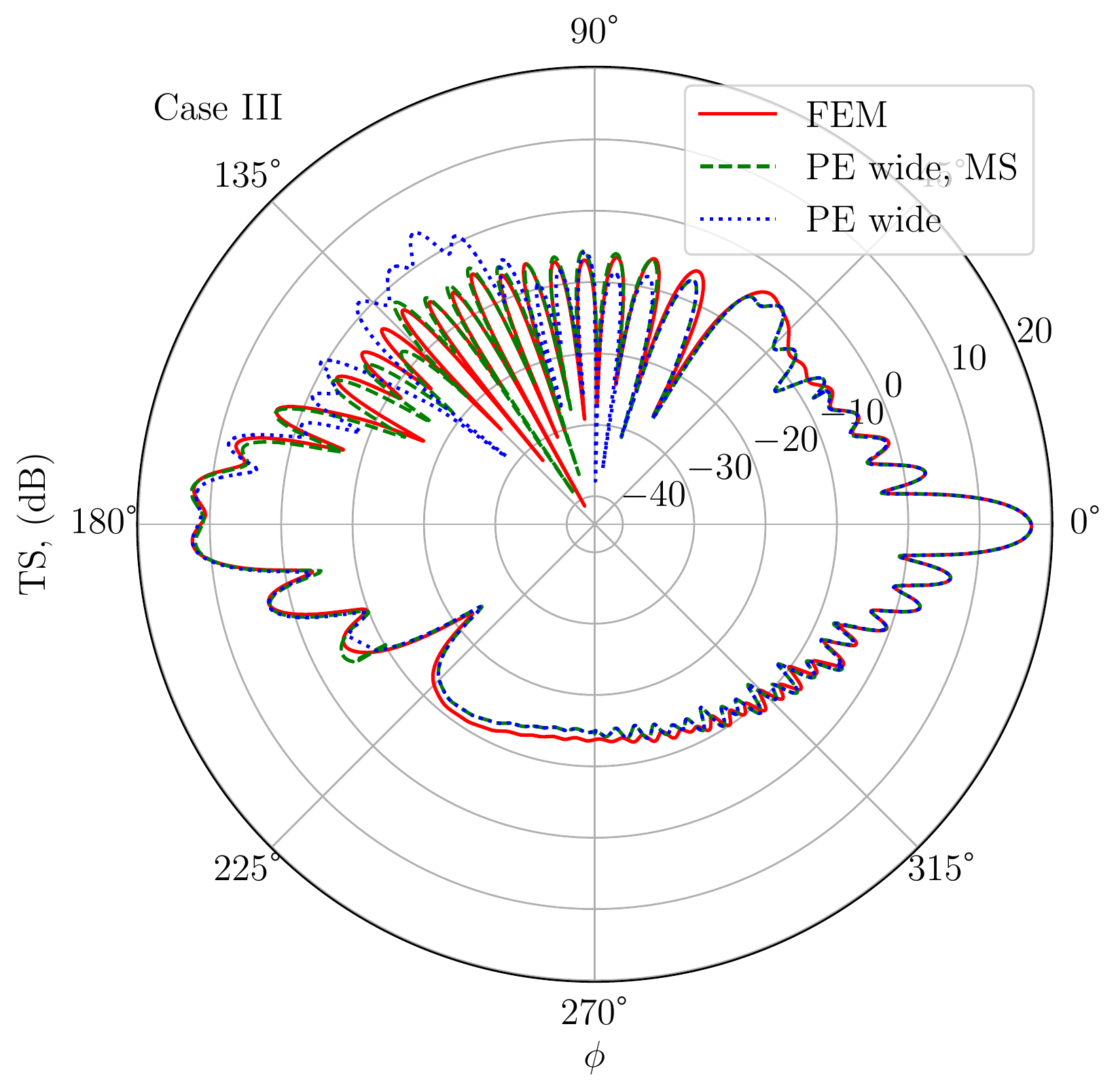}
\end{subfigure} 
\caption{Target strength of a soft L-shape for a plane wave of frequency 2500~Hz ($ka \approx 31$) for three different sets of parameters (see text). Red solid lines are FEM calculations, blue dotted are wide-angle PE, and green dashed are wide-angle PE with multiple-scattering contributions.}\label{fig:LB_res_2500}
\end{figure}

The discrepancy in target strength between the wide-angle formulation (dotted blue) and the FEM (solid red) is shown in Fig.~\ref{fig:LB_res}. 
If we modify the boundary conditions on the object by using the forward-scattered field from the horizontal portion of the L-shape (incident on the vertical part) as an additional source on the vertical part, then we can recover the features missed by the original method; the PE with the modified boundary data agrees much more with the FEM (compare dashed green and solid red lines). This agreement remains at higher frequencies; the results for 2500~Hz are shown in Fig.~\ref{fig:LB_res_2500}. 

We note that, if one is looking at the entire angular spectrum of the far-field pressure, this multiple-scattering approach {\em does not} require an additional PE run. For this L-shape, the scattered field from the forward-direction march is combined with the original incident field as a ``modified'' incident field, which is used to source the boundary conditions on the relevant vertical portions of the object when marching in the other (backward) directions.

\FloatBarrier

\begin{figure}[h!]
\begin{subfigure}[t]{.45\textwidth}
\includegraphics[width=\textwidth]{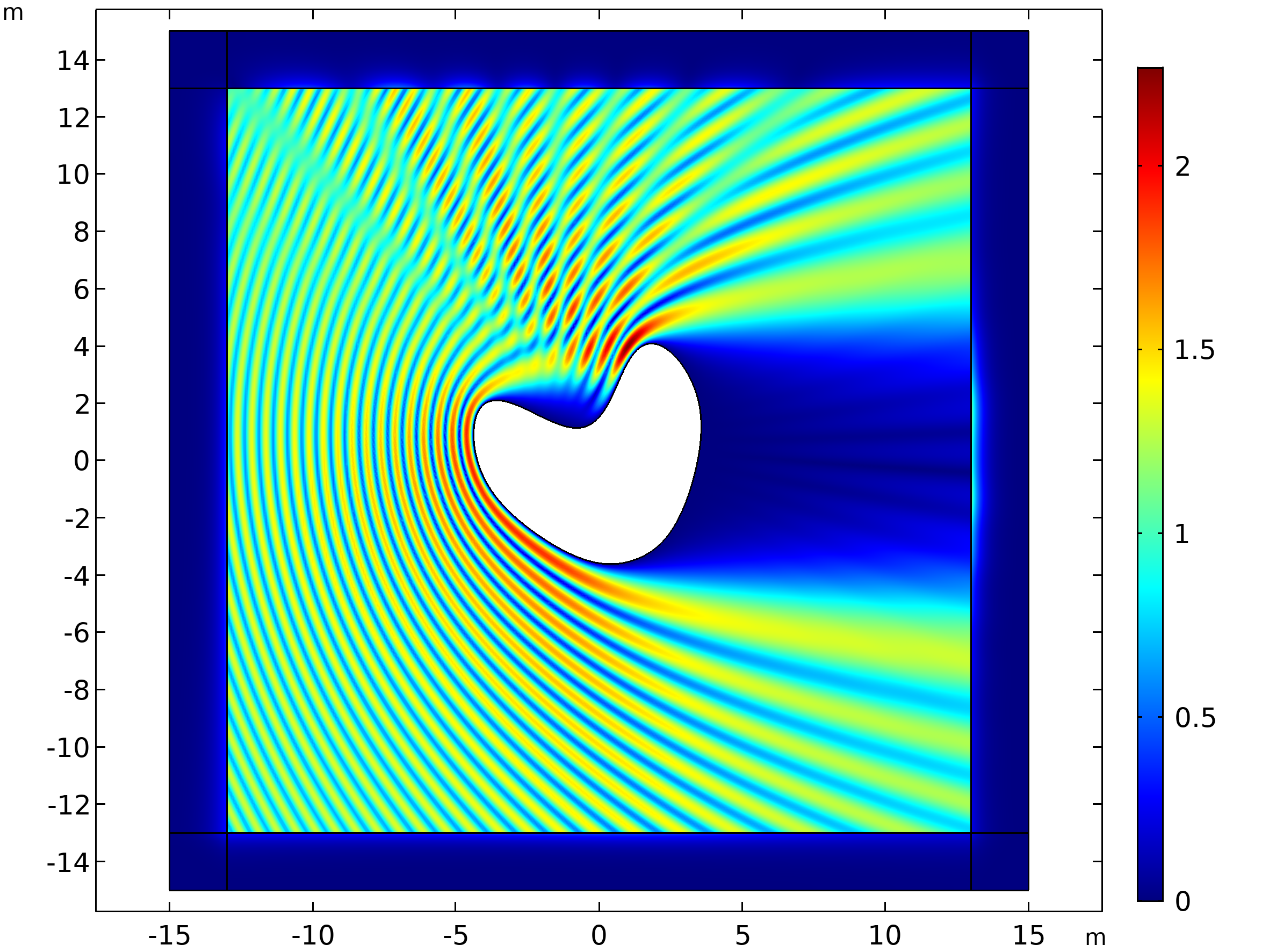}
\caption{}
\label{fig:beanIIIrot_FEM}
\end{subfigure}\qquad
\begin{subfigure}[t]{.4\textwidth}
\includegraphics[width=\textwidth]{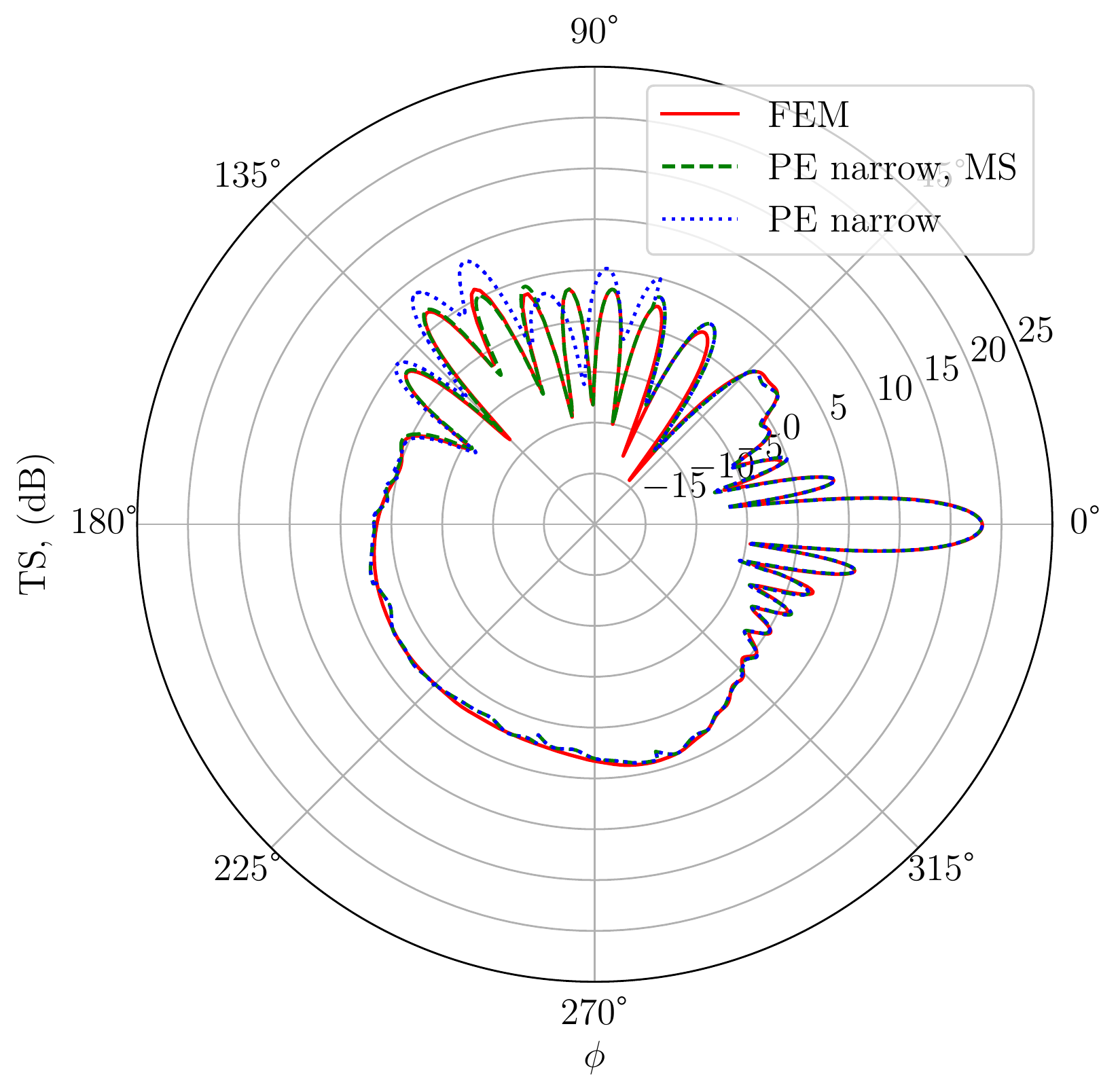}
\caption{} \label{fig:beanIIIrot20}
\end{subfigure}
\caption{Calculations for a plane wave of frequency 1500~Hz incident on a soft bean rotated by 20$^\circ$. (a) Full-field pressure (absolute value) computed using FEM. (b) Target strength computed using FE and PE methods.  Red solid lines are FEM calculations, blue dotted are narrow-angle PE, and green dashed are narrow-angle PE with multiple-scattering contributions.}
\end{figure}

In the ocean environment, realistic scattering situations (experiments) involve backscattering from incident pressure waves grazing an undulating ocean floor. 
One could suppose that the ocean floor could be modeled by a series of the bean shaped objects studied in this paper. 
As such, we can test the multiple-scattering algorithm for an incident grazing wave (taken in this case to be at an angle of 20$^\circ$) onto the Case III bean with soft boundary conditions. 
The FEM solution is shown in Fig.~\ref{fig:beanIIIrot_FEM}. 
We can clearly see the shadow zone caused by the leading lobe of the bean, which then modifies the field incident on the trailing lobe. 
Figure~\ref{fig:beanIIIrot20} shows the result of the narrow-angle PE calculation with and without multiple-scattering modification of the incident field boundary condition. 
The narrow-angle PE accurately captures the scattering in the perpendicular and backward directions perfectly when including the multiple-scattering contribution. 

It is important to note that the use of the wide-angle approximation was not necessary to capture the multiple-scattering phenomena in this case. 
This is because the discrepancy between the standard PE method and the FEM  around the $\phi=2\pi/3$ direction is caused by the leading lobe modifying the field incident on the trailing lobe and is not due to a portion of the incident field being scattered outside of the paraxial cone.
In the former case, it is necessary to modify the sourcing fields on the boundary of the object when marching in the $\phi=2\pi/3$ direction to take into account this shadowing.
In instances where the latter applies, such as in the first example with the bean, it is more computationally efficient to use the wide-angle PE.

 \begin{figure}[h!]
\begin{subfigure}[t]{.45\textwidth}
\includegraphics[width=\textwidth]{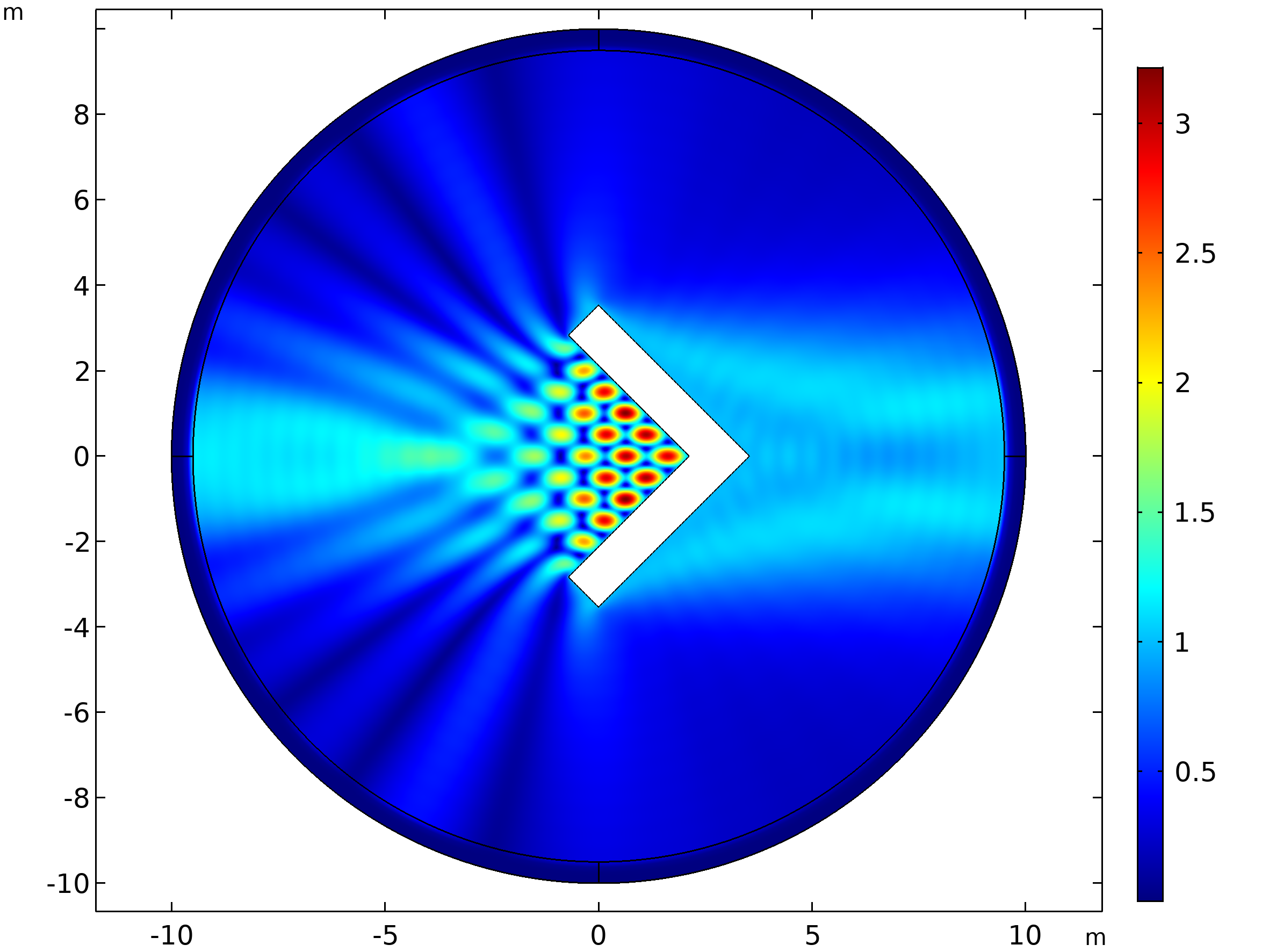}
\caption{}
\label{fig:V_FEM}
\end{subfigure}\qquad
\begin{subfigure}[t]{.4\textwidth}
\includegraphics[width=\textwidth]{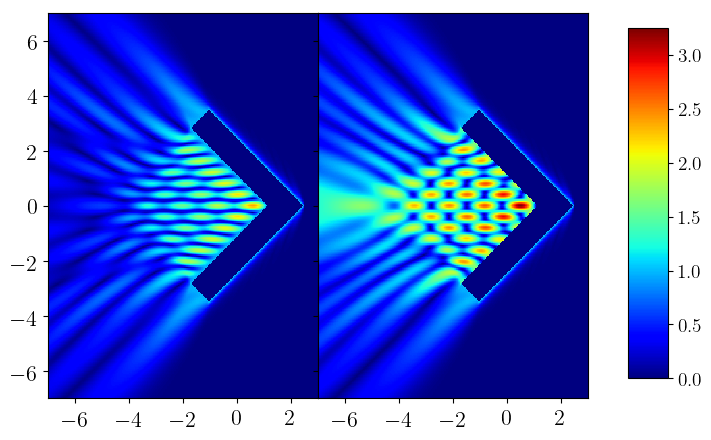}
\caption{} \label{fig:V_PE_comp}
\end{subfigure}\qquad
\begin{subfigure}[t]{.4\textwidth}
\includegraphics[width=\textwidth]{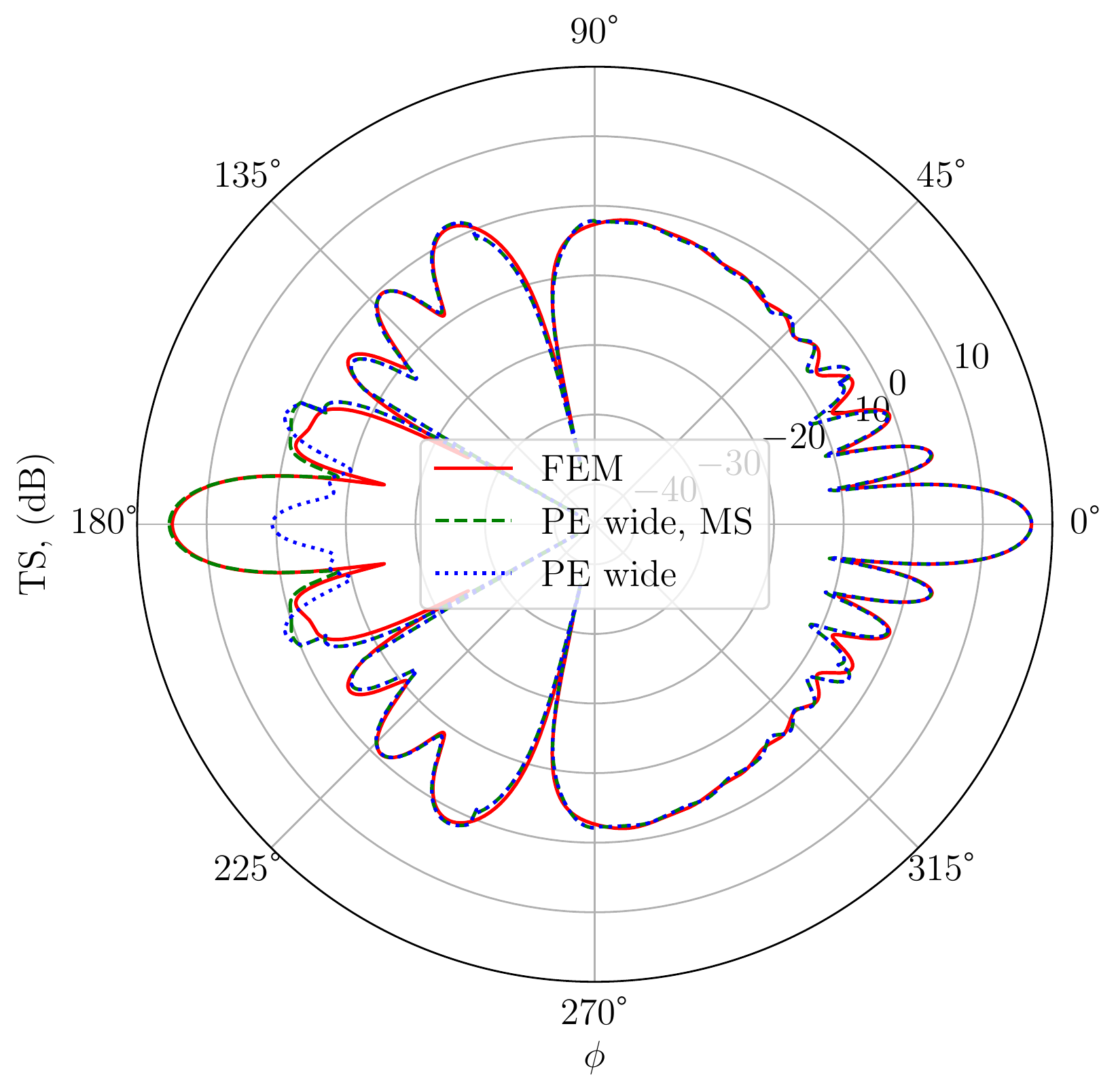}
\caption{} \label{fig:V_PE}
\end{subfigure}
\caption{Calculations for a plane wave of frequency 1500~Hz incident on a forward-pointing chevron with soft boundary conditions. (a) Full-field pressure (absolute value) computed using FEM. (b) Backscattered field (absolute value) computed using the PE method without (left) and with (right) multiple-scattering modification of the incident field. (c) Target strength computed using FE and PE methods.  Red solid lines are FEM calculations, blue dotted are wide-angle PE, and green dashed are wide-angle PE with multiple-scattering contributions.}
\end{figure}

\FloatBarrier

Finally, as a most extreme case, we can look at a forward-pointing chevron shape. The finite-element result for a plane wave incident on this object with soft boundary conditions is shown in Fig.~\ref{fig:V_FEM}. Clearly there are strong multiple-scattering effects, particularly in the backscattering direction. To source the correct scattered field in the backward direction, we use the scattered field calculated from the $\pi/2$ and $3\pi/2$ paraxial directions as additional incident sources on the ``legs.'' These effects are most prominent in the backscattered direction; the difference is shown in Fig.~\ref{fig:V_PE_comp}. The left plot shows backscattering (i.e. marching in the leftward direction) without multiple-scattering effects, while the right plot shows the backscattered field including multiple scatterings. By comparing the right plot of Fig.~\ref{fig:V_PE_comp} to Fig.~\ref{fig:V_FEM}, we see that the multiple scatterings give the correct interference pattern and backscattering peak enhancement. Figure~\ref{fig:V_PE} shows the target strength calculations of the PE with and without the multiple-scattering correction and the FEM benchmark. By including the multiple-scattering effects, the PE completely reproduces the backscattering peak that was absent in the original calculation. We note that, once again, if one is looking at the entire angular spectrum of the far-field pressure, then this multiple-scattering approach does not require any additional PE runs.

\section{Conclusions}

We have shown that the multisector parabolic equation scattering method yields accurate and efficient results for target strength calculations of a variety of scatterers in two and three dimensions.
Computational times are comparable to finite-element methods at lower frequencies or smaller objects, and are significantly faster at larger $ka$.  
We have shown how wide-angle and multiple-scattering approaches allow accurate modeling of the target strength of concave scatterers without a large increase in computational cost. 
The promising results of the multiple-scattering approach suggest that further development --- such as using iterative methods for multiple scatterings --- could yield good results for multiple objects in close proximity and for scatterers with more complex shapes and structures than those studied in this paper.

\vspace{1in}
{\bf Acknowledgements.}
This work is sponsored by the Office of Naval Research (ONR). AR thanks M. D. Collins, M. D. Guild, and J. F. Lingevitch for useful discussion and feedback. AR is supported through NRL's Jerome and Isabella Karle Fellowship Program. 
\FloatBarrier

\FloatBarrier

\appendix

\section{Soft and hard circles and spheres}\label{app:repl}

\begin{figure}[h!]
\begin{subfigure}[]{.33\textwidth}
\includegraphics[width=\textwidth]{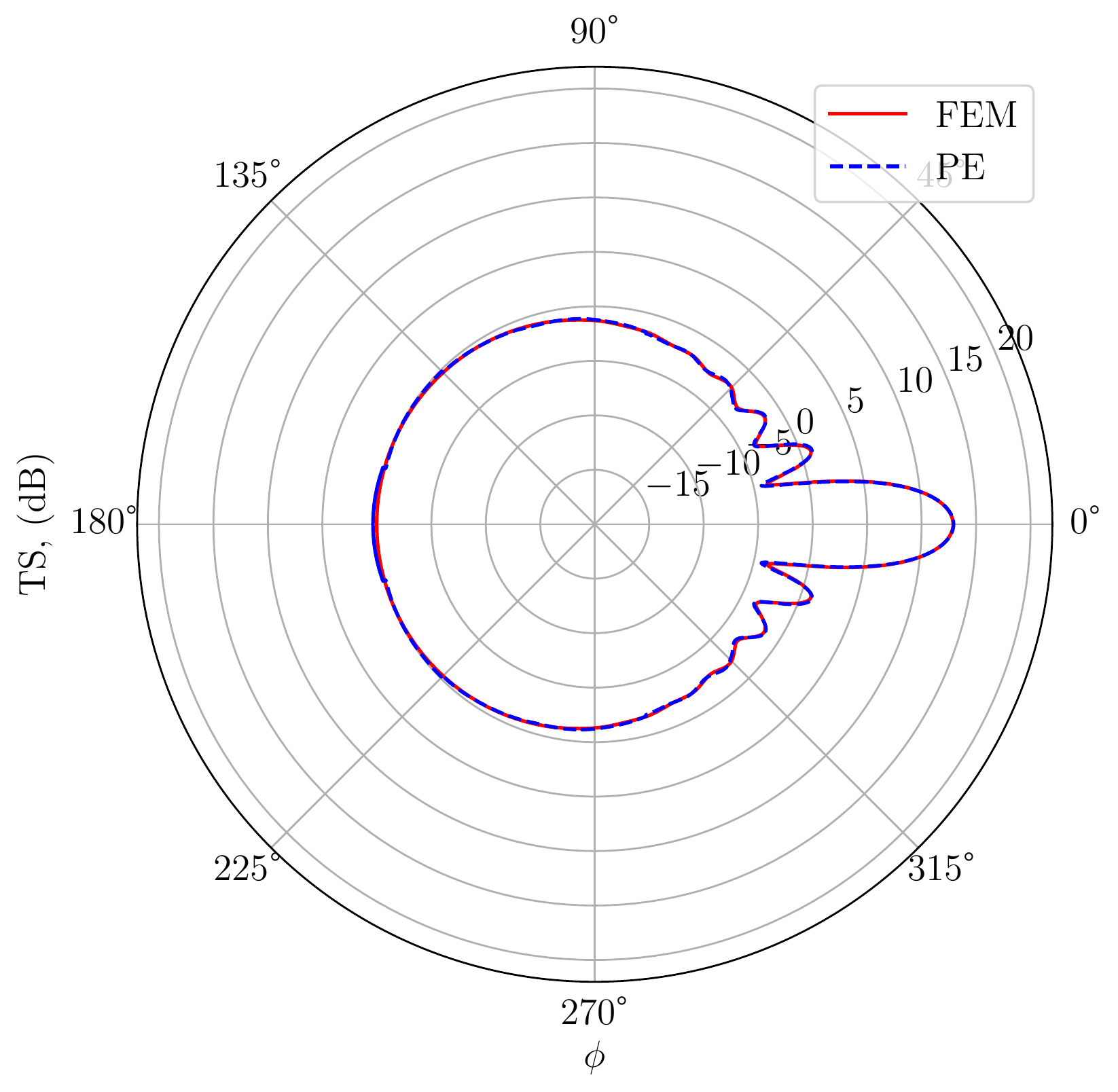}
\caption{}
\end{subfigure}\\
\begin{subfigure}[]{.33\textwidth}
\includegraphics[width=\textwidth]{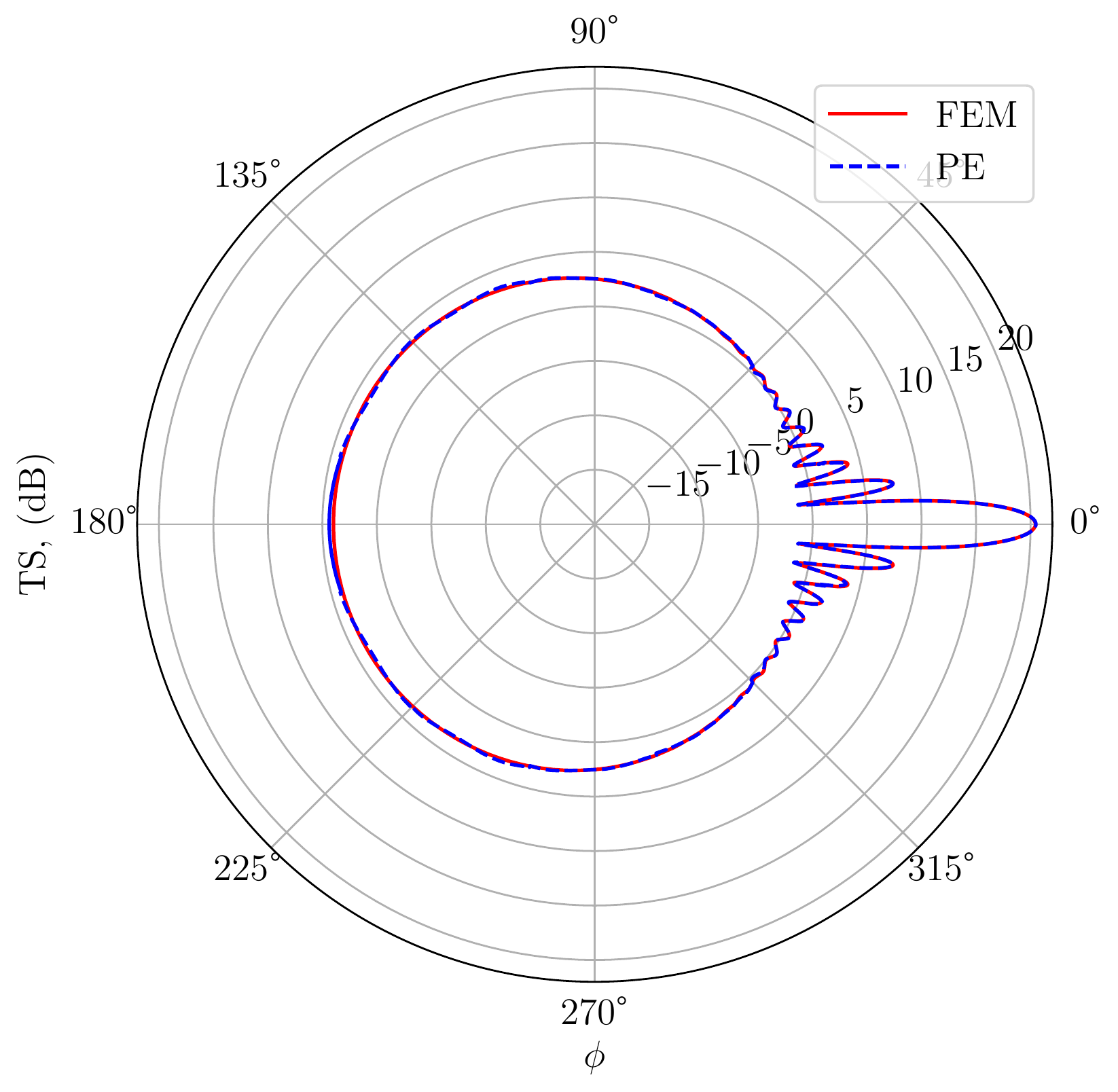}
\caption{}
\end{subfigure} 
\caption{Two-dimensional target strength calculations for a plane wave scattered from acoustically soft circles for (a) $ka = 4 \pi$ and (b) $ka = 10 \pi$. Dashed blue lines are from the multisector PE method, and solid red lines are finite-element results.}\label{fig:scircle}
\end{figure}

\begin{figure}[t!]
\begin{subfigure}[]{.33\textwidth}
\includegraphics[width=\textwidth]{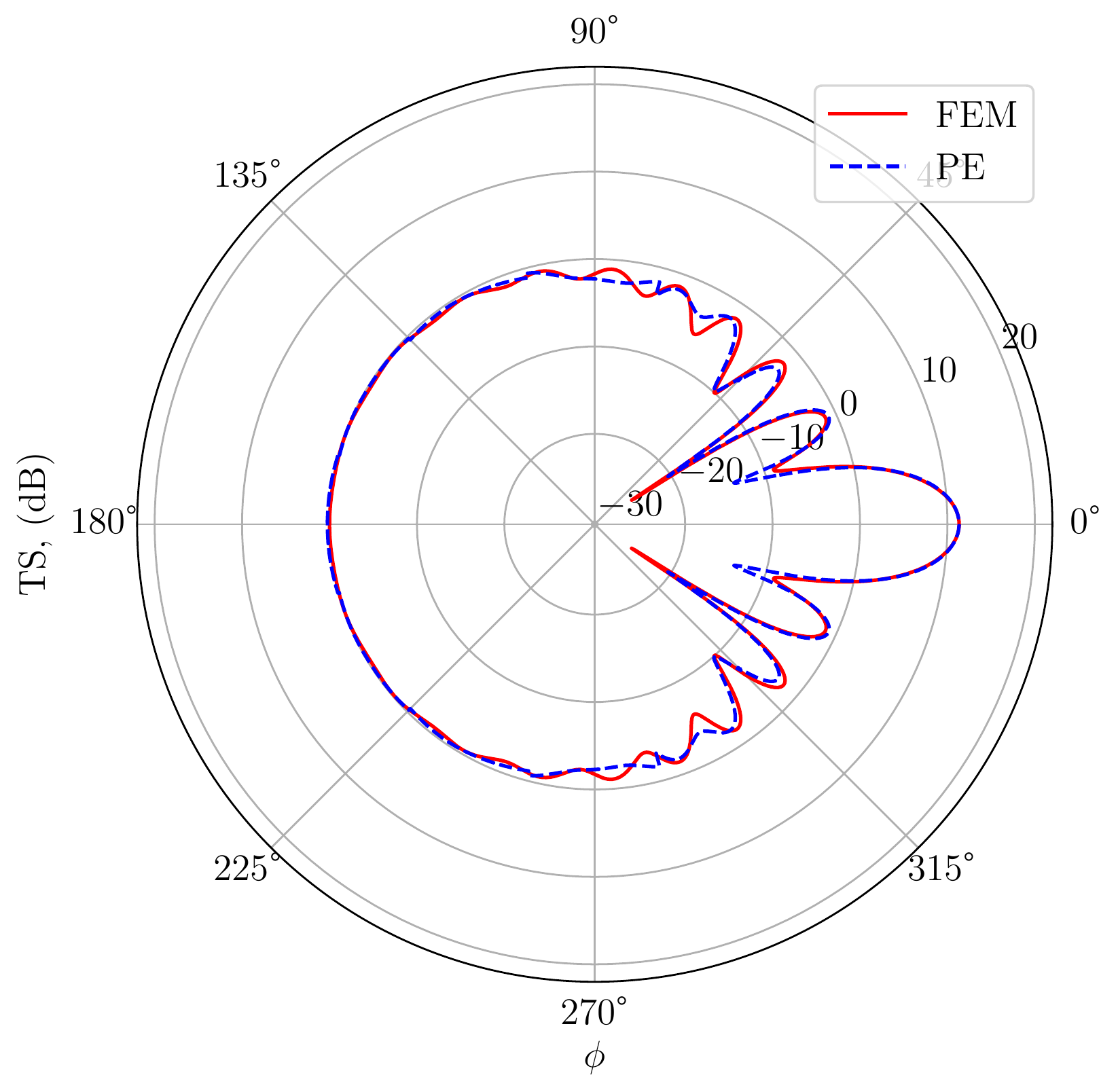}
\caption{}
\end{subfigure}\\
\begin{subfigure}[]{.33\textwidth}
\includegraphics[width=\textwidth]{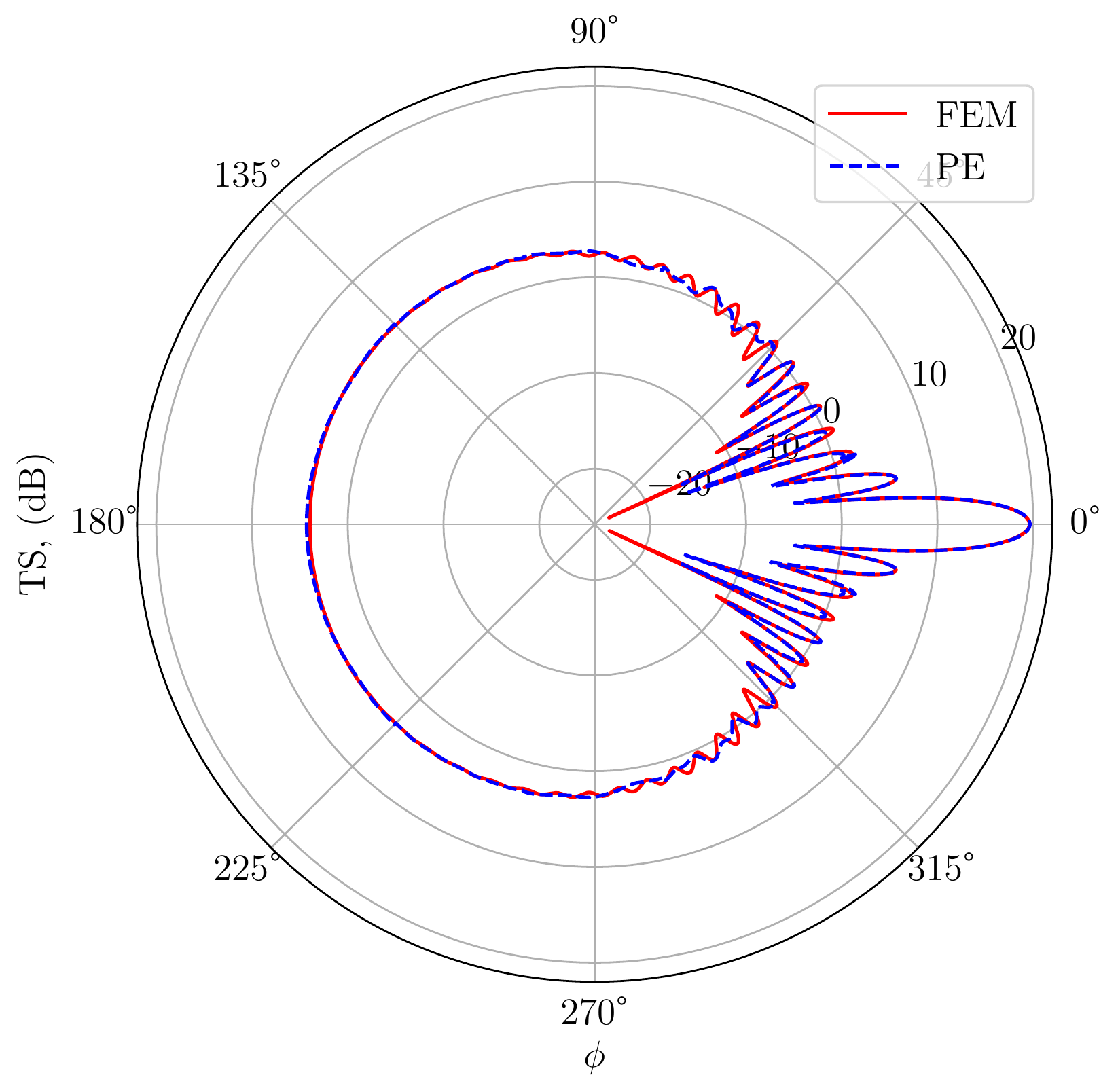}
\caption{}
\end{subfigure}
\caption{Two-dimensional target strength calculations for a plane wave scattered from acoustically hard circles for (a) $ka = 4 \pi$ and (b) $ka = 10 \pi$. Dashed blue lines are from the multisector PE method, and solid red lines are finite-element results.}\label{fig:rcircle}
\end{figure}

\begin{figure}[t!]
\begin{subfigure}[b]{.33\textwidth}
\includegraphics[width=\textwidth]{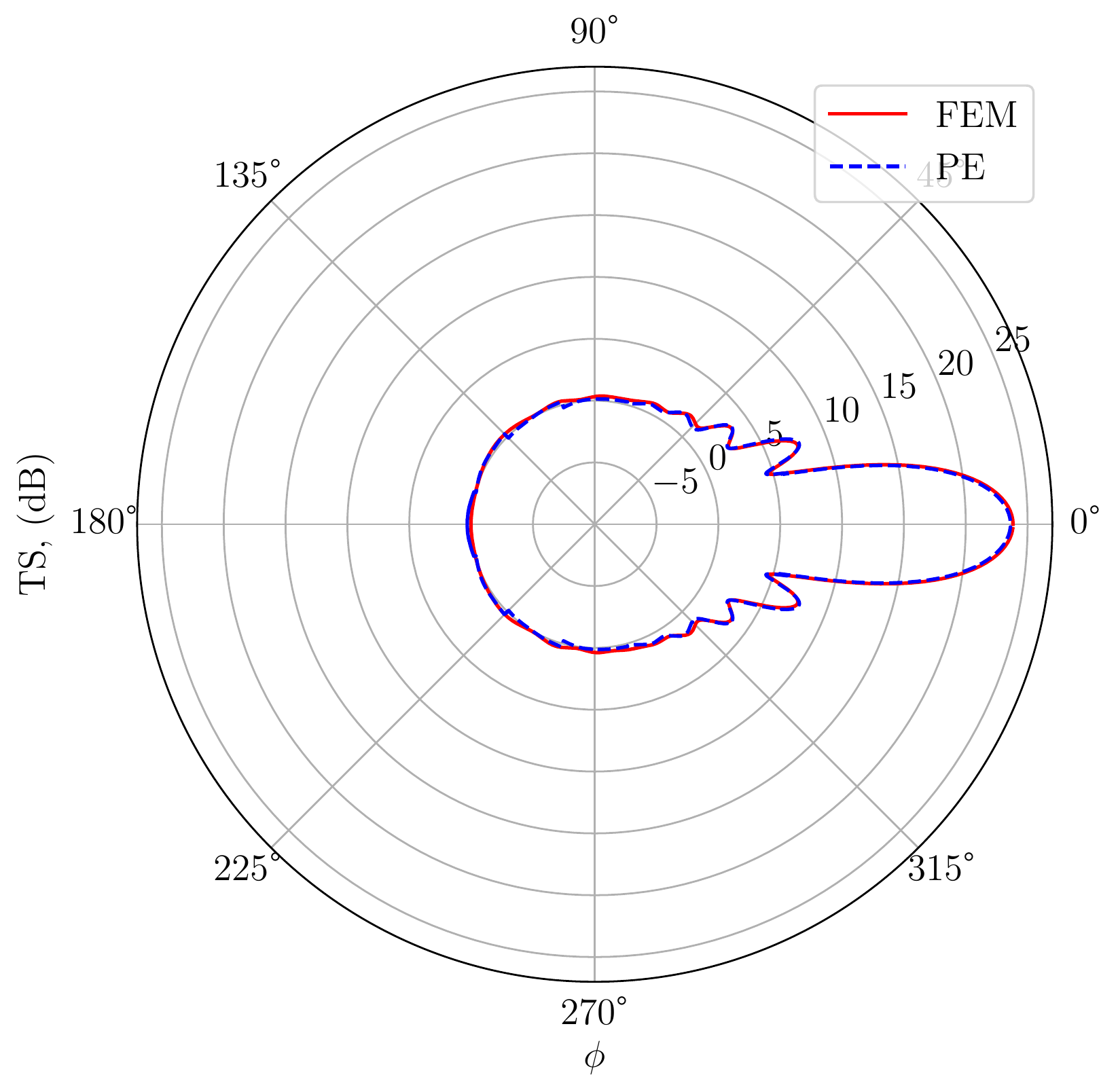}
\caption{}
\end{subfigure} \\ 
\begin{subfigure}[b]{.33\textwidth}
\includegraphics[width=\textwidth]{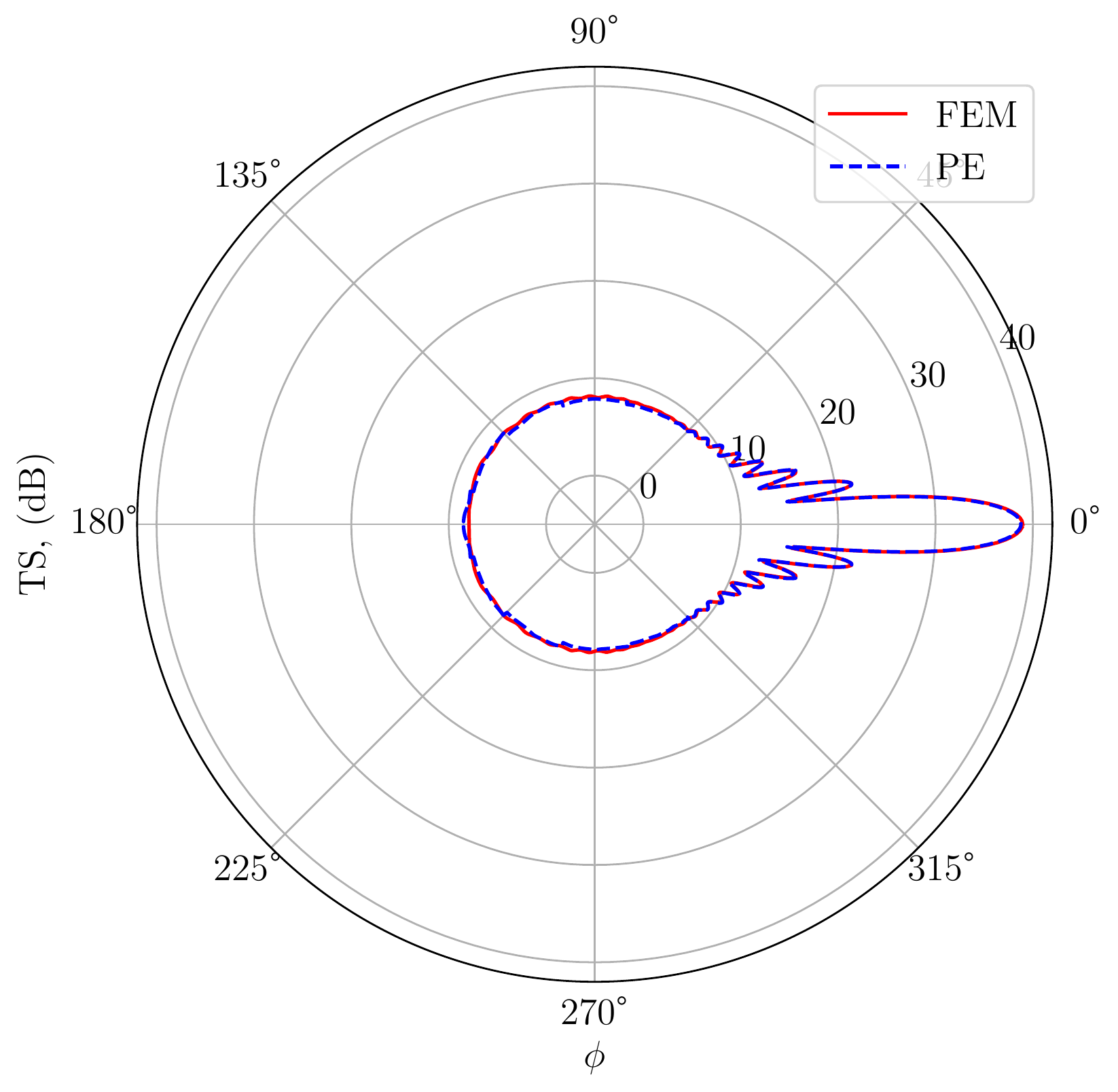}
\caption{}
\end{subfigure}
\caption{Three-dimensional target strength of spheres with soft boundary conditions for plane-wave incidence for (a) $ka = 4 \pi$ and (b) $ka = 10 \pi$. Dashed blue lines are from the multisector PE method, and solid red are finite-element results.}\label{fig:ssphere}
\end{figure}

\begin{figure}[h!]
\begin{subfigure}[b]{.33\textwidth}
\includegraphics[width=\textwidth]{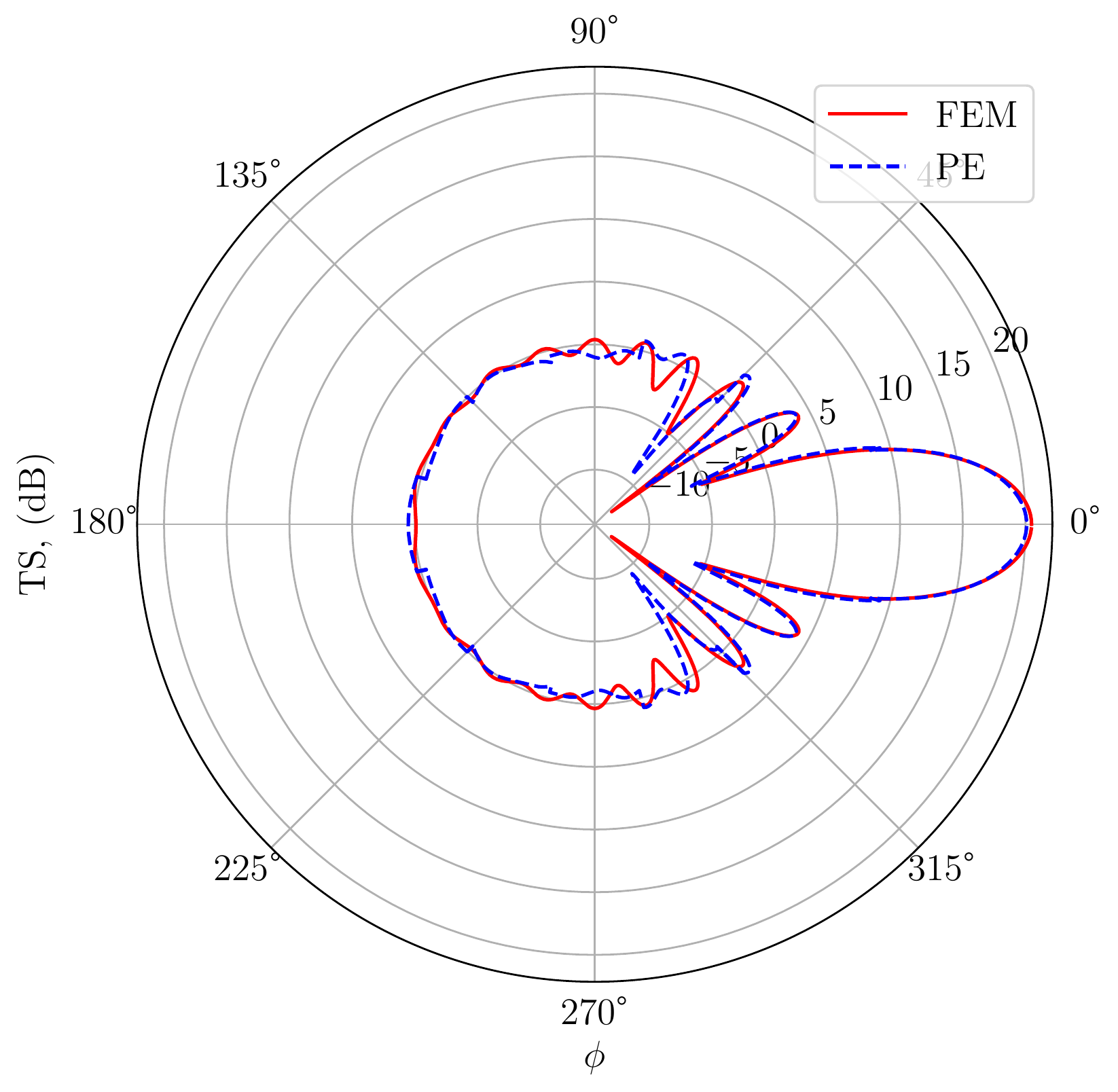}
\caption{}
\end{subfigure}
\begin{subfigure}[b]{.33\textwidth}
\includegraphics[width=\textwidth]{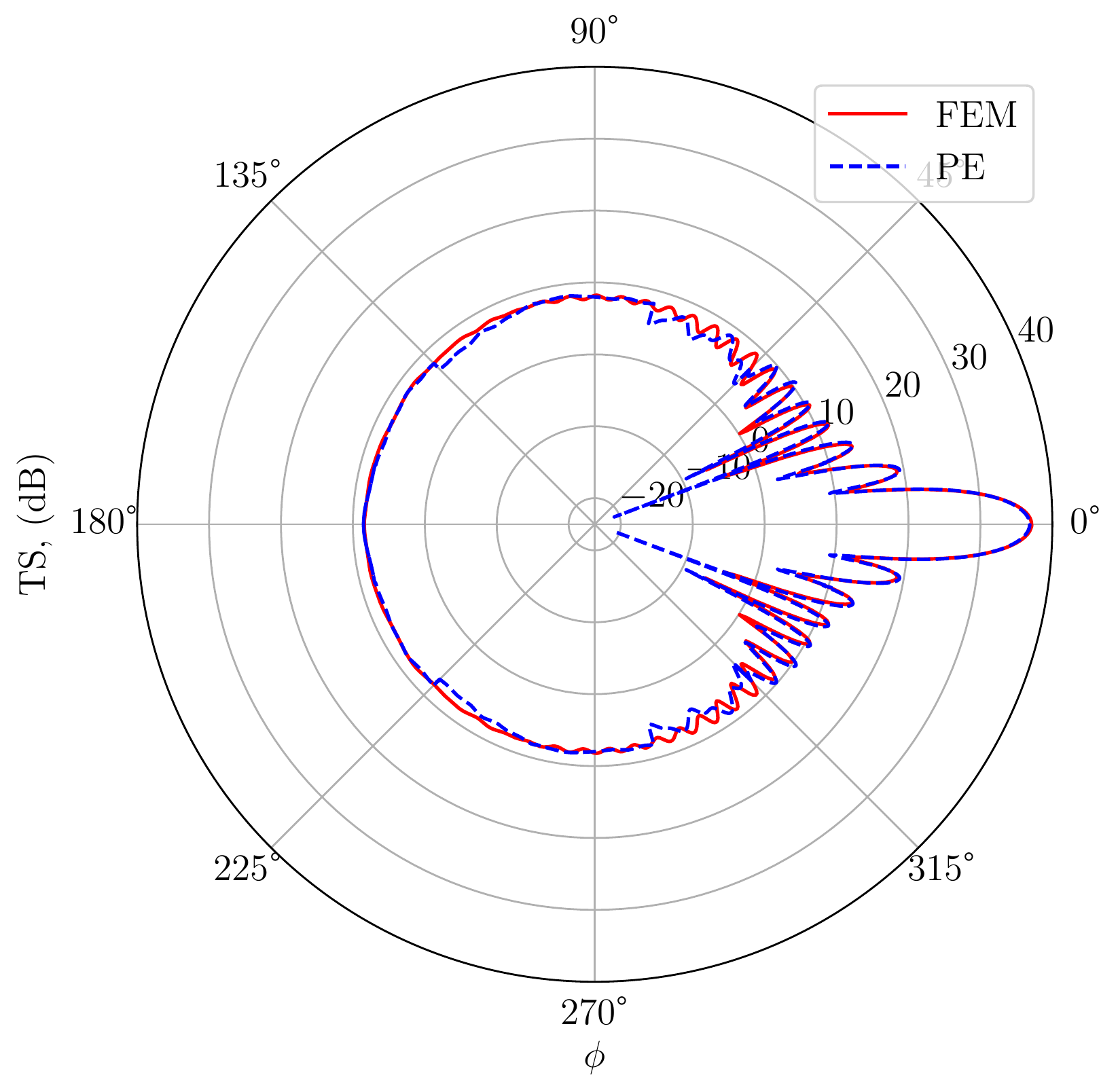}
\caption{}
\end{subfigure}
\caption{Three-dimensional target strength of spheres with hard boundary conditions for plane-wave incidence for (a) $ka = 4 \pi$ and (b) $ka = 10 \pi$. Dashed blue lines are from the multisector PE method, and solid red are finite-element results.}\label{fig:rsphere}
\end{figure}

In this appendix, we replicate the results of Ref.~\cite{levy1998target} for soft and hard circles (2D) and spheres (3D). 
The target strength calculation results are shown  in Figs.~\ref{fig:scircle},~\ref{fig:rcircle},~\ref{fig:ssphere},and ~\ref{fig:rsphere} for soft circles, hard circles, soft spheres, and hard spheres, respectively, with (a) for $a = 2$ m ($ka = 4 \pi$) and (b) for $a = 5$ m ($ka = 10 \pi$). 
The numerical results from the PE calculation match very well with the finite-element calculation, though there are slightly more discrepancies in the hard case with smaller $ka$. 
This discrepancy is possibly due to the fact that PE properly induces creeping waves on a hard object, but those that travel more than once around the circumference of the object are not accurately captured by the PE method \cite{levy1998target}.

\FloatBarrier

\end{document}